\documentclass[preprint,nofootinbib,aps,superscriptaddress,eqsecnum]{revtex4-1} 
\pdfoutput=1
\usepackage{graphicx,rotate}
\usepackage{amsmath}
\usepackage{aurical}
\usepackage[T1]{fontenc}
\usepackage{amsfonts}
\usepackage{amssymb}
\usepackage{ulem}
\usepackage{comment}
\usepackage{feynmp}
\graphicspath{{./Diagrams/}}

\usepackage[bookmarks=true,bookmarksopen=true,
                    bookmarksnumbered=true,
                    colorlinks=true,linkcolor=red,
                    citecolor=red]{hyperref}

\usepackage{caption}
\usepackage{subcaption}
\usepackage{color}

\definecolor{darkcyan}{cmyk}{1,0,0,0.4}

\definecolor{darkred}{cmyk}{0,1,1,0.4}

\definecolor{darkgreen}{cmyk}{1,0,1,0.4}

\allowdisplaybreaks
\def\bea{\begin{eqnarray}}
\def\eea{\end{eqnarray}}
 \def\be{\begin{equation}}
\def\ee{\end{equation}}
\def\nn{\nonumber}
\def\mpl{M_{\rm Pl}}
\newcommand{\beq}{\begin{equation}}
\newcommand{\eeq}{\end{equation}}

\usepackage[utf8]{inputenc}
\DeclareUnicodeCharacter{2212}{-}

\begin{document}
 
 
\title{A model of light pseudoscalar dark matter}

\author{Shreyashi Chakdar}
\email{schakdar@holycross.edu}
\affiliation{Department of Physics,College of the Holy Cross, Worcester, MA 01610, USA}

\author{ Dilip Kumar Ghosh}
\email{tpdkg@iacs.res.in}
\affiliation{School of Physical Sciences, Indian Association for the Cultivation of Science,
2A $\&$ 2B Raja S.C. Mullick Road, Kolkata 700032, India}

\author{P. Q. Hung}
\email{pqh@virginia.edu}
\affiliation{Department of Physics, University of Virginia, Charlottesville, VA 22904-4714, USA}

\author{\small Najimuddin Khan} 
\email{nkhan.ph@amu.ac.in }
\affiliation{\small Department of Physics, Aligarh Muslim University, Aligarh-202002, India}

\author{Dibyendu Nanda}
\email{dnanda@kias.re.kr}
\affiliation{School of Physical Sciences, Indian Association for the Cultivation of Science,
2A $\&$ 2B Raja S.C. Mullick Road, Kolkata 700032, India}
 
\begin{abstract}
The EW-$\nu_R$ model was constructed in order to provide a seesaw scenario operating at the Electroweak scale $\Lambda_{EW} \sim 246$  GeV, keeping the same SM gauge structure. In this model, right-handed neutrinos are non-sterile and have masses of the order $\Lambda_{EW}$. They can be searched for at the LHC along with heavy mirror quarks and leptons, the lightest of which has large decay lengths. The model also incorporates a rich scalar sector, consistent with various experimental constraints, predicts a $\sim 125$ GeV scalar with the SM Higgs characteristics satisfying the current LHC Higgs boson data. The seesaw mechanism requires the existence of a complex scalar which is singlet under the SM gauge group. The imaginary part of this complex scalar denoted by $A^{0}_s$ is proposed to be the sub-MeV dark matter candidate in this manuscript.  We find that the sub-MeV scalar can serve as a viable non-thermal feebly interacting massive particle (FIMP)-DM candidate. This $A_s^0$ can be a naturally light sub-MeV DM candidate due to its nature as a pseudo-Nambu-Goldstone (PNG) boson in the model. We show that the well-studied freeze out mechanism falls short in this particular framework producing DM overabundance.  We identify that the freeze in mechanism produce the correct order of relic density for the sub-MeV DM candidate satisfying all applicable constraints.  We also discuss the allowed parameter space arising from the current indirect search bounds for this sub-MeV DM. 

\end{abstract}
 
\pacs{}
\maketitle

\section{Introduction and Framework}
\label{sec:intro}

The various astronomical and cosmological evidences, like the galaxy 
rotation curves \cite{Zwicky:1933gu,Rubin:1970zza}, 
gravitational lensing \cite{Clowe:2006eq},
the bullet cluster \cite{Massey:2010hh} etc., vindicate the 
existence of the dark matter (DM). The Planck Collaboration \cite{Aghanim:2018eyx} 
using the precise map of the cosmic microwave background (CMB) indicates that DM contributes almost $26\%$ to the 
mass/energy budget of the Universe. 
These observations can be explained by postulating the existence 
of particles that interact either very weakly with ordinary particles or only 
through gravity or any other forces potentially not part of the Standard Model (SM). 
The limitation of the standard model of particle physics to 
account for a viable DM candidate opens up the pathway of various plausible
scenarios beyond the standard model (BSM) to resolve the DM puzzle.\\
The most sought after candidate for the DM in any BSM scenario 
is the so-called weakly interacting massive particle (WIMP) with mass 
close to the Electroweak scale and weak couplings with the 
SM particles. In this scenario, once WIMPs are thermally produced 
at the very early Universe, they remain in thermal equilibrium with the SM bath up to a certain 
temperature  depending on its mass and interaction strength. Eventually they decouple from the thermal bath at some temperature $T_f$,
which is commonly called the {\it Freeze-out} temperature, where the DM 
interaction rate drops below the expansion rate of the Universe, governed
by the Hubble parameter $H$. At that point, the DM freezes-out from the 
SM thermal bath providing the observed DM relic density, which is affected only
by the expansion of the Universe. It is interesting to note that in the 
WIMP scenario, the theoretical prediction for the DM thermal relic density coincides with the observed DM relic abundance 
($\Omega h^2 \sim 0.12$~\cite{Aghanim:2018eyx}), popularly known as the {\it WIMP-miracle}.
However, various null results coming from the WIMP searches at LHC~\cite{Giagu:2019fmp,Aaboud:2018xdl,Aaboud:2017phn,Aaboud:2017dor,Aaboud:2017bja}, at the spin independent and dependent WIMP-nucleon
scattering experiments at LUX \cite{Akerib:2016vxi}, PANDAX-II \cite{Cui:2017nnn}, XENON-1T \cite{Aprile:2018dbl},
PICO \cite{Amole:2019fdf} etc. and from the indirect detection coming from FERMI-LAT \cite{Daylan:2014rsa} MAGIC \cite{Ahnen:2016qkx} and 
PLANCK \cite{Aghanim:2018eyx} experiments have already excluded a significant part of the parameter space in the WIMP-nucleon scattering cross-section vs WIMP mass plane, casting doubt on the simplest WIMP hypothesis.
As the various WIMP searches continue to provide null results, one approach is to design and implement completely new search techniques for WIMP detection different from the aforementioned scattering experiments \cite{Lin:2019uvt} or to deviate from the WIMP paradigm altogether towards an alternative scenario, where the DM particle has very feeble interactions with SM particles and never enter the thermal bath. 
In this scenario, the DM obtain their relic abundance very slowly through the decay and/or the annihilation of the bath particles, known as the {\it Freeze-in} process, providing a completely different explanation for the DM conundrum \cite{Hall:2009bx,Bernal:2017kxu}. 
For such cases, the production of DM from the decays will contribute dominantly if the same couplings are involved in both decay and annihilation. Due to such feeble interactions, it is extremely challenging to detect the DM in the present direct search experiments. However, many new experimental techniques with low-threshold direct detection\cite{Essig:2011nj, Essig:2017kqs, Emken:2017erx, Green:2017ybv, Essig:2015cda,  Essig:2012yx} have been proposed having the capability to probe FIMPs in the near future. Particularly, for a MeV scale feebly interacting massive particle (FIMP) with interaction with electron could be tested by next generation experiments\cite{Essig:2011nj,Essig:2017kqs, Bernal:2017mqb} constraining the typical DM-electron cross sections. In general for a light DM, the number density have to be very large to satisfy the observed relic abundance enhancing their detection rate.\\
In this paper, we present a framework for a sub-MeV DM arising naturally from the theoretical construction of our model. It is to be noted that our detailed knowledge of nucleosynthesis puts a strong constraint on the thermal production of such sub-MeV DM. In particular, this kind of DM could be overabundant if produced by a thermal freeze-out mechanism and motivated by this fact, we have analyzed the non-thermal production of such MeV-scale DM. In building such a framework for the sub-MeV DM, it is legitimate to ask how natural is it to have such a light (sub-MeV) particle and the ways it could satisfy the aforementioned experimental constraints.  This work is based on a well studied framework (the EW-$\nu_R$ model) proposed by one of us \cite{Hung:2006ap} for non-sterile right-handed neutrinos with Majorana masses being proportional 
to the Electroweak scale $\Lambda_{EW}= 246.21 \,$ GeV and as a result, can be produced and hunted at the Large Hadron Collider (LHC). 
To see how the EW-$\nu_R$ model generates a light DM particle, a brief summary of the seesaw mechanism of the model is in order. 
The model contains mirror fermions and, for the purpose of this introduction, it is sufficient to discuss one 
generation of SM leptons: $\psi_L=(\nu_\ell, \ell)_L^T$; $\ell_R$, and mirror leptons: 
$\psi^{M}_R= (\nu_\ell^M, \ell^{M})_R^T$; $\ell^{M}_L$. The Majorana mass for right-handed neutrinos is obtained by 
coupling $\ell^{M}_R$ to a complex Higgs triplet $\tilde{\chi}$ as $ y_M \, 
\psi^{M,T}_R  \sigma_2 (\tau_2 \ \tilde{\chi}) \psi^M_R$.
With $\langle 
\tilde{\chi} \rangle=v_M$, one obtains $M_R = \frac{y_M v_M}{\sqrt{2}} $. Right-handed neutrinos 
in this model, being non-sterile, have to be heavier than $M_Z/2$ constrained by the 
$Z$-width data, implying that $v_M \propto 
\Lambda_{EW}$ severely affects the custodial symmetry which ensures that 
$M_W = M_Z \cos \theta_W$ at tree-level. Custodial symmetry is restored by the 
introduction of real triplet $\xi$ having the same VEV as $\tilde{\chi}$. 
(It turns out that this real triplet also provides a solution for a 
topologically stable, finite energy Electroweak monopole 
\cite{Hung:2020vuo, Ellis:2020bpy}.) The Dirac mass term in this seesaw 
mechanism comes from the coupling of a SM left-handed lepton doublet, a 
mirror right-handed lepton doublet with a {\it complex singlet scalar} 
$\Phi_s$:  $ y_{s\ell} \,\bar{\ell}_L \ \Phi_s \ell_R^M 
+ {\rm h.c.}$, giving $m_D = y_{s\ell} \ v_s$ when 
$\langle \Phi_s \rangle= v_s$. As we show below, the imaginary part of this 
complex singlet $\Phi_s$ which is a pseudo-Nambu-Goldstone (PNG) boson, $A_s^0$,
can be used as a light DM candidate, for the simple reason that $A_s^0$ would be massless when a global symmetry present in the EW-$\nu_R$ model is spontaneously broken. $A_s^0$ acquires a mass when there is a explicit breaking term in the scalar potential. As one has encountered similar situations in various other scenarios such as chiral symmetry breaking, the explicit breaking term is characterized by some mass scale which is usually assumed to be much smaller than the scale of spontaneous symmetry breaking (SSB). For example, hadronic $SU(2)_L \times SU(2)_R$ is an approximate symmetry with the SSB scale $\Lambda_{QCD} \approx 300 {\rm MeV}$ and the scales of explicit breaking are the up and down quarks masses, $2.3$ MeV and $4.8 $ MeV. As we shall see below,  the SSB scale of the global symmetry in this framework is of the order of the Electroweak scale and the explicit breaking scale is the mass of $A_s^0$ which will be assumed to be in the sub-MeV region making this PNG boson $A_s^0$ an ideal sub-MeV DM candidate.\\
It will be shown that the production of $A_s^0$ through the freeze-out mechanism is not favored here since it will result in overabundance. We will further show that the Freeze-in mechanism is the most attractive alternate possibilities in this scenario. As $A_s^0$ interacts extremely feebly with other particles (FIMP), it can be non-thermally produced through the Freeze-in mechanism, yielding the correct relic density and satisfying current constraints from the indirect searches. The model contains a large parameter space exhibiting the aforementioned behavior of the DM candidate. From a particle physics point of view, this DM scenario has a very interesting implication concerning the seesaw mechanism of the EW-$\nu_R$ model: the symmetry breaking scales proportional to the Dirac and Majorana masses as described in the previous paragraphs are found to be comparable in sizes, avoiding the kind of hierarchy found in a generic seesaw mechanism.\\
The organization of the paper is as follows. In Section II, 
we present the framework for the rich scalar sector for the EW-$\nu_R$ model.
Section III contains the influence of various theoretical and experimental constraints on 
the model parameters. In Section IV we present our selection of benchmark points and discuss the 
various LHC bounds examining the stability of this kind of light dark matter candidate. 
Section V discusses how $A_s^0 (\equiv {\rm Im (\Phi_s)})$, a pseudo-Nambu Goldstone boson can successfully play the role of a potential sub-MeV dark matter in this framework and section VI includes the mechanism for the FIMP dark matter candidate $A_s^0$ producing the correct relic density. Section VII discusses the constraints pertaining to the indirect searches. 
The summary and implications are presented in Section VIII.

\section{Extended Scalar sector of the model }
\label{sec:scalar_sect}
The main idea of the EW-$\nu_R$ model~\cite{Hoang:2014pda} containing mirror 
fermions including Majorana masses for right-handed neutrinos proportional to the Electroweak scale 
with an extended scalar sector is very appealing. Unlike the Standard Model, 
the framework is not only left-right symmetric, but each left handed fermion multiplet is accompanied by
new right handed fermion multiplet of opposite chirality. 
The framework contains a rich scalar sector incorporating four doublets (two belonging to the Two Higgs doublet model-THDM like $\Phi_1,\Phi_2$, two for mirror sector $\Phi_{1M}, \Phi_{2M}$), two triplets $\tilde{\chi}$, $\xi$ and one complex singlet $\Phi_s$ is represented by
\allowdisplaybreaks
\begin{eqnarray}
\Phi_1 &=& \begin{pmatrix}
 \phi_{1}^{0,*} & \phi_{1}^+ \\ 
 \phi_{1}^- &  \phi_{1}^{0} 
\end{pmatrix},~
\Phi_{1M}= \begin{pmatrix}
 \phi_{1M}^{0,*} & \phi_{1M}^+ \\ 
 \phi_{1M}^- &  \phi_{1M}^{0} 
\end{pmatrix},~ 
\Phi_2 = \begin{pmatrix}
 \phi_{2}^{0,*} & \phi_{2}^+ \\ 
 \phi_{2}^- &  \phi_{2}^{0} 
\end{pmatrix},~
\Phi_{2M}= \begin{pmatrix}
 \phi_{2M}^{0,*} & \phi_{2M}^+ \\ 
 \phi_{2M}^- &  \phi_{2M}^{0} 
\end{pmatrix},\nn\\
~ \tilde{\chi} &=& \begin{pmatrix}
\chi^+/\sqrt 2 & \chi^{++}\\
\chi^0 & -\chi^+/\sqrt 2
\end{pmatrix}, ~ \xi = \begin{pmatrix}
 \xi^+, \xi^0,\xi^-
\end{pmatrix},~
\Phi_s
\label{eq:fieldSA}
\end{eqnarray}
The Higgs potential consisting of these scalars has a global 
$SU(2)_L \times SU(2)_R$ symmetry, under which the triplet and doublet 
scalars transform as (3,3) and (2,2)~\footnote{The transformation ($SU(2)_L$ triplet, $SU(2)_R$ triplet)$\equiv (3,3)$ and the doublet transformation is denoted by $(2,2)$.}. The Electroweak symmetry is 
spontaneously broken by the VEVs of the neutral component of the doublet and
triplet scalars. We denote VEVs of the scalar fields $\Phi_1, \Phi_2, 
\Phi_{1M}, \Phi_{2M} ,\chi $ and $ \Phi_s $ as 
$v_1, v_2, v_{1M},v_{2M}, v_{M}$ and $v_s$  respectively, where, $\chi$ field
is the combination of two triplet scalars $\tilde{ \chi} $ and $\xi $ (see
eqns.\ref{eq:fieldS}-\ref{eq:fieldS3}). The standard model vacuum expectation
value is given by, $v_{\rm SM} \equiv \sqrt {v^2_1 + v^2_2 + v^2_{1M}+v^2_{2M} +8 v^2_M} \approx 246 $ GeV. 
After the spontaneous Electroweak symmetry breaking, the 
VEVs are aligned in such a manner that there
still remains an unbroken $SU(2)_D $ custodial symmetry, ie. $SU(2)_L \times 
SU(2)_R \to SU(2)_D $ and we get six $SU(2)_D$ singlet 
CP-even Higgs like scalars $(H^0_1, H^0_2, H^0_{1M},H^0_{2M}, 
H^0_s, H^{0\prime}_1 )$, forming a $6\times 6$ matrix 
${\mathcal M}^{2}_{{\cal H}}$.
We obtain the 
physical eigenstates $\widetilde{H}^{''''}, $ $\, 
\widetilde{H}^{'''},$ $\, \widetilde{H}^{''}, $ $\, \widetilde{H}^{'},$ $\, 
\widetilde{H},$ $ \, \widetilde{H}_s$, in descending order of mass 
$(M_{\widetilde {H}^{''''}} > M_{\widetilde {H}^{'''}} >
M_{\widetilde {H}^{''}} > M_{\widetilde {H}^{'}} > M_{\widetilde {H}} > 
M_{\widetilde {H}_S} )$
after diagonalizing ${\mathcal M}^2_{{\cal H}}$
by an orthogonal matrix ${\Fontskrivan O}_H$. 
Among these new physical scalars, $\widetilde{H} $ behaves as the SM 
like Higgs boson with a mass of 125 GeV.    
This SM like Higgs boson state is expressed as a superposition of 6 weak eigenstates as:
$\widetilde{H} = {\Fontskrivan O}_H^{51} \, H_1^0 + {\Fontskrivan O}_H^{52} 
\, H_2^0 + {\Fontskrivan O}_H^{53} \, H_{1M}^0 + {\Fontskrivan O}_H^{54}\, 
H_{2M}^0 + {\Fontskrivan O}_H^{55} \, H_s^0 + {\Fontskrivan O}_H^{56}\,  
H_{1}^{0'} $.
The mixing angles ${\Fontskrivan O}^{51}_H$ 
and ${\Fontskrivan O}^{52}_H$ control the  
$\widetilde {H}$ couplings with various standard model particles and the SM 
like $SU(2)_L$ doublet scalars $H^0_1$ and $H^0_2$ respectively. 
The detailed analysis on the extended 
scalar sector of this present framework have been discussed in 
the Appendix~\ref{app:A}.

\section{Theoretical and Experimental constraints on the Model}
\label{sec:constraints}
The additional charged fermions and scalars in this model can lead to 
non-trivial implications on various existing experimental observations. 
Moreover, mathematical consistency of the model demands that various model 
parameters should satisfy certain theoretical conditions. In this section we discuss the various theoretical and experimental constraints crucial to the framework discussed in this paper. 
\begin{itemize}
\item {\bf Perturbativity :} The various quartic couplings in the Higgs potential are assumed to be perturbative: $\mid \lambda_i \mid < 4 \pi $, 
where $\lambda_i$s are defined in the section ~\ref{sec:scalar_sect}. The 
Yukawa couplings are taken to be $< \sqrt{4 \pi }$. 
\item {\bf Constraints from the Electroweak precision observables:} 
The additional mirror fermions and heavy $SU(2)$ doublet and triplet 
scalars contribute to the electroweak precision observables in this framwork, namely 
the oblique S,T, \& U parameters~\cite{Peskin:1991sw}.
It should be noted that the constraints arising from these parameters
have been presented in the 
Ref.~\cite{Hoang:2013jfa} in an earlier rendition of the model where only two 
Higgs doublets were incorporated (in addition to the two triplets and the singlet). 
As shown in Ref.~\cite{Hoang:2013jfa}, the positive contributions in the oblique (S and T) parameters stemming from the new mirror fermions can be offset by the negative contributions from the triplets present in the framework. Additional Higgs doublets will not alter this picture and we preserve the features of the earlier version of the model.                                   
Nevertheless, in our numerical analysis, we have applied the T-parameter constraints on the mass difference for the doubly, singly charged and neutral scalars and the the mirror $up-down$ fermions of the $SU(2)$ fermion  doublets,
$\Delta M_{5,ij} \equiv \mid M_{i} -M_{j} \mid $ 
(where $i\neq j$, $i,j=H^{\pm\pm}_5,H^{\pm}_5 $ and $H^{0}_5$),  
$\Delta M_3 \equiv \mid M_{H^{\pm}_3} -M_{H^{0}_3} \mid $ and 
$\Delta M_f \equiv \mid M_{f_{MF}^{u,\nu_R}} -M_{f_{MF}^{d,l}} \mid $  to be less than $\sim 50 $ GeV with the assumption of a light 125 GeV SM like Higgs boson and 173.1 GeV top quark.
\item {\bf Constraints from the Lepton Flavor Violating processes:} 
The presence of mirror leptons would lead to additional contribution to 
$\mu \to e \gamma $ at the one loop level and $\mu \rightarrow 3e$ as well 
as $\tau \rightarrow 3 \ell$ processes~\cite{Miyazaki:2007zw} at 
the tree level due to the charged lepton mixing through the
Yukawa interaction $ y_{s\ell} {\bar \psi_L} \psi^M_R \Phi_s $. Among all the aforementioned LFV processes, the most stringent limit ${\rm BR} (\mu^+ \to e^+ \gamma ) 
< 4.2 \times 10^{-13}$ at $90\%$ C.L.~\cite{Baldini:2018nnn,Dohmen:1993mp} comes from the MEG experiment at PSI on the $\mu \to e \gamma $ process.
On the other hand for the $\mu$ to $e$ conversion in the nuclei, the strongest experimental upper limit on branching ratio is provided by the SINDRUM II experiment for titanium targets: 
${\rm BR}(\mu^- + {\rm Ti} \to e^- + {\rm Ti}) < 4.3 \times 10^{-12}$  
at $90\%$ C.L. ~\cite{TheMEG:2016wtm,Dohmen:1993mp}.
The current experimental upper limit 
on ${\rm Br }(\mu \to e \gamma )$ sets the constraint of 
$y_{s\ell} \leq  10^{-4} $ for the mirror lepton mass 
$M_{f_{MF}} \sim 100 -  800$ GeV~\cite{Hung:2017voe,Hung:2015hra,Hung:2007ez}. It is also important to note that according to previous studies performed by one us the most stringent constraints placed on the additional couplings $y_s \sim y_{sq} \sim y_{su} \sim y_{sd}$ are given by
$y_s  < 0.1 y_{s\ell}$~\cite{Hung:2017pss, Hung:2017exy} coming from the solution of the strong CP problem. 
\item {\bf The Higgs signal strength :}  
It is to be noted that the experimentally measured properties of the 
$125$-GeV scalar particle discovered at the LHC so far tend to favor 
the characteristics of the SM Higgs boson. Since we assume one of the
CP-even scalars $\tilde H $ fulfilling the role of the SM like Higgs,  it is necessary that  we examine whether the production and decay of $\tilde H$ in our scenario is in agreement with the current experimental data. 
The agreement can be probed by estimating the Higgs signal in the 
$x$th decay of $\tilde H$ as 
$\mu_{x} = \frac{\sigma(pp 
\rightarrow \widetilde{H} )_{\rm BSM}}{\sigma(pp \rightarrow h )_{\rm SM}
} \, \frac{BR(\widetilde{H} \rightarrow x )_{\rm BSM}}
{BR(h \rightarrow x )_{\rm SM}}$, where, $x = \gamma \gamma, W^+W^-, ZZ, 
b{\bar b} $ and $\tau^+ \tau^-$.
Theoretically obtained Higgs signal strengths depend very strongly on 
the choice of model parameters as shown in Table~\ref {table-BP}. In the next section while describing the model parameters, we will elaborate the implication of the Higgs signal strength on various benchmark points.
%
\end{itemize}
\begin{table}[h!]
\begin{center}\scalebox{0.6}{
\begin{tabular}{|c|c|c|c|c|c|c|c|c|c|c|c|c|c|c|c|c|c|c|c|c|c|c|c|c|c|}
\hline
\hline
&\multicolumn{15}{c|}{Benchmark Points}&\multicolumn{9}{c|}{}\\
\cline{2-16}
&\multicolumn{6}{c|}{VEV of the scalar fields (GeV)} &\multicolumn{9}{c|}{ Scalar quartic couplings $\lambda$'s }&\multicolumn{9}{c|}{ Masses of the scalar fields (GeV)}\\
&\multicolumn{6}{c|}{} & \multicolumn{9}{c|}{ }& \multicolumn{9}{c|}{ }\\
\cline{2-25} 
& $~v_{1}~$&$~v_{2}~$ &$~v_{1M}~$&$~v_{2M}~$ & $v_M$ & $v_s$ & $\lambda_{1a}$ & $\lambda_{1b}$ & $\lambda_{2a}$&$\lambda_{2b}$&$\lambda_{3}$&$\lambda_{4}$& $\lambda_{5}$&$\lambda_{8}$& $\lambda_{s}$&  $M_{\widetilde{H}^{''''}}$  & $M_{\widetilde{H}^{'''}}$  & $M_{\widetilde{H}^{''}}$  & $M_{\widetilde{H}^{'}}$  &   $M_{\widetilde{H}}$ & $M_{\widetilde{H}_s}$   &$m_5$  & $m_{3,H^\pm,H^0_3}$ & $m_{3,\rm All~others}$ \\
\hline
\hline
~~~BP-1~~~&$140$&$145$&$43.5$&$43.5$&$45$&$10^4$&$0.09$&$0.1$&$9.0$&$9.0$&$9.0$&$2.9$&$9.0$&$9.0$&$10^{-15}$&$1126.12$&$ 607.15$&$ 369.85$&$ 352.90$&$ 124.16$&$ 0.00161$&$1279.4$&$738.66$&$972.59$\\
$52.04\%\,\Phi_1$, $47.95 \%\,\Phi_2$&&&&&&&&&&&&&&&&&&&&&&&&\\
\hline
\hline
~~~BP-2~~~&$138$&$142$&$51.07$&$51.07$&$45$&$10^4$&$0.1$&$0.1$&$9.0$&$9.0$&$9.0$&$2.9$&$9.0$&$9.0$&$10^{-15}$&$1130.13$&$ 610.94$&$ 433.36$&$ 402.58$&$ 125.18$&$ 0.00161$&$1279.4$&$738.66$&$972.34$\\
$51.52\%\,\Phi_1$, $48.47 \%\,\Phi_2$&&&&&&&&&&&&&&&&&&&&&&&&\\
\hline
\hline
~~~BP-3~~~&$152$&$145$&$42.99$&$42.99$&$40$&$10^4$&$0.001$&$0.1$&$9.0$&$9.0$&$9.0$&$0.5$&$9.0$&$9.0$&$10^{-15}$&$622.02$&$ 454.13$&$ 364.76$&$ 337.63$&$ 125.82$&$ 0.00161$&$1279.4$&$738.66$&$987.95$\\
$51.52\%\,\Phi_1$, $48.47 \%\,\Phi_2$&&&&&&&&&&&&&&&&&&&&&&&&\\
\hline
\hline
~~~BP-4~~~&$130$&$135$&$68.19$&$68.19$&$45$&$10^4$&$0.116$&$0.1$&$9.0$&$9.0$&$9.0$&$2.9$&$9.0$&$9.0$&$10^{-15}$& $1142.13$ & $ 624.92$ & $ 534.13$ & $ 463.67$ & $ 125.23$ & $ 0.00161$ & $ 1279.4$ & $ 738.66$ & $ 972.34$ \\
$53.69\%\,\Phi_1$, $46.31 \%\,\Phi_2$&&&&&&&&&&&&&&&&&&&&&&&&\\
\hline
\hline
~~~BP-5~~~&$130$&$140$&$62.95$&$62.95$&$45$&$10^4$&$0.11$&$0.11$&$9.0$&$9.0$&$9.0$&$2.9$&$9.0$&$9.0$&$10^{-15}$& $1150.57$ & $ 635.59$ & $ 578.61$ & $ 481.12$ & $ 124.23$ & $ 0.00161$ & $ 1279.4$ & $ 738.66$ & $ 972.34$ \\
$52.03\%\,\Phi_1$, $47.97 \%\,\Phi_2$&&&&&&&&&&&&&&&&&&&&&&&&\\
\hline
\hline
\end{tabular}}
\end{center}
\caption{ \it Extended scalar sector parameters (scalar field VEV's and 
quartic couplings $\lambda's$) and the corresponding scalar 
masses in GeV for five representative benchmark points.}
\label{table-BP}
\end{table}

\vspace{0.8cm}
\section{Benchmark points in the model}
\label{sec:benchmark}
Here we discuss the significance of the choice of our model benchmark points as shown in Table~\ref{table-BP} in the context of FIMP like dark matter phenomenology. The masses and phenomenology of the 
additional scalars and mirror fermions are strongly dependent on the
vacuum expectation value (VEV) of the scalar fields (GeV) 
and scalar quartic couplings $\lambda's$ ( $\lambda_{1a}$, $\lambda_{1b}$, $\lambda_{2a}$, $\lambda_{2b}$, $\lambda_{3}$, 
$\lambda_{4}$, $\lambda_{5}$, $\lambda_{8}$ and $\lambda_{s}$).
To analyze the dark matter phenomenology, we are interested in the region of parameter space allowed by various theoretical and experimental constraints discussed in the previous section\footnote{We performed a scan on the parameter space randomizing the input parameters 
(vevs, quartic couplings etc) to obtain the Higgs-like scalar mass 
$M_{\widetilde{H}} 
\approx 125$ GeV. Out of the one million data points used to scan with 
$M_{\widetilde{H}} \approx 125$ GeV, we obtained a set of data 
points satisfying all the theoretical as well as experimental constraints as discussed in the text with most of the points being discarded by the LHC Higgs signal strength data.}.


In our analysis, we choose the model parameters (scalar quartic couplings, VEVs, etc.) to accommodate the second lightest component $\widetilde{H}$ 
acting as the SM like Higgs boson with dominant contributions coming from the Higgs doublets $\Phi_1$ and $\Phi_2$. Thus, the VEVs $v_1$, $v_2$ and the scalar mixing elements $O^{51}_{H}$ and $O^{52}_{H}$ control the SM like Higgs boson couplings with the SM fermions and gauge bosons. 
The other scalars $(\Phi_{1M, 2M}, \chi)$ interact with the SM particles 
through the mixing with the doublet scalars and for the benchmark 
points shown in Table~\ref{table-BP}, the mixing angles are negligibly small. This leads to a scenario where these scalars have highly suppressed couplings with the SM particles. As a consequence of this, one can have heavy scalar with mass $\sim 300$ GeV satisfying the aforementioned experimental  constraints. In our parameter space scan, we have varied all the quartic couplings (parameters of the scalar potential) within their perturbative range. 

In this framework, the Yukawa couplings of the SM like Higgs boson 
is analogous to the Type-II 2HDM\cite{Lee:1973iz, 
WahabElKaffas:2007xd, Branco:2011iw} in both (SM and Mirror fermions) sectors. The first THDM like 
doublet $\Phi_1$ interacts with the SM charged leptons and down-type quarks 
whereas the second THDM like doublet $\Phi_2$ couples to the up-type quarks (see eqn.~\ref{eq:LYukawa}). Here, we assume that the Higgs-125 GeV scalar 
$\widetilde{H}$ is mostly generated from the real part of the doublets
 $\Phi_1$ and $\Phi_2$. Hence, the contributions on the Higgs signal strength coming from the other scalar multiplets due to the mixings can be taken to be negligible. The decay rate of $\widetilde{H}$ into two lightest CP-even scalar 
$\widetilde{H}_s$ (and two dark matter $A_s^0$) fields can contribute to the 
Higgs invisible decay width depending on mixing, i.e., the value of the 
quartic coupling $\lambda_{4a}$ and the VEVs ($v_1,v_2,v_{1M}, v_{2M},v_M$ 
and $v_s$) of the scalar fields.  
We exercise caution while choosing the values for these VEVs as they can alter the mixing and masses of all the model particles. 
In particular, the VEV $v_2$ can significantly modify the 
$\widetilde{H} t\bar{t}$ yukawa coupling $ Y_{{\tilde H}t \bar t}$, 
which in turn can alter the one loop effective Higgs to gluon-gluon ($ \widetilde{H} gg$) coupling away from its Standard Model value. Thus, the $Y_{{\tilde H}t {\bar t}}$ coupling modifier $\kappa_t$ can be parameterized as 
$\kappa_t = \left( \frac{y_{t,\rm New}}{y_{t\rm, SM}} \right) 
\equiv \left( \frac{v_{\rm SM}}{v_2} \, {\Fontskrivan O}_H^{52} \right)$, 
the ratio of the top Yukawa coupling for ${\widetilde H}$ in this model
relative to the SM value. As a result of this, the production cross-section for the Higgs-125 GeV will be different in this framework in comparison to the SM and the deviation is approximately proportional to $\kappa_t^2$: 
$ R_\sigma \equiv \frac{\sigma(pp \rightarrow \widetilde{H} )_{\rm BSM}}
{\sigma(pp \rightarrow h )_{\rm SM}}
\approx \frac{\sigma(gg \rightarrow \widetilde{H} )_{\rm BSM}}
{\sigma(gg \rightarrow h )_{\rm SM}} $  
$\propto \kappa_t^2$. For all the benchmark points shown in 
Table~\ref{table-BP}, the mixing angle 
${\Fontskrivan O}_H^{52} \approx 0.69 $ and the 
minimization condition for the complete scalar potential imply 
that the individual values of both $v_1 $ and $v_2$ must be 
smaller than $v_{\rm SM}$. This results in the estimated value of  
$R_\sigma > 1 $, leading to an enhanced production cross-section for 
the SM like $\widetilde{ H}$ in the gluon fusion channel 
compared to the SM. However, the Higgs signal strengths in the $\gamma \gamma, WW $ and $ZZ$ final 
states $\mu_{\gamma\gamma}$ and $\mu_{WW,ZZ}$ are measured very precisely at the LHC and provide stringent limits on the model parameters for any beyond the standard model scenario that contribute non-trivially to these final states~\cite{Sirunyan:2018koj}. Hence, to satisfy these constraints the enhancement in the Higgs production rate $(g g \to {\widetilde H})$ must be compensated by the branching ratio  
$BR(\widetilde{H} \rightarrow xx )_{\rm BSM} $, where $x=\gamma,W,Z$. 
We find that for the benchmark points (see Table~\ref{table-BP}) the branching ratio ${\rm Br}({\widetilde H} \rightarrow \gamma\gamma )$ is ($5-18\%$) lower compared to the SM-value (see Table~\ref{table-BR}) thus making the model predicted $\mu_{\gamma\gamma}$ values consistent with the observed data~\cite{Sirunyan:2018koj}. We also note that the contributions coming from the other heavy charged scalars at the one-loop level do not impact the 
$\Gamma (\tilde H \to \gamma \gamma )$ due to the small quartic coupling 
$\lambda_{4a}$\footnote{The Higgs quartic coupling 
$\lambda_{4a} \leq 10^{-9}$ is required for a FIMP like dark matter 
production via freeze-in mechanism in this scenario.}. Also, due to the six custodial $SU(2)_D$ CP-even scalar mixings,
the ${\rm Br} ({\widetilde H} \to WW,ZZ)$ is approximately 
$15-30\%$ lower than the corresponding branching ratios in the SM. 
As a consequence of this, the Higgs $( {\widetilde H})$ signal strengths 
for $\mu_{WW,ZZ}$ channels are consistent with the experimental values within $1\sigma $ range for our choice of benchmark points.
While scanning the parameter space, we have found that the values of $v_{1M}=v_{2M}\approx v_M \gtrsim 70$ GeV could violate the Higgs signal strength data for the following reasons. 
In this scenario, we have total four scalar doublets and two triplets, 
and their corresponding vacuum expectation values satisfy the constraint:
${v_{\rm SM}}\equiv\sqrt{v_{1}^2+v_{1M}^2+v_{2}^2+v_{2M}^2+8 v_{M}^2}= 246.21 
~{\rm GeV}$. 
For the aforementioned values of $v_{1M}, v_{2M} $ and $v_{M}$, 
the VEVs $v_1$ and $v_2$ are bound to be tiny to satisfy the condition
$v_{\rm SM} = 246.21 $ GeV . Such low values for $v_1$ and $v_2$ 
would substantially increase the Yukawa couplings $Y_{u,d, \ell}$ which 
in turn will elevate the tree-level SM-like Higgs partial width in the 
$b\bar{b},\tau\bar{\tau}$ final states. In addition to producing a surge in these tree-level decay widths, the loop induced $\gamma\gamma $ and 
gluon-gluon partial widths could also rise as a consequence of this. The increase in these partial decay modes eventually add up in the total decay width of the SM-like Higgs boson. As a result of this inflated total decay width, all branching ratios for the SM-like Higgs boson are altered significantly, violating the experimentally measured 
Higgs signal strength data. 
\begin{table}[h!]
\begin{center}\scalebox{0.6}{
\begin{tabular}{|c|c|c|c|c|c|c|c|c|c|c|c|c|c|c|c|c|c|c|c|c|c|c|c|c|c|}
\hline
\hline
&\multicolumn{7}{c|}{Benchmark Points and Branching of SM like Higgs}\\
\cline{2-8}
& $ \,Br(\widetilde{H} \rightarrow b\bar{b})$ \, & $ \,Br(\widetilde{H} \rightarrow \tau\bar{\tau})$ \, & $ \,Br( \widetilde{H} \rightarrow WW^*)$ \,&$ \,Br(\widetilde{H} \rightarrow ZZ^*)$ \,&$ \,Br(\widetilde{H} \rightarrow \gamma\gamma)$ \,& $ \,Br(\widetilde{H} \rightarrow \widetilde{H}_s \widetilde{H}_s, A^0_s A^0_s)$ \,&$ \,Br(\widetilde{H} \rightarrow {\rm Other~BSM})$ \,\\
\hline
\hline
SM~&$ \,5.66\,\times \, {10}^{-01}$ \,&$ \,6.21\,\times \, {10}^{-02}$ \,&$ \,2.26\,\times \, {10}^{-01}$ \,&$ \,2.81\,\times \, {10}^{-02}$ \,&$ \,2.28\,\times \, {10}^{-03}$ \,&--&-- \\
\hline
\hline
BP-1~&$ \,6.91\,\times \, {10}^{-01}$ \,&$ \, 8.56\,\times \, {10}^{-02}$ \,&$ \,1.98\,\times \, {10}^{-01}$ \,&$ \, 2.46\,\times \, {10}^{-02}$ \,&$ \, 1.96\,\times \, {10}^{-03}$ \,&$ \,<1\,\times \, {10}^{-06}$ \,&$ \,<1\,\times \, {10}^{-06}$ \,\\
\hline
\hline
BP-2~&$ \,6.98\,\times \, {10}^{-01}$ \,&$ \, 8.64\,\times \, {10}^{-02}$ \,&$ \,1.92\,\times \, {10}^{-01}$ \,&$ \, 2.38\,\times \, {10}^{-02}$ \,&$ \, 1.95\,\times \, {10}^{-03}$ \,&$ \,<1\,\times \, {10}^{-06}$ \,&$ \,<1\,\times \, {10}^{-06}$ \,\\
\hline
\hline
BP-3~&$ \,6.71\,\times \, {10}^{-01}$ \,&$ \, 8.31\,\times \, {10}^{-02}$ \,&$ \,2.12\,\times \, {10}^{-01}$ \,&$ \, 2.63\,\times \, {10}^{-02}$ \,&$ \, 2.11\,\times \, {10}^{-03}$ \,&$ \,<1\,\times \, {10}^{-06}$ \,&$ \,<1\,\times \, {10}^{-06}$ \,\\
\hline
\hline
BP-4~&$ \,7.25\,\times \, {10}^{-01}$ \,&$ \, 8.98\,\times \, {10}^{-02}$ \,&$ \,1.63\,\times \, {10}^{-01}$ \,&$ \, 2.02\,\times \, {10}^{-02}$ \,&$ \, 1.94\,\times \, {10}^{-03}$ \,&$ \,<1\,\times \, {10}^{-06}$ \,&$ \,<1\,\times \, {10}^{-06}$ \,\\
\hline
\hline
BP-5~&$ \,7.23\,\times \, {10}^{-01}$ \,&$ \, 8.95\,\times \, {10}^{-02}$ \,&$ \,1.64\,\times \, {10}^{-01}$ \,&$ \, 2.03\,\times \, {10}^{-02}$ \,&$ \, 1.93\,\times \, {10}^{-03}$ \,&$ \,<1\,\times \, {10}^{-06}$ \,&$ \,<1\,\times \, {10}^{-06}$ \,\\
\hline
\hline
\end{tabular}}
\end{center}
\caption{ \it The Branching fraction of the SM Higgs and 
the $ \,125 $ \, GeV $\widetilde{H}$ are shown for five benchmark points of   
Table.~\ref{table-BP}. }
\label{table-BR}
\end{table}

\begin{table}[h!]
\begin{center}\scalebox{0.8}{
\begin{tabular}{|c|c|c|c|c|c|c|c|c|c|c|c|c|c|c|c|c|c|c|c|c|c|c|c|c|c|}
\hline
\hline
Signal Strength&\multicolumn{5}{c|}{Benchmark Points and Signal strength of SM like Higgs}\\
\cline{2-6}
 & $~~~~~\mu_{ b\bar{b}}~~~~~$ & $~~~~~\mu_{  \tau\bar{\tau}}~~~~~$ & $~~~~~\mu_{ WW}~~~~~$&$~~~~~\mu_{ ZZ}~~~~~$&$~~~~~\mu_{  \gamma\gamma}~~~~~$\\
\hline
$\mu_{\rm Best-Fit}$~&$2.51^{+2.43}_{-2.01}$&$1.05 ^{+0.53 }_{-0.47 }$&$ 1.35^{+0.35 }_{-0.21 }$&$ 1.22^{+0.23 }_{-0.21 }$&$ 1.16^{+0.21 }_{-0.18 }$\\
\hline
$\mu_{\rm BP-1}$~&1.70&1.91&1.214&1.211&1.19\\
\hline
\hline
$\mu_{\rm BP-2}$~&1.81&2.03&1.239&1.236&1.25\\
\hline
\hline
$\mu_{\rm BP-3}$~&1.42&1.59&1.114&1.111&1.10\\
\hline
\hline
$\mu_{\rm BP-4}$~&1.85&2.06&1.03&1.029&1.23\\
\hline
\hline
$\mu_{\rm BP-5}$~&2.06&2.30&1.16&1.15&1.22\\
\hline
\hline
\end{tabular}}
\end{center}
\caption{ \it 125-GeV Higgs boson signal strengths for five benchmark points 
of Table.~\ref{table-BP}. The experimental best fit $\mu_{\rm Best-Fit}$ 
are the CMS combined measurements of the Higgs boson couplings at 
13 TeV run of the LHC with 
$35.9~{\rm fb}^{-1}$ data~\cite{Sirunyan:2018koj}. 
The error bars shown on $\mu_{\rm Best-Fit}$ are at the one sigma level.}
\label{table-BR1}
\end{table}
In Table ~\ref{table-BR} we show different branching ratios (BR) for 
$\widetilde{H}$ and also provide the corresponding SM higgs boson BR values for direct comparison. For our choice of benchmark points, 
the $\widetilde{H}$ branching ratios in various SM two body final 
states are consistent with that of the current LHC Higgs boson data. 
It should also be noted that the invisible decay of $\widetilde{H}$ 
is negligible small in this scenario. The various Higgs signal strengths calculated for our choice of benchmark points and the corresponding  experimental best-fit values are shown in Table \ref{table-BR1}. For all the chosen benchmark points, the Higgs signal strengths are consistent with the experimental bounds within $2\sigma $~\cite{Sirunyan:2018koj}.

\section {The sub-MeV dark matter candidate in the model}
\label {sec:darkmatter}

Let us now turn our attention to the possibility of accommodating a sub-MeV dark matter candidate in this framework. In this scenario, it is noted that the imaginary part of the complex singlet scalar does not mix with the other scalars and as a result can serve as a viable dark matter candidate $A^0_s\equiv Im( \Phi_s)$.
The transformation of the following fields can be expressed as
$\Phi_{1,2}\rightarrow e^{-2 i \alpha_{\rm SM}}\,\Phi_{1,2},\Phi_{1M,2M} \rightarrow e^{2 i \alpha_{\rm MF}}\,\Phi_{1M,2M},~  \tilde{\chi} \rightarrow e^{-2 i \alpha_{\rm MF}}\, \tilde{\chi} ,\,\xi \rightarrow \xi$ and $\Phi_s \rightarrow e^{- i (\alpha_{\rm SM} + \alpha_{\rm MF})}\,\Phi_s$.
The $\lambda_{5,6}$'s terms, in the potential (see eqn.~\ref{eq:pot}), break the $U(1)_{\rm SM} \times U(1)_{\rm MF}$ symmetries explicitly.
Total three `massless' Nambu-Goldstone bosons can be obtained after spontaneous breaking of $SU(2)_L \times U (1)_Y \rightarrow U (1)_{em}$ by imposing the condition $\lambda_{5a}=\lambda_{5b}=\lambda_{6a}=\lambda_{6b}=\lambda_{7a}=\lambda_{7b}=\lambda_{7ab}=\lambda_{7Mab}=\lambda_{7aMb}=\lambda_{7abM}=\lambda_5$.
Similarly the first line of $\lambda_{5c}$ term  in the potential (see eqn.~\ref{eq:pot}) is $U(1)_{\rm SM} \times U(1)_{\rm MF}$ conserving, and the second line explicitly violates these symmetries.
All these terms help to get exact minimization of the scalar potential.
In the absence of the $\lambda_{5c}$ term, one can obtain an additional `massless' neutral Nambu-Goldstone, i.e., the complex singlet type pseudoscalar (PSS) can remain massless.
The $U(1)_{\rm SM} \times U(1)_{\rm MF}$ breaking $\lambda_{5c}$ terms help us to get 
non-zero sub-MeV mass for the singlet-type complex pseudoscalar field $A^0_s$.
At the tree-level the mass of the complex singlet scalar is given by
\begin{eqnarray}
M_{A^0_s}^2 = 8 \, \lambda_{5c} \, (v_{1}+v_{2}) (v_{1M}+v_{2M}).
\label{eq:AsMass}
\end{eqnarray}
For the chosen BPs, the numerical values of $(v_{1}+v_{2}) (v_{1M}+v_{2M})$ remain almost same and the dark matter mass only depends on the quartic coupling $\lambda_{5c}$.  
The Higgs portal coupling ($A^0_s A^0_s\,\widetilde{H}$) is also proportional to the coupling $\lambda_{5c},\lambda_{4a}$ and VEVs (see the eqns.~\ref{eq:sig2}). 
In the following sections, we discuss the various bounds on the model parameter space~\cite{Hung:2006ap,Chakdar:2016adj,Hung:2017exy,Hung:2017voe} coming from terrestrial and laboratory-base experiments.

Motivated by the possibility of a viable dark matter candidate in this 
scenario, we investigate the mass ranges and the corresponding stability conditions 
for the $A^0_s$. If the dark matter $A^0_s$ is heavy, it can decay into two fermions at tree-level and the corresponding decay width is given by
\beq
\Gamma (A_s^0 \rightarrow ff )=  \frac{N^c_f M_{A_s^0} y_f^2 }{8 \pi} (1-\frac{4 m_i^2}{M_{A_s^0}^2})^{\frac{1}{2}},
\eeq
where, $y_f \approx  \sqrt{2}  \frac{ y_{si}^2 v_s}{ y_i^M v_{2M}}, i=u,d,\ell$ (see eqns~\ref{eq:Asll} and \ref{eq:Asqq})  for $v_M=v_{1M}=v_{2M}$.
One can also find the decay life time in this case to be
\bea
\tau_{A_s^0} = \frac{6.5821 \times 10^{-25}}{\Gamma^{\rm Total}_{A_s^0}  \,\,\, [\rm in~ GeV]} \,\, \rm seconds
\eea
Here we use the natural unit conversion ${\rm 1~GeV^{-1}=6.5821 \times 10^{-25} ~seconds}$. If $M_{A_s^0}> 2 m_e$, it can decay only into two electrons. The stability of the DM demands that $\tau_{A_s^0} > \tau_U$ where,
$\tau_U\approx 4.35 \times 10^{17}$ seconds is the lifetime of the Universe. For the chosen parameters, $M_{A_s^0}= 1.023$ MeV,  
$\Big(\frac{v_{2M}}{v_s}\Big)^2=1.848\times 10^{-5}$ and $y_e^M\simeq 
\sqrt{4 \pi}$, we find the limit on the coupling $y_{s \ell}$ to stabilize the dark matter to be
\beq
y_{s\ell}<5.14091 \times 10^{-11}.
\eeq
 \begin{figure}[h!]
 \begin{center}
 {\includegraphics[width=2.8in,height=1.8in, angle=0]{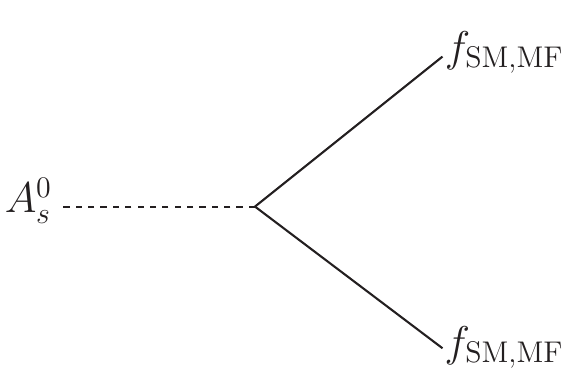}}
 \hskip 1cm
 {\includegraphics[width=2.8in,height=1.8in, angle=0]{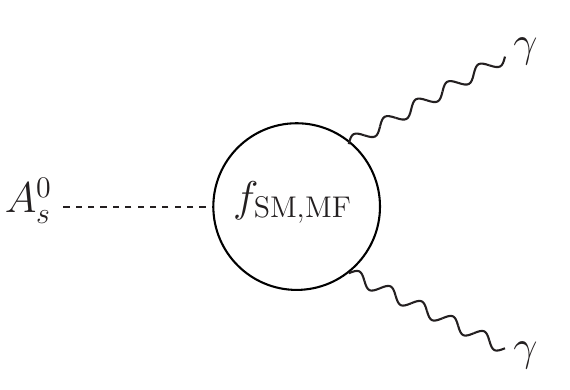}} 
 \caption{ \it Possible two body (tree- {\rm \& } one-loop) decay
Feynman diagram of $A_s^0$.}
 \label{fig:decay}
 \end{center}
 \end{figure}
Furthermore, motivated from the theoretical framework present in the current scenario, we are interested in the sub-MeV DM mass range. As the dark matter is assumed to be light (sub-MeV), it is unable to decay into two fermions at tree-level, however there is a possibility of it decaying to two photons through the SM and mirror charged particles (see Fig.~\ref{fig:decay}).
The decay width of the dark matter decaying into two photons is given by
\begin{equation}
\Gamma (A_s^0 \rightarrow \gamma \gamma ) = \Gamma (A_s^0 \rightarrow \gamma \gamma )^{\rm SM~ fermions} + \Gamma (A_s^0 \rightarrow \gamma \gamma )^{\rm MF~ fermions},
\end{equation}
where,
\bea
\Gamma (A_s^0 \rightarrow \gamma \gamma )^{\rm SM~ fermions} &=& {\alpha^2 M_{A_s^0}^3\over 256\pi^3
v^2_{\rm SM}}\left|\sum_{f}N^c_fQ_f^2y_f
F_{1/2}(\chi_f)\right|^2\\
\Gamma (A_s^0 \rightarrow \gamma \gamma )^{\rm MF~ fermions} &=&  {\alpha^2 M_{A_s^0}^3\over 256\pi^3
v^2_{\rm SM}}\left|\sum_{f}N^c_f Q_f^2y_f^M
F_{1/2}(\chi_f)\right|^2,
\eea
here, $\chi_i=M_{A_s^0}^2/4m_i^2$. $Q_{f}$ denotes electric charges of the corresponding particles. $N_f^c$ is the color factor. In the limit 
$y_{sd}\approx y_{sq}\approx y_{su} = y_s$, $y_f=-y_f^M \approx \sqrt{2}  
\frac{ y_{si}^2 v_s}{ y_i^M v_{2M}} $ (see eqns~\ref{eq:Asll} and \ref{eq:Asqq}) denote $A_s^0$ couplings to $f_i\bar{f_i}$ and $f_{i}^M\bar{f}_{i}^M$ where, $i=u,d,\ell$. The loop function $F_{1/2}$ is defined as
\begin{align}
F_{1/2}(\chi)&=2[\chi+(\chi-1)f(\chi)]\chi^{-2}\, ,
\label{eq:loopfn}
\end{align}
where,
\beq
f(\chi)= \bigg\{\begin{array}{ll}
(\sin^{-1}\sqrt{\chi})^2\,,\hspace{60pt}& \chi\leq 1\\
-{1\over4}[\ln{1+\sqrt{1-\chi^{-1}}\over1-\sqrt{1-\chi^{-1}}}-i\pi]^2\,,\quad
&\chi>1
\label{eq:ftau}
\end{array}\;,\;
\eeq
As previously stated, to solve the strong CP problem in this framework the corresponding couplings are of the order $y_s < 0.1 y_{s\ell}$ ~\cite{Hung:2017pss,Hung:2017exy} as a result of which the decay $\Gamma (A_s^0 \rightarrow \gamma \gamma)$ channel through SM-quarks are suppressed.
Additionally, as the mass of the mirror-fermions are large ($\mathcal{O}(150)$ GeV), the decays through the mirror-fermions are negligible compared to other channels. It is to be noted that the couplings of $A_s^0$ with two charged scalars or charged gauge bosons are absent (protected from the $U(1)_{\rm SM}\times U(1)_{\rm MF}$ symmetries) in this model. As a result, the decay $\Gamma (A_s^0 \rightarrow \gamma \gamma )$ through the SM charged lepton loop becomes the dominant one and gives rise to the most stringent bound on the coupling $y_{s\ell}$.
Hence we obtain
\bea
\Gamma_{tot} (A_s^0 \rightarrow \gamma \gamma ) &=& { \, \alpha^2 M_{A_s^0}^3\over 256\pi^3
v^2_{\rm SM}}  \, \frac{y_{s\ell}^4}{(y_\ell^M)^{2}}\, \frac{v_s^2}{v_{2M}^2} \, \left|
F_{1/2}(\chi_{e}) + F_{1/2}(\chi_{\mu}) + F_{1/2}(\chi_{\tau})\right|^2.
\eea
The decay through electron loop is the dominant one as $F_{1/2}(\chi_{e}) >> F_{1/2}(\chi_{\mu}) >> F_{1/2}(\chi_{\tau}) $ for $M_{A_s^0}\leq 1$ MeV. Hence
\bea
\Gamma_{tot} (A_s^0 \rightarrow \gamma \gamma ) &\approx& { \, \alpha^2 M_{A_s^0}^3\over 256\pi^3
v^2_{\rm SM}}  \, \frac{y_{s\ell}^4}{(y_\ell^M)^{2}}\, \frac{v_s^2}{v_{2M}^2} \, \left|
F_{1/2}(\chi_{e})\right|^2
\eea
and using this we can obtain the life-time of the pseudoscalar given by
\bea
\tau_{A_s^0} & =& \frac{6.582 \times 10^{-25}}{ \Gamma_{tot} (A_s^0 \rightarrow \gamma \gamma )\, \, [\rm in~ GeV] }     \,\,\,\, \,\,  \rm seconds
\eea
The dark matter stability condition imposes an upper limit on the couplings. Taking $\Big(\frac{v_{2M}}{v_s}\Big)^2=1.848\times 10^{-5}$ and $(y_\ell^M)^2\sim 4\pi$, the limit on $y_{s\ell}$ can be obtained as 
 \begin{figure}[h!]
 \begin{center}
 {\includegraphics[width=4in,height=3.2in, angle=0]{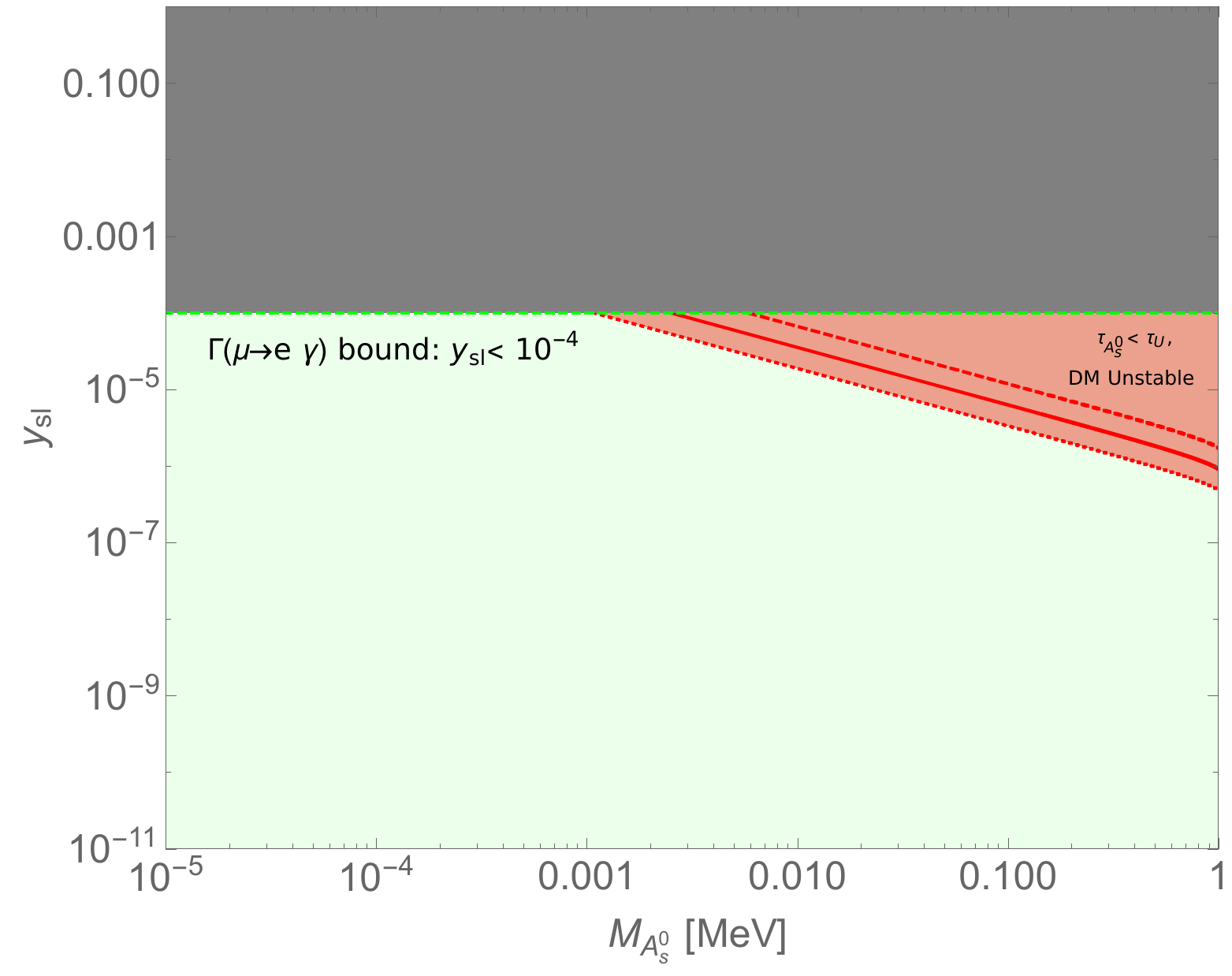}}  
 \caption{ \it Plot shows the exclusion region in $M_{A_s^0}-y_{s\ell}$ 
plane. The dark matter is stable in the region below the red-line given by, $\tau_{A_s^0}>\tau_U$ where the three red lines correspond to  $y^M_u=1$ (dashed), $\sqrt{4\pi}$ (solid) and $4 \pi$ (dotted) respectively. The gray-region is excluded from the $\mu \rightarrow e \gamma$ constraints and $\mu-e$ conversion ~\cite{Hung:2017voe} implying $y_{s\ell} < 10^{-4}$. 
}
 \label{fig:gslMAs}
 \end{center}
 \end{figure}
\bea
y_{s\ell}^4 & \lesssim & \frac{6.582\times 10^{-25} }{ \tau_U} \, {256 \pi^3
v^2_{\rm SM} \, (y_u^M)^2 \over \alpha^2 M_{A_s^0}^3  |\,F_{1/2}(\chi_e)|^2 } \, \Big(\frac{v_{2M}}{v_s}\Big)^2\nn\\
y_{s\ell} & \lesssim & {7.255 \times 10^{-9} \over M_{A_s^0}^{3/4}  |\,F_{1/2}(\chi_e)|^{1/2}}~~{\rm [GeV]}^{3/4}.
\label{eq:boundloop}
\eea
We take $|F_{1/2}(\chi_e)|=1.922$ for $M_{A_s^0}=1$ MeV and obtain the bound on $y_{s\ell}$
\bea
y_{s\ell}&<& \,\, 9.305 \times 10^{-7}. \,
\eea
Similarly taking $|F_{1/2}(\chi_e)|=1.33$ for $M_{A_s^0}=1$ keV, we get
\bea
y_{s\ell}&<& \,\, 1.986 \times 10^{-4}.\,
\eea

It is to be noted that the parameter space is constrained by $\mu \rightarrow e \gamma$  and $\mu - e$ conversion implying $y_{s\ell} \leq 10^{-4}$~\cite{Hung:2017voe}. 
In Fig.~\ref{fig:gslMAs}, we present the allowed parameter space in the $y_{s\ell}-M_{A_s^0}$ plane. The gray-region is excluded from the $\mu \rightarrow e \gamma$ decay and $\mu - e$ conversion constraints~\cite{Hung:2017voe}. The indirect searches for the light dark matter can also place stringent constraints on the model parameter space which will be discussed in the later sections. It is to be noted that the  direct detection limits~\cite{Aprile:2018dbl} for such light dark matter masses may not be applicable in this framework. We will discuss the details of the relic density analysis (through the successful implementation of Freeze-in mechanism) in the upcoming section.

\section{FIMP-like Dark Matter Density}
\label{sec:fimp}
From the discussion in the previous section, it is established that the 
pseudoscalar singlet (PSS) $A_s^0$ can be a viable candidate for the dark matter in this framework.
However, a decaying MeV scale dark matter candidate poses a viable limitation due to the fact that one has to implement fine-tuning~\cite{Rott:2014kfa,Fiorentin:2016avj} to stabilize the dark matter as the lifetime of the DM particles has to be at least larger than the age of the Universe~\cite{Audren:2014bca,Aartsen:2014gkd}.  
In this model, we observe a viable, stable dark matter candidate in the interesting mass range of $\mathcal{O}(<1)$ MeV with the corresponding quartic coupling related to the dark matter mass to be in the range $\lambda_{5c}<\mathcal{O}(10^{-12})$. 
%
It is possible that for the low dark matter mass $\mathcal{O}(<1)$ MeV,   the relic density at the right ballpark $\Omega h^2=0.1198\pm 0.0026$~\cite{Aghanim:2018eyx} is still produced through the well established Freeze-out mechanism with the choice of the large Higgs portal and other couplings, but it will violate the perturbative-unitarity limits.
For example, very large Higgs portal couplings $\lambda's=\mathcal{O}(500)$ (here $\lambda's$ are $\lambda_{4a},~\lambda_{5c} $ and $ \lambda_{s}$) for $M_{A_s^0}=0.5$ MeV and $\lambda's=\mathcal{O}(1500)$ for $M_{A_s^0}=100$ keV are needed to obtain the thermally averaged annihilation cross-section $\sim 2.0 \times 10^{-26}~{\rm cm^3/s}$.


We observe that the non-thermally produced PSS can serve as a viable (depending on the parameters $\lambda_{5c},\lambda_{4a},\lambda_{s}$, $y_{s}$, $y_{s\ell}$ and VEVs, see the Section~\ref{sec:scalar_sect}) $\mathcal{O}(<1)$ MeV dark matter candidate satisfying the dark matter relic density constraints. As the dark matter interacts with other particles very weakly (feebly), for such very weakly interacting particles, called feebly interacting massive particles or FIMPs, we can invoke the non-thermal, so-called Freeze-in mechanism. This mechanism needs feeble interactions which could be one of the reasons to have aforementioned tiny couplings existing in this framework. 
 We examine the possibility of the dark matter sector getting populated through decay or annihilation of other heavy particles until the number density of the corresponding heavy particle species becomes Boltzmann-suppressed. To carry this out, we need to solve the Boltzmann equation that dictates the final relic abundance for the dark matter candidate $A_s^0$. The production of the dark matter resulting from the decay of any mother particle ($\widetilde{H}_i\,(i=1,..6), f_{MF}$) is in thermal equilibrium at early universe and is given by 
\beq
{\Gamma \over H } \geq 1,
\eeq
where, $\Gamma$ is the relevant decay width and $H$ is the Hubble parameter given by~\cite{Plehn:2017fdg,Hall:2009bx,Biswas:2016bfo}
\beq
H(T) = \left( g^* \, \frac{\pi^2}{90} \, \frac{T^4}{\mpl^2} \right)^{1/2},
\label{eq:Hub}
\eeq
where, $\mpl=1.2\times 10^{19}$ GeV is the Planck mass and $T$ is the temperature ($1~ \text{GeV}=1.16 \times 10^{13}~ \text{Kelvin}$).
 \begin{figure}[h!]
 \begin{center}
  {\includegraphics[scale=.45]{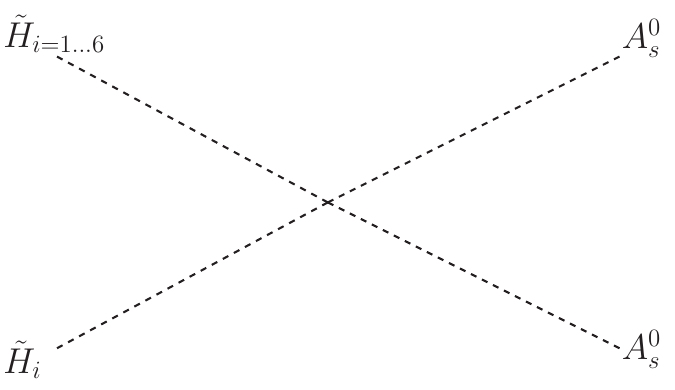}} 
 {\includegraphics[scale=.45]{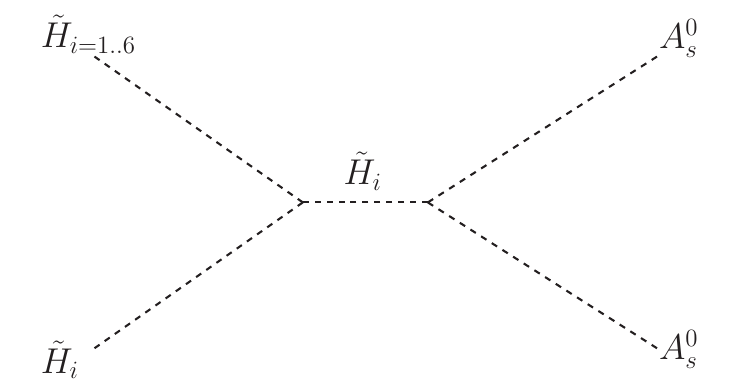}}
 {\includegraphics[scale=.45]{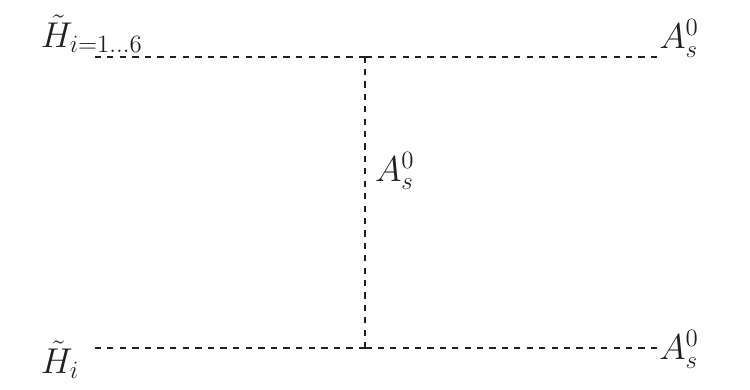}}\\
 {\includegraphics[scale=.45]{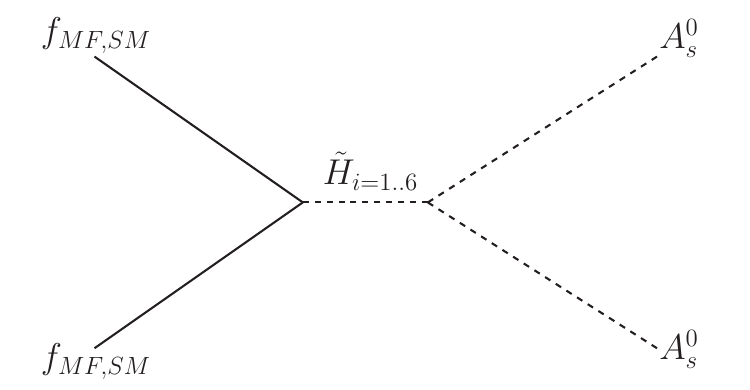}} 
 {\includegraphics[scale=.45]{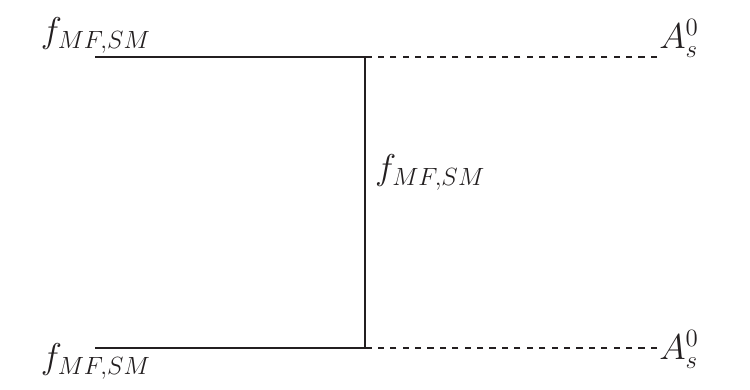}}
 \caption{ \it   Annihilation-production diagrams for the dark matter from the Higgs, SM and mirror fermion.}
  \label{fig:anniPDM}
 \end{center}
 \end{figure}
If the production of the mother particles occur mainly from the annihilation of other particles in the
thermal bath, $\Gamma$ will be replaced by~\cite{Plehn:2017fdg,Hall:2009bx,Biswas:2016bfo}
\beq
\Gamma = n_{eq} <\sigma v>,
\eeq
where $n_{eq}$ is their equilibrium number density and is given by~\cite{Plehn:2017fdg}
\beq
\begin{aligned}
    &n_{eq} &=
  &\left\{ \begin{array}{l} \vspace{0.3cm} g^* \left( \frac{M\, T}{2 \pi}\right)^{3/2} \, e^{-M/T},  ~~~~~~~{\rm for ~non\text{-}relativistic ~states}~~T<<M\\
     \frac{\zeta_3}{\pi^2} g^* T^3,  ~~~~~~~{\rm for~relativistic~boson ~states}~~T>>M\\
     \vspace{0.5cm}
    \frac{ 3}{4}\, \frac{\zeta_3}{\pi^2} g^* T^3 ,  ~~~~~~~{\rm for~relativistic~fermion ~states}~~T>>M
    \vspace{-0.2cm}\end{array}
       \right.
       \label{eq:n}
\end{aligned}
\eeq
where the Riemann zeta function has the value $\zeta_3=1.2$ and $g^*=208.5$ (for $T>>M$) is the effective degrees of freedom in this framework and $M$ stands for the mass of the particle.
Here $<\sigma v>$ is the thermally averaged annihilation cross-section for the particles in the thermal bath and can be written as~\cite{Gondolo:1990dk,Plehn:2017fdg} 
\beq
<\sigma_{xx} v> = \frac{  2 \pi^2 T \, \int_{4 M^2}^\infty ds \sqrt{s} \, (s-4 M^2)  \, K_1(\frac{\sqrt{s}}{T})  \sigma_{xx}   }{   \left( 4 \pi M^2 T K_2(\frac{M}{T})  \right)^2        },
\eeq 
where $\sigma_{xx}$ is the production annihilation cross-section of the mother particles ($x=\widetilde{H}_{i=1,..6}, f_{MF}$) from other particles in the thermal bath (see the production annihilation diagrams in Fig.~\ref{fig:anniPDM})and $K_{1,2}$ is the modified Bessel function of functions of order 1 and 2 respectively. 
 \begin{figure}[h!]
 \begin{center}
 {\includegraphics[width=1.5in,height=1.2in, angle=0]{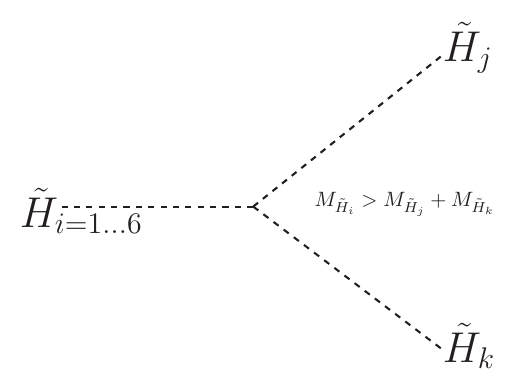}} 
 {\includegraphics[width=1.5in,height=1.2in, angle=0]{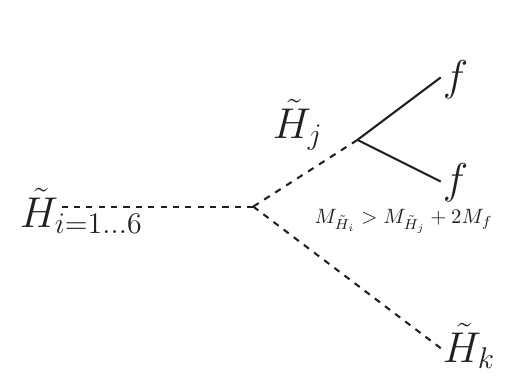}}
 {\includegraphics[width=1.5in,height=1.2in, angle=0]{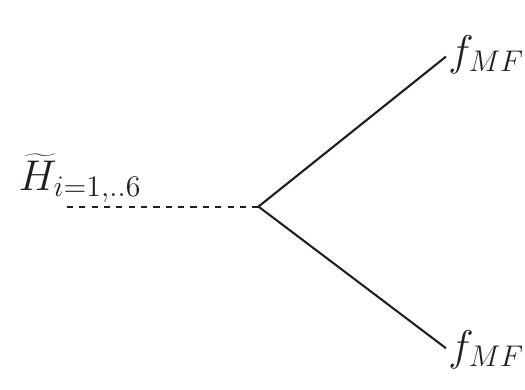}}
 {\includegraphics[width=1.5in,height=1.2in, angle=0]{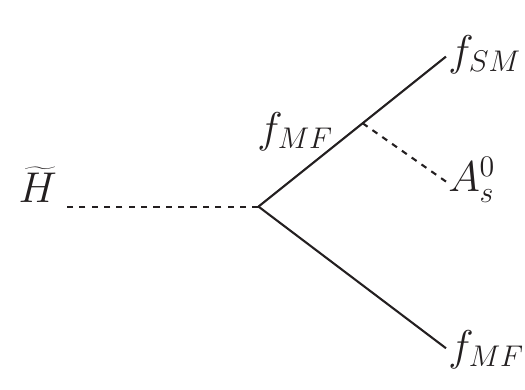}} 

 \caption{ \it  Decay diagrams contributing to the relic density. Decay-production diagrams for the heavy Higgs and mirror fermion help in thermal equilibrium in the early universe.}
  \label{fig:decayP}
 \end{center}
 \end{figure}
 \begin{figure}[h!]
 \begin{center}

  {\includegraphics[width=1.5in,height=1.2in, angle=0]{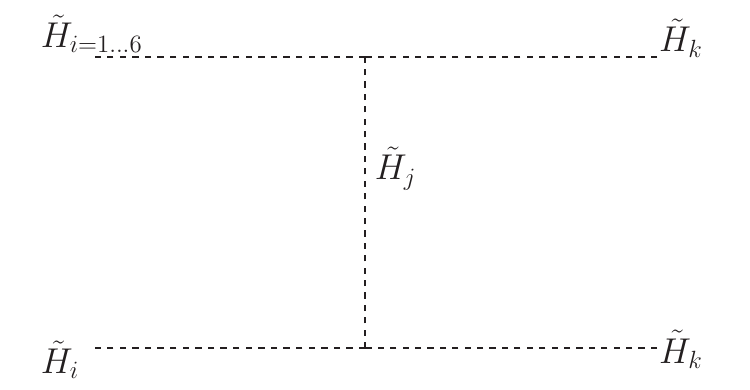}} 
 {\includegraphics[width=1.5in,height=1.2in, angle=0]{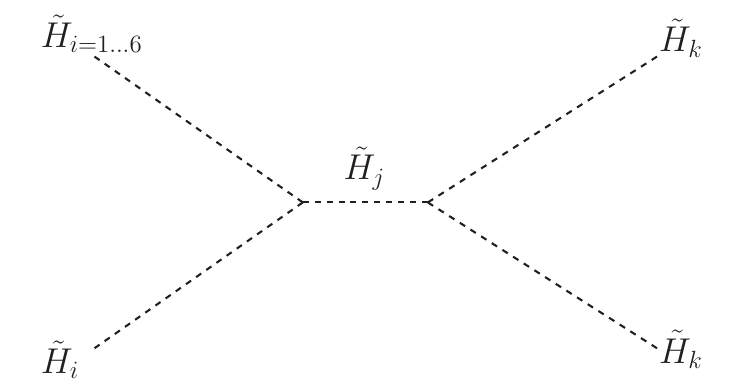}}
 {\includegraphics[width=1.5in,height=1.2in, angle=0]{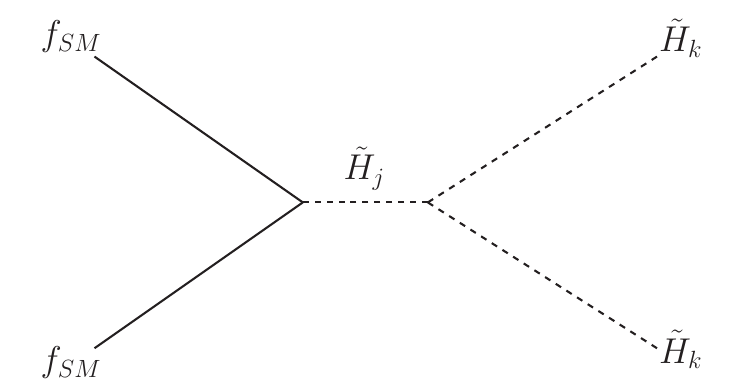}}
 {\includegraphics[width=1.5in,height=1.2in, angle=0]{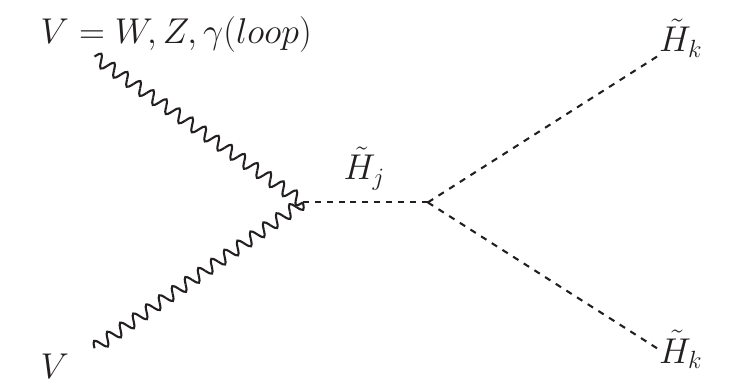}} \\
 {\includegraphics[width=1.5in,height=1.2in, angle=0]{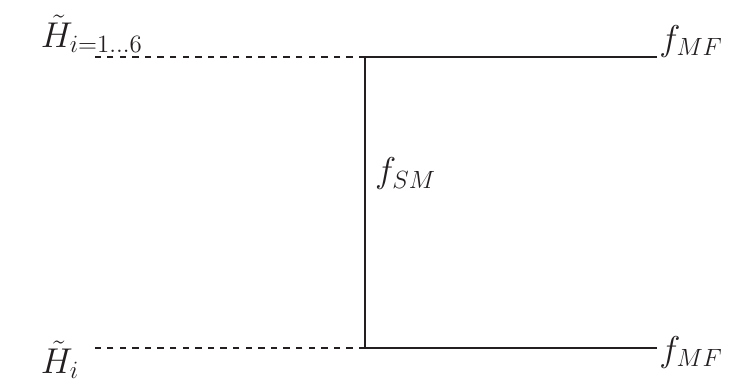}} 
 {\includegraphics[width=1.5in,height=1.2in, angle=0]{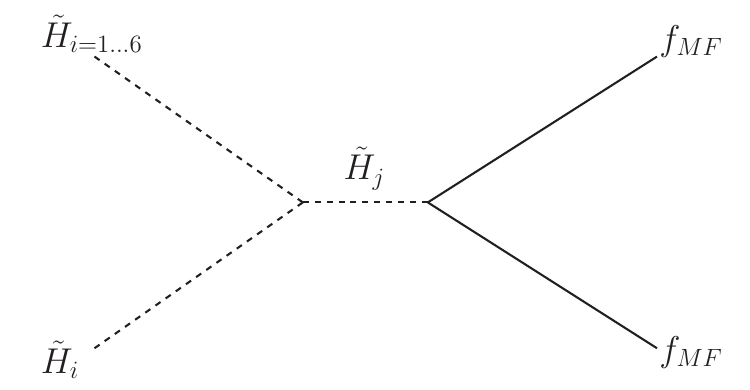}}
 {\includegraphics[width=1.5in,height=1.2in, angle=0]{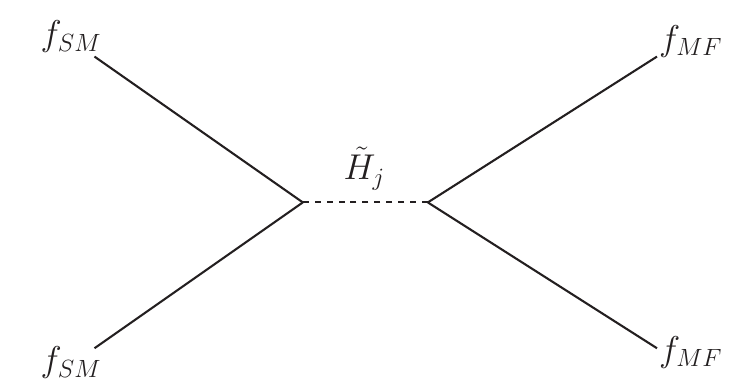}}
 {\includegraphics[width=1.5in,height=1.2in, angle=0]{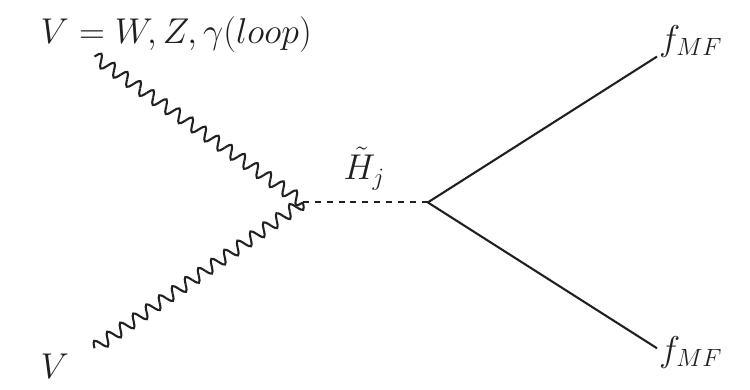}}\\
 {\includegraphics[width=1.5in,height=1.2in, angle=0]{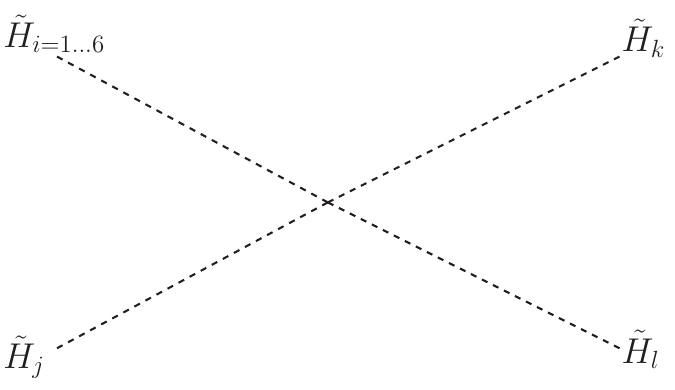}}
 {\includegraphics[width=1.5in,height=1.2in, angle=0]{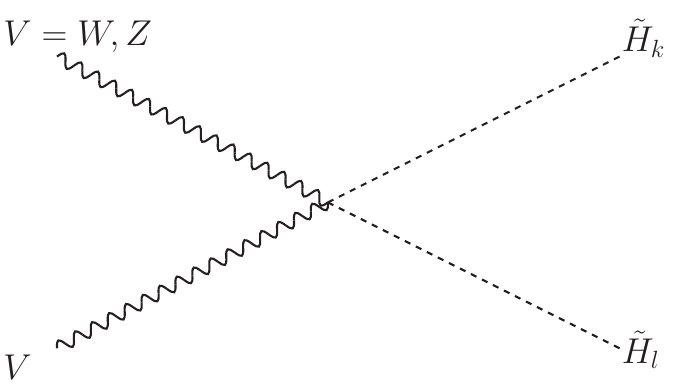}}

 \caption{ \it  Annihilation-production diagrams for the heavy Higgs and mirror fermions help in thermal equilibrium in the early universe. It is to be noted that there are many other similar diagrams that can contribute in the production of the heavy Higgs and mirror fermions.}
  \label{fig:anniP}
 \end{center}
 \end{figure}

In this picture, we present various possible decay (see Fig.~\ref{fig:decayP}) and annihilation (see Fig.~\ref{fig:anniP}) diagrams for the production of the heavy Higgs and mirror fermions that facilitate thermal equilibrium in the early universe. The 2-body and 3-body decay widths for the heavy Higgs ($\widetilde{H}_{i}$) are given by
\bea
\Gamma(\widetilde{H}_i\rightarrow f_{MF} f_{MF} )&=& \frac{M_{\widetilde{H}_i} \, y_M^2}{8 \pi} \, \left( 1- 4 \frac{M_{MF}^2}{M_{\widetilde{H}_i}^2} \right)^{3/2}\\
\Gamma(\widetilde{H}_i \rightarrow \widetilde{H}_j \widetilde{H}_k )&=& \frac{ y_{\widetilde{H}_i \widetilde{H}_j \widetilde{H}_k}^2}{16 \pi M_{\widetilde{H}_i}^3} \, \left( (M_{\widetilde{H}_i}^2 -M_{\widetilde{H}_j}^2-M_{\widetilde{H}_k}^2)^2 - 4\, M_{\widetilde{H}_j}^2 M_{\widetilde{H}_k}^2 \right)^{1/2}\\
\Gamma(\widetilde{H}\rightarrow f_{MF} f_{SM}  A_s^0 ) &=& \frac{1}{2 \pi^3} \, \frac{1}{32 M_{\widetilde{H}}^3} \, |\mathcal{M}|^2 dE_1 dE_3,~~{\rm where},~~|\mathcal{M}|^2 \propto \frac{ y_M^2 \, y_{s\ell}^2}{M_{MF}^2}.
\eea
Here $E_{1,2,3}$ are the energies of the final state particle for three-body decay~\cite{Zyla:2020zbs}.
There are other processes and diagrams that can also contribute to the production of the heavy Higgs and mirror fermions which for simplicity, have been ignored in our calculation. We have closely followed the Ref.~\cite{Djouadi:1995gv} to calculate the three body decay widths and find the three-body decay rates to be always suppressed by additional propagator mass $M_{MF}^2$ and the coupling $y_{si}^2$. 
It is to be noted that it is not challenging to obtain ${\Gamma \over H }>> 1$ due to the large decay width of the heavy particles in the early universe, resulting the heavy fermions being in thermal equilibrium with the thermal bath particles.
Also to note that the lighter CP-even Higgs could also be produced from the decay of the heavier CP-even Higgs with the corresponding decay width $\mathcal{O}(10)$ GeV satisfying the condition ${\Gamma \over H }>> 1$ in this case too. Hence the lighter CP-even Higgs may also remain in thermal equilibrium with the thermal bath particles. 

Similarly the scattering diagrams shown in Fig.~\ref{fig:anniP}
also satisfy the condition $\frac{n_{eq} <\sigma v>}{H} >>1 $ in the model parameter space we are interested in.
In this framwork, the dark matter can be produced from the decay of the mirror fermions, heavy scalars and from the annihilation of the other particles. It has already been discussed in existing literatures~\cite{Borah:2018gjk,Hall:2009bx,Biswas:2016bfo,PeymanZakeri:2018zaa,Herms:2019mnu,DEramo:2020gpr,Das:2021zea,Das:2021qqr,Pandey:2017quk,Biswas:2017tce,Yaguna:2011qn,Borah:2019bdi} that if same couplings are involved in both decays as well as scattering processes then the former has the dominant contribution to DM relic density over the latter one. 

Considering all these discussion, we take into account that the dark matter candidate is stable and can be produced only from the decay of the mirror fermions and heavy Higgses in this framework. 
 The Boltzmann equation for the dark matter can be written as~\cite{Plehn:2017fdg,Hall:2009bx,Biswas:2016bfo}
{\small
\begin{eqnarray}
\hspace{-0.7cm} \frac{dn}{dt}+3 H n &=& -  \sum_i S(X_{Heavy,i} \rightarrow  A_s^0  A_s^0, f_{\rm SM} A_s^0 ),
\label{eq:boltz1}
\end{eqnarray}}
where $X_{Heavy,i}=\widetilde{H}_{j=1..6} $ ($\widetilde{H}_{j=1..6}$ stand for the six physical CP-even mass eigenstates $\widetilde{H}^{''''}, $ $\, 
\widetilde{H}^{'''},$ $\, \widetilde{H}^{''}, $ $\, \widetilde{H}^{'},$ $\, 
\widetilde{H},$ and $ \, \widetilde{H}_s$.), $ f_{MF}$. We have summed over all these heavy particle contributions. The notation $f_{\rm SM} $ stands for SM fermions.
Here the decay-based source term $S$ can be written as
\bea
S = \Gamma (X_{Heavy,i} \rightarrow A_s^0  A_s^0, f_{\rm SM}  A_s^0) \, \frac{K_1(\frac{m_{X_{Heavy,i}}}{T})}{K_2(\frac{m_{X_{Heavy,i}}}{T})} \, n^{eq}_{Heavy,i}
\eea
where $K_{1,2}$ is the modified Bessel function of the first and second kind. For $x=\frac{m_{\widetilde{H}^{''''}}}{T}$ and $Y=\frac{n}{T^3}$, the eqn.~\ref{eq:boltz1} now reads~\cite{Plehn:2017fdg}
\bea
\frac{dY(x)}{dx} = \sum_i \frac{g_{X_{Heavy,i}}}{2 \pi^2} \frac{\Gamma (X_{Heavy,i} \rightarrow A_s^0  A_s^0, f_{\rm SM}  A_s^0) }{H (x\approx 1)} x^3 K_1(x), \eea
where $g_{X_{Heavy,i}}$ is the degrees of freedom of the heavy particle. We can integrate the dark matter production over the entire thermal history and find the
final yield $Y(x_0)$ with the help of the appropriate integral~\cite{Plehn:2017fdg,Hall:2009bx,Biswas:2016bfo}
\bea
Y(x_0)=\frac{45 M_{Pl}}{6.64 \, \pi ^4 \, g_*^S \sqrt{g^\rho}} \,\sum_i  \frac{g_{X_{Heavy,i}}}{ M_{X_{Heavy,i}}^2} \Gamma (X_{Heavy,i} \rightarrow A_s^0  A_s^0, f_{\rm SM}  A_s^0 ) \, \int_{0}^{\infty} x^3 K_{1} (x) dx
\eea
$g_*^S$ and $ g^\rho$ are the effective relativistic of degrees of freedom for the entropy density and energy density respectively.
The relic density now can be written as~\cite{Plehn:2017fdg,Hall:2009bx,Biswas:2016bfo}
\bea
\Omega h^2 &=& \frac{h^2}{3\, H_0^2\, M_{Pl}^2} \, \frac{ M_{A_s^0}}{28} T_0^3 \, Y(x_0)\nn\\
 &\approx& 1.09 \times 10^{27} \, M_{A_s^0}  \,\sum_i  \frac{g_{X_{Heavy,i}} \, \Gamma (X_{Heavy,i} \rightarrow A_s^0  A_s^0, f_{\rm SM}  A_s^0 ) }{ M_{X_{Heavy,i}}^2} 
 \label{eq:totalOmega}
\eea

 \begin{figure}[h!]
 \begin{center}
 {\includegraphics[scale=.6]{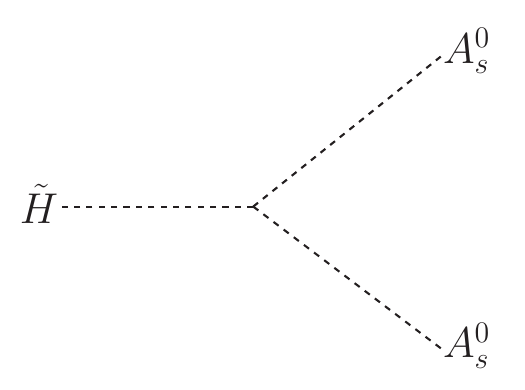}} 
 {\includegraphics[scale=.6]{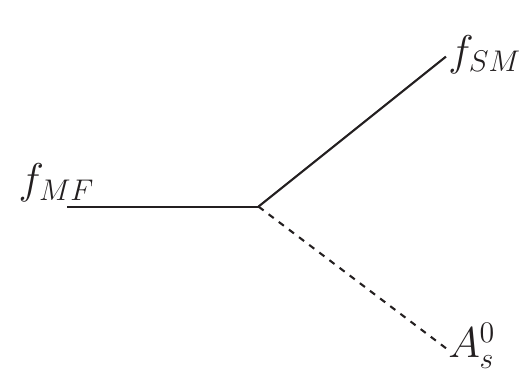}}
 {\includegraphics[scale=.6]{SFFFvertex}} 

 \caption{ \it   Dark matter production diagrams from the decay of the heavy particles contributing to the relic density. }
  \label{fig:decayHAsAs}
 \end{center}
 \end{figure}

We now use eqn.~\ref{eq:totalOmega} to calculate the relic density in this scenario.
The main DM production diagrams from the decay widths of the heavy particles are shown in Fig.~\ref{fig:decayHAsAs}.
The partial decays of the heavy Higgs into the pair of dark matter and thee mirror fermion decaying into a SM fermion and single dark matter are given by
\bea
\Gamma(\widetilde{H}_i \rightarrow  A_s^0  A_s^0) &=& \frac{ y_{\widetilde{H}_i A_s^0  A_s^0}^2 }{32 \pi M_{\widetilde{H}_i} } \, \left( 1 -  \frac{M_{A_s^0}^2}{M_{\widetilde{H}_i}^2}  \right)^{1 \over 2}\\
\Gamma(f_{\rm MF} \rightarrow  f_{\rm SM}  A_s^0) &=& \frac{M_{f_{\rm MF}}}{8 \pi } \, y_{f_{MF} \, f_{SM} \,  A_s^0}^2 .
\label{eq:decayMF}
\eea
with the corresponding coupling strengths given in terms of the mixing angles in the scalar sector (see eqns.~\ref{eq:pot}-\ref{eq:MhhMx})
\begin{equation} { 
\begin{aligned}
y_{\widetilde{H}_s \, A_s^0 A_s^0} &= {\Fontskrivan O}_H^{61} \, y_{H_1^0 \, A_s^0 A_s^0} + {\Fontskrivan O}_H^{62} \, y_{H_2^0 \, A_s^0 A_s^0}  + {\Fontskrivan O}_H^{63} \, y_{H_{1M}^0 \, A_s^0 A_s^0} + {\Fontskrivan O}_H^{64} \, y_{H_{2M}^0 \, A_s^0 A_s^0} + {\Fontskrivan O}_H^{65} \, y_{H_s^0 \, A_s^0 A_s^0}, \\
y_{\widetilde{H} \, A_s^0 A_s^0} &= {\Fontskrivan O}_H^{51} \, y_{H_1^0 \, A_s^0 A_s^0} + {\Fontskrivan O}_H^{52} \, y_{H_2^0 \, A_s^0 A_s^0}  + {\Fontskrivan O}_H^{53} \, y_{H_{1M}^0 \, A_s^0 A_s^0} + {\Fontskrivan O}_H^{54} \, y_{H_{2M}^0 \, A_s^0 A_s^0} + {\Fontskrivan O}_H^{55} \, y_{H_s^0 \, A_s^0 A_s^0} ,  \\
y_{\widetilde{H}^{'} \, A_s^0 A_s^0} &= {\Fontskrivan O}_H^{41} \, y_{H_1^0 \, A_s^0 A_s^0} + {\Fontskrivan O}_H^{42} \, y_{H_2^0 \, A_s^0 A_s^0}  + {\Fontskrivan O}_H^{43} \, y_{H_{1M}^0 \, A_s^0 A_s^0} + {\Fontskrivan O}_H^{44} \, y_{H_{2M}^0 \, A_s^0 A_s^0} + {\Fontskrivan O}_H^{45} \, y_{H_s^0 \, A_s^0 A_s^0} ,  \\
y_{\widetilde{H}^{''} \, A_s^0 A_s^0} &= {\Fontskrivan O}_H^{31} \, y_{H_1^0 \, A_s^0 A_s^0} + {\Fontskrivan O}_H^{32} \, y_{H_2^0 \, A_s^0 A_s^0}  + {\Fontskrivan O}_H^{33} \, y_{H_{1M}^0 \, A_s^0 A_s^0} + {\Fontskrivan O}_H^{34} \, y_{H_{2M}^0 \, A_s^0 A_s^0} + {\Fontskrivan O}_H^{35} \, y_{H_s^0 \, A_s^0 A_s^0} ,  \\
y_{\widetilde{H}^{'''} \, A_s^0 A_s^0} &= {\Fontskrivan O}_H^{21} \, y_{H_1^0 \, A_s^0 A_s^0} + {\Fontskrivan O}_H^{22} \, y_{H_2^0 \, A_s^0 A_s^0}  + {\Fontskrivan O}_H^{23} \, y_{H_{1M}^0 \, A_s^0 A_s^0} + {\Fontskrivan O}_H^{24} \, y_{H_{2M}^0 \, A_s^0 A_s^0} + {\Fontskrivan O}_H^{25} \, y_{H_s^0 \, A_s^0 A_s^0} ,  \\
y_{\widetilde{H}^{''''} \, A_s^0 A_s^0} &= {\Fontskrivan O}_H^{11} \, y_{H_1^0 \, A_s^0 A_s^0} + {\Fontskrivan O}_H^{12} \, y_{H_2^0 \, A_s^0 A_s^0}  + {\Fontskrivan O}_H^{13} \, y_{H_{1M}^0 \, A_s^0 A_s^0} + {\Fontskrivan O}_H^{14} \, y_{H_{2M}^0 \, A_s^0 A_s^0} + {\Fontskrivan O}_H^{15} \, y_{H_s^0 \, A_s^0 A_s^0}  , \\
y_{\widetilde{H}_i \widetilde{H}_i \, A_s^0 A_s^0} &= \lambda_{4a}, \,\, y_{f_{MF} \, f_{SM} \,  A_s^0} = y_s \, ({\rm f ~is~ quark}), \, y_{s\ell} \, ({\rm f ~is~ lepton}),
\end{aligned}}
\label{eq:sig2}
\end{equation}
where,
\vspace{-0ex}
\begin{equation}
\begin{aligned}
     y_{H_1^0 \, A_s^0 A_s^0}   &  =     \lambda_{4a} v_{1} + 2 \lambda_{5c} v_{1M} + 2 \lambda_{5c} v_{2M} ,\\
        y_{H_2^0 \, A_s^0 A_s^0}   &  =    2 \lambda_{5c} v_{1M} +  \lambda_{4a} v_{2} + 2 \lambda_{5c} v_{2M},
\\
      y_{H_{1M}^0 \, A_s^0 A_s^0}   &  =    2 \lambda_{5c} v_{1} +  \lambda_{4a} v_{1M} + 2 \lambda_{5c} v_{2},
\\
       y_{H_{2M}^0 \, A_s^0 A_s^0}   &  =   2 \lambda_{5c} v_{1} + 2 \lambda_{5c} v_{2} +  \lambda_{4a} v_{2M},
 \\
        y_{H_s^0 \, A_s^0 A_s^0}   &  =    2 \lambda_{s} v_{s},\,\, 
       y_{H_1^{0'} \, A_s^0 A_s^0}     =   0.
\end{aligned}
\label{eq:sig1}
\end{equation}
where $H_1^0$,$~H_2^0$,$~H_{1M}^0$,$~H_{2M}^0$,$~ H_s^0$ and $H_{1}^{0'}$ is the unphysical scalar fields (before mixing, see the Section~\ref{sec:scalar_sect} for details). The CP-even scalar incorporating both the triplet scalars $H_1^{0'} = \sqrt{\frac{2}{3}} \, \chi^{0r}+ \sqrt{\frac{1}{3}} \, \zeta^0$ does not have the direct coupling to PSS dark matter, i.e., $y_{H_1^{0'} \, A_s^0 A_s^0}=0$.
The other scalars are unable to decay to PSS dark matter due to conservation of the $U(1)_{\rm SM} \times U(1)_{\rm MF}$ symmetry, charge, etc.
Hence the initial DM density coming from the decay scenario mainly depends on the decay of these CP-even scalars and mirror fermions. 
One can see from these eqns.~\ref{eq:sig2} and~\ref{eq:sig1} that the decay of these heavy (physical) scalar fields and mirror fermions can be controlled by the $\lambda_{5c},\lambda_{4a},\lambda_{s}, y_s, y_{s\ell}$ and VEVs with the mass of the dark matter mainly depending on the $\lambda_{5c}$ and VEVs (see eqn.~\ref{eq:AsMass}). 
 \begin{figure}[h!]
 \begin{center}
 {\includegraphics[width=2.8in,height=2.8in, angle=0]{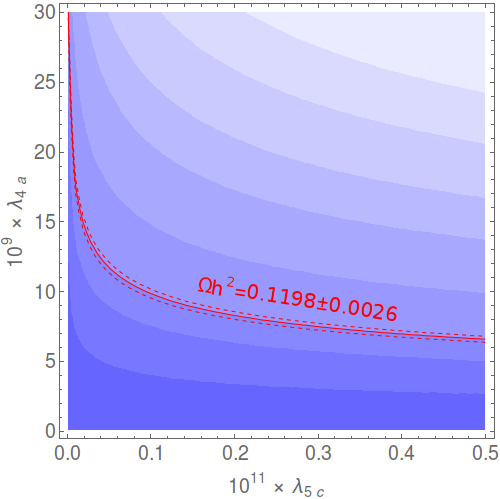}}
 \hskip 0.2cm
 {\includegraphics[width=2.8in,height=2.8in, angle=0]{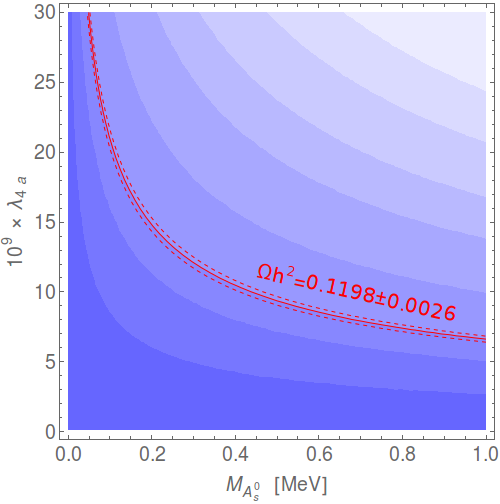}} 
 \caption{ \it Plots show the variation of parameters $\lambda_{5c}$ and $\lambda_{4a}$ in left panel and similarly show the dark matter mass against $\lambda_{4a}$ variation in right panel. These plots are generated for BP-3 as in Table~\ref{table-BP}. In both plots the red solid line represents $\Omega h^2 = 0.1198$ and the dashed red lines correspond to the $3\sigma$ variation in $\Omega h^2$. The lighter region corresponds to higher values of $\Omega h^2$. For both these plots the contribution from the mirror fermion decay $f_{MF}\rightarrow f_{SM} A_s^0$ is sub-dominant and has been neglected.}
  \label{fig:Relic}
 \end{center}
 \end{figure}

As an example, let us first neglect the contribution from the decay of the mirror fermions $f_{MF}\rightarrow f_{SM} A_s^0$ (we consider $y_{s\ell}<<10^{-9}$) and consider the benchmark point BP-3 (see Table~\ref{table-BP}). We choose $\lambda_s=10^{-15}$, hence the lightest CP-even scalar field will not be able to decay to the dark matter ($M_{\widetilde{H}_s}< 2 M_{A_s^0}$). 
The other CP-even states including the 125 GeV scalar field could decay into the dark matter which increases the abundance of the dark matter. Using BP-3 and $\lambda_{5c}=3.2\times 10^{-12}$ and $\lambda_{4a}=7.377\times 10^{-9}$, we obtain the dark matter mass as $M_{A_s^0}=0.808$ MeV  and find the numerical values of the coupling strengths $y_{\widetilde{H}_s \, A_s^0 A_s^0} = 1.630 \times 10^{-11}$, $y_{\widetilde{H} \, A_s^0 A_s^0} = -1.021 \times 10^{-6}$, $y_{\widetilde{H}^{'} \, A_s^0 A_s^0} = 3.390 \times 10^{-6}$, $y_{\widetilde{H}^{''} \, A_s^0 A_s^0} =  1.059 \times 10^{-21} $, $y_{\widetilde{H}^{'''} \, A_s^0 A_s^0} = -9.424\times 10^{-6}$ and $y_{\widetilde{H}^{''''} \, A_s^0 A_s^0} = 3.669\times 10^{-6}$. Finally, we obtain the relic density to be $\Omega h^2 = 0.1198$. 
We show the variation of the parameters $\lambda_{5c}$ and $\lambda_{4a}$ in Fig.~\ref{fig:Relic}(left) and similarly show the dark matter mass against $\lambda_{4a}$ variation in Fig.~\ref{fig:Relic}(right). In both plots the red solid line represents $\Omega h^2 = 0.1198$ and the dashed red lines correspond to the $3\sigma$ variation in $\Omega h^2$ with the darker region corresponding to the lower values of $\Omega h^2$.

\begin{figure}[h!]
 \begin{center}
 {\includegraphics[width=4in,height=3.2in, angle=0]{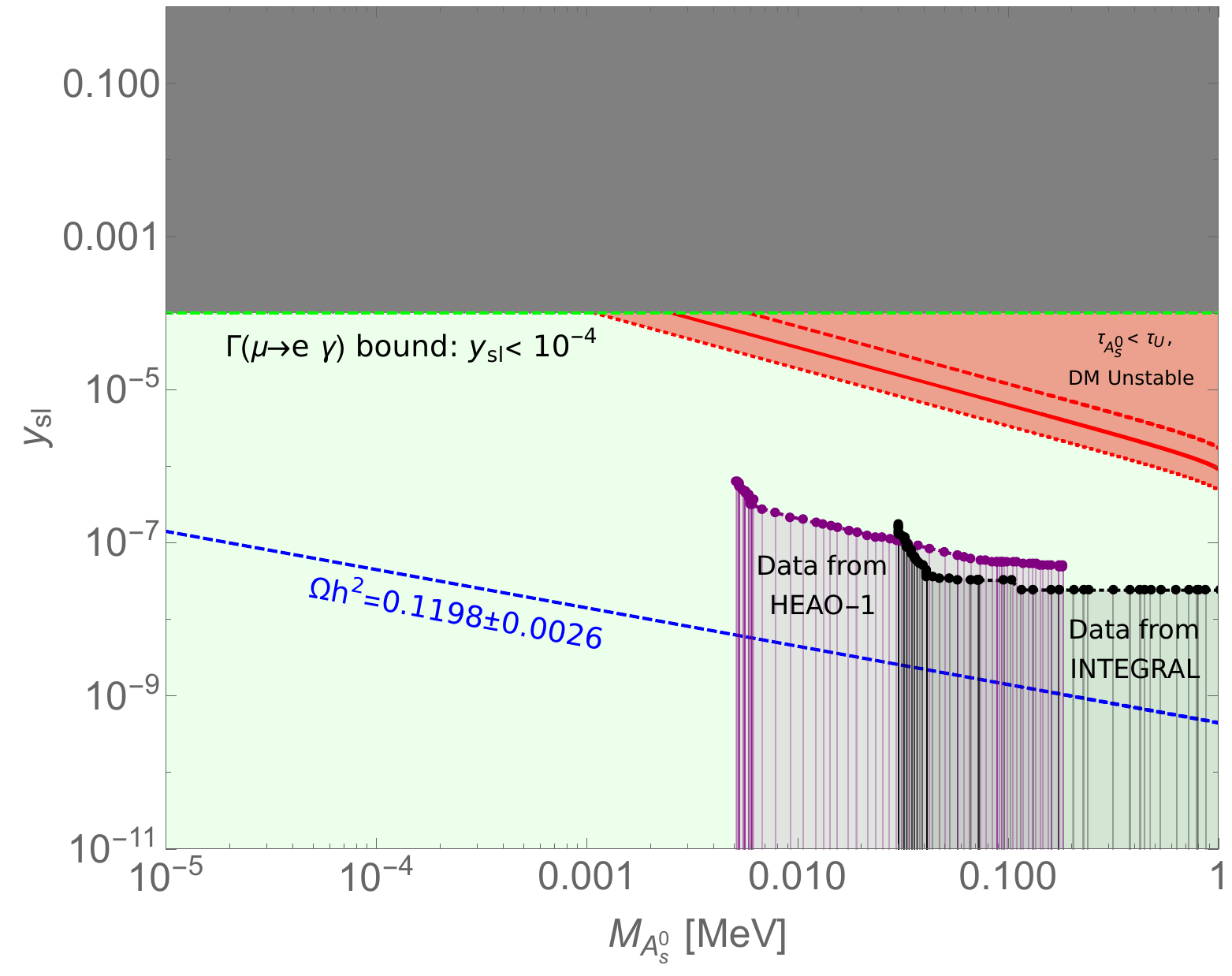}}  
 \caption{ \it The same $y_{s\ell}-M_{A_s^0}$ plot as in Fig.~\ref{fig:gslMAs} showing the relic density constraint and including the indirect detection bounds applicable in the parameter space. The blue dashed line indicates the relic density $3\sigma$ band $\Omega h^2=0.1198\pm 0.0026$. The region above the purple and black shaded lines are ruled out from the HEAO-1~\cite{Gruber:1999yr} and INTEGRAL~\cite{Bouchet:2008rp} indirect detection experiments. }
  \label{fig:gslMAs1}
 \end{center}
 \end{figure}

We find that if we neglect the contribution coming from the scalar fields, the dark matter abundance could increase from the decay of the mirror fermions $f_{MF}\rightarrow f_{SM} A_s^0$. In this case, the contribution from the mirror quarks will also be negligibly small; as the corresponding Yukawa couplings are very small ($y_s \sim y_{sq} \sim y_{su} \sim y_{sd} < 0.1 y_{s\ell}$~\cite{Hung:2017pss,Hung:2017exy}). We obtain the relic density in the right ballpark for $y_{s\ell}\sim 4.428 \times 10^{-9}$ for $M_{A_s^0} = 10 $ keV and $y_{s\ell}\sim 4.429  \times 10^{-10}$ with $M_{A_s^0} = 1 $ MeV. 
The variation of dark matter mass against $y_{s\ell}$ is shown in Fig.~\ref{fig:gslMAs1}. The blue dashed line indicates the $3\sigma$ relic density $\Omega h^2=0.1198 \pm 0.0026$ band~\cite{Aghanim:2018eyx}. The line above this blue line will overclose the Universe. {The indirect detection bounds form HEAO and INTEGRAL experiments will be discussed in the section~\ref{sec:dm_indirect}. }

 \begin{figure}[h!]
 \begin{center}
 {\includegraphics[width=2.8in,height=2.8in, angle=0]{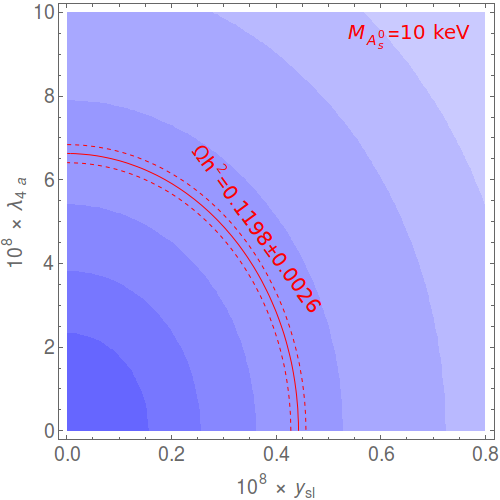}}
 \hskip 0.2cm
 {\includegraphics[width=2.8in,height=2.8in, angle=0]{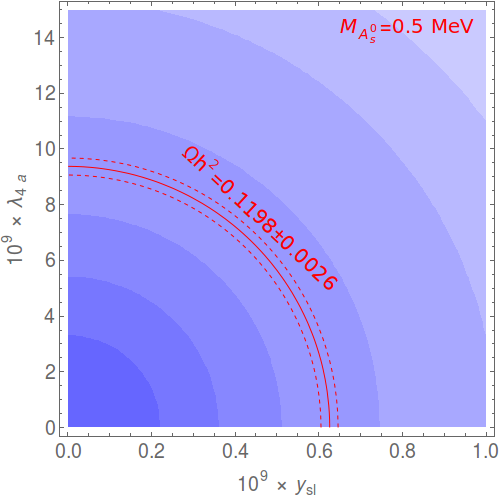}} 
 \caption{ \it  Plots show the variation of parameters $\lambda_{4a}$ and $y_{s\ell}$ for two different dark matter masses $M_{A_s^0} = 10$ keV (left) and $0.5$ MeV (right). These plots are also generated for BP-3 as in Table~\ref{table-BP}. In both plots the red solid line represents $\Omega h^2 = 0.1198$ and the dashed red lines correspond to the $3\sigma$ variation in $\Omega h^2$. The lighter region corresponds to higher values of $\Omega h^2$.}
 \label{fig:RelicAll}
 \end{center}
 \end{figure}

\begin{figure}[h!]
 \setlength{\belowcaptionskip}{-24pt}
 \begin{center}
 {\includegraphics[width=2.8in,height=2.6in, angle=0]{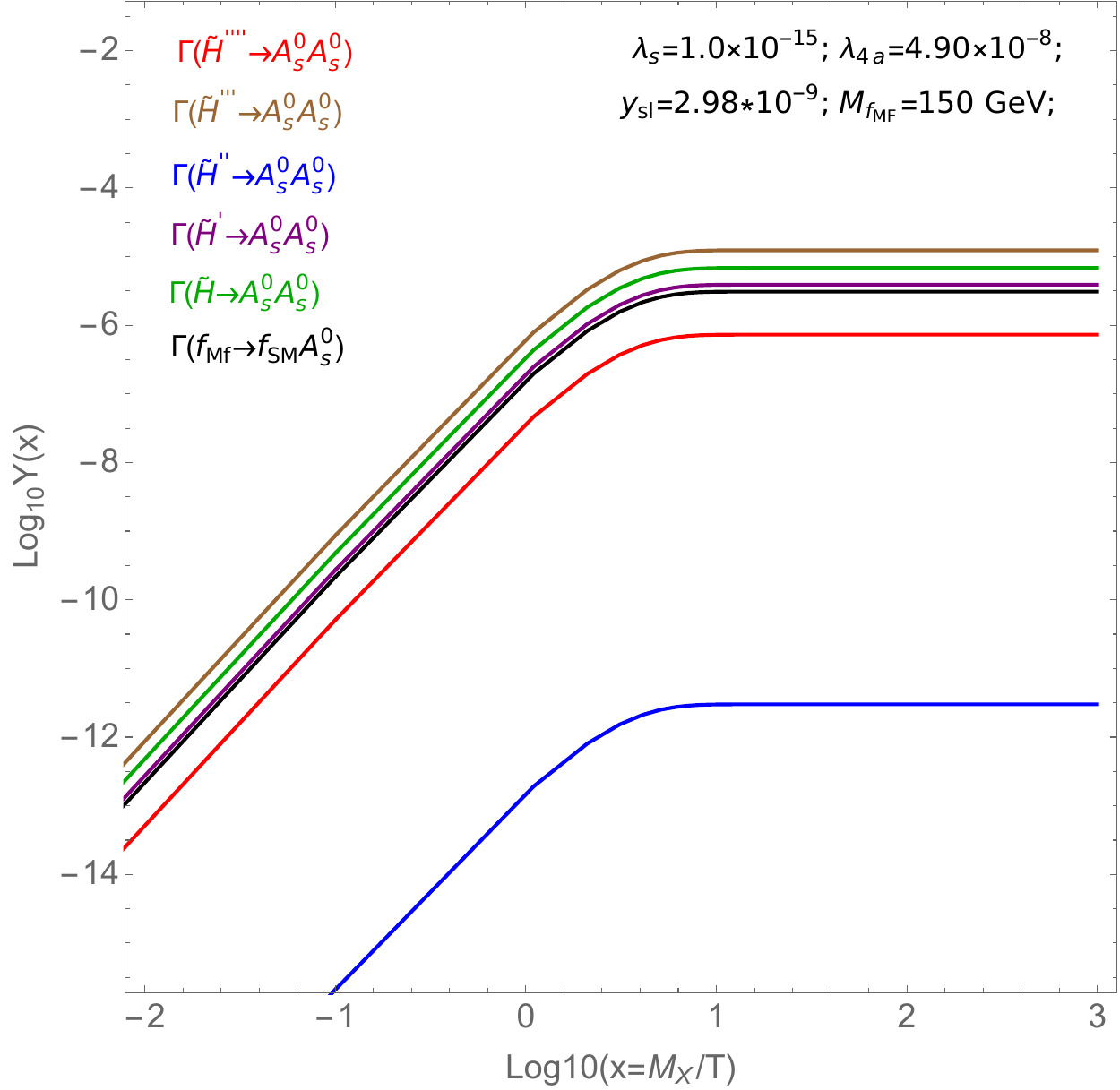}} 
 \caption{ \it   Plot shows the variation of the yield $Y(x)$ against $x$ for all the contribution coming from the heavy Higgs and mirror fermion decays. The Freeze-in mechanism effect can be clearly seen as the initial DM density is zero and increases during the cooling of the Universe and attaining a constant value after a certain temperature. We scale the mirror fermion decay contribution (black line) by a factor $1/10$ to distinguish from the others and used $M_X=100$ {\rm GeV} mass scale factor.
  }
 \label{fig:YieldTemp}
 \end{center}
 \end{figure}

We now consider all these aforementioned contributions in the relic density calculation. Increasing $\lambda_{4a}$ increases the contributions coming from the heavy Higgs decays whereas large values of $y_{s\ell}$ increases the contributions from the mirror fermions. We present such variations for two different dark matter masses $M_{A_s^0} = 10$ keV and $0.5$ MeV respectively in Fig.~\ref{fig:RelicAll}.
The contribution is almost equal for $\lambda_{4a}\sim 4.90\times 10^{-8}$ and $y_{s\ell} \sim 2.98 \times 10^{-9}$ with dark matter mass $M_{A_s^0} = 10$ keV and similarly for $\lambda_{4a}\sim 7.447\times 10^{-9}$ and $y_{s\ell} \sim 3.78 \times 10^{-10}$ with dark matter mass $M_{A_s^0} = 0.5$ MeV. The dark matter abundance increases as a result of the decay of the mirror fermions $f_{MF}\rightarrow f_{SM} A_s^0$. In this case, the contributions from the mirror quarks are negligibly small as $y_s \sim y_{sq} \sim y_{su} \sim y_{sd} < 0.1 y_{s\ell}$ ~\cite{Hung:2017pss,Hung:2017exy}. 
We have also shown the evolution of the dark matter with the temperature of the Universe in Fig.~\ref{fig:YieldTemp} for the following parameters: $\lambda_s=10^{-15}$, $\lambda_{4a}\sim 4.90\times 10^{-8}$ and $y_{s\ell} \sim 2.98 \times 10^{-9}$ and $M_{A_s^0}=10$ keV. The plot clearly represents the significance of the Freeze-in mechanism in this framework, i.e., the initial DM density being zero and increasing during the cooling of the Universe. After a certain temperature (T $\sim \, \mathcal{O}$(100 GeV) as shown in Fig.~\ref{fig:YieldTemp}) the dark matter density becomes constant. 

We have also looked into the bounds coming from the free streaming length $l_{fs}$ which will denote whether the dark mater will behave as hot, warm or cold. The $l_{fs}> 2$ Mpc region stands for hot dark matter region and can create challenges for the structure formation~\cite{Dev:2013yza}. We avoid these regions (dark matter mass $<<1$ keV) in this analysis and calculate the free streaming length~\cite{Choi:2020kch} for the dark matter with mass rage $10^{-5}-1$ MeV and find it to be consistently less than $10$ kpc in the parameter space referred to in the Fig.~\ref{fig:gslMAs1}. Hence we conclude that in this scenario $A^0_s$ behaves as a cold dark matter candidate~\cite{Dev:2013yza}.

In summary, in this section we have discussed the Freeze-in scenario for the dark matter candidate possible in this framework and have discussed the stability bounds on the dark matter in the previous section. Now as we recognize that the galactic and extra-galactic diffuse $X-ray$ or $\gamma-ray$ may come from the decay or annihilation of a dark matter through the loop placing stringent constraints on the model parameter space of any dark matter model~\cite{Essig:2013goa}, in the upcoming section we devote a discussion on the possible additional bounds coming from the available data from various indirect dark matter search experiments.

\section{Indirect Detection of Dark Matter}
\label{sec:dm_indirect}
Let us now divert our attention towards the various dark matter detection prospects plausible in this framework. One interesting possibility is to look for the excess of photon or charged particles from different directions of our sky coming from the annihilation or the decay of dark matter particles. In this scenario, the dark matter decay is less constrained from the early Universe cosmology and the galactic and extra-galactic diffuse $X-ray$ or $\gamma-ray$ background puts the most stringent bounds on the model parameters for the decaying DM~\cite{Essig:2013goa}. \\
The observations of the Fermi Large Area Telescope (Fermi-LAT) put limits on the weak-scale DM (mass $\mathcal{O}(100)$ MeV), having decay lifetime of $\tau_{A_s^0}>10^{26}\,s$, many orders of magnitude larger than the age of the Universe~\cite{Ackermann:2012qk, Cirelli:2009dv}. 
Although the usual gamma-ray constraints from the Fermi-LAT do not apply for the $<\mathcal{O}(100)$ MeV masses, several other satellite-based experiments such as HEAO-1~\cite{Gruber:1999yr}, INTEGRAL~\cite{Bouchet:2008rp}, COMPTEL~\cite{kappadath}, EGRET~\cite{Strong:2004de} are sensitive to photons with energies well below $\mathcal{O}(100)$ MeV. 
In this particular framework with dark matter mass less than $\mathcal{O}(1)$ MeV, only HEAO-1~\cite{Gruber:1999yr} and INTEGRAL~\cite{Bouchet:2008rp} can put stringent constraints on the dark matter parameters. In Fig.~\ref{fig:gslMAs1}, we show~\footnote{We extract these data from Ref.~\cite{Essig:2013goa} where the authors have put bounds on the lifetime of a scalar dark matter decaying to two photons. We translate it into the interaction strength ($y_{s\ell}$),
i.e. the interaction between dark matter, SM and mirror leptons, to constrain our model parameter space.} the limits coming from HEAO-1~\cite{Gruber:1999yr} and INTEGRAL~\cite{Bouchet:2008rp} in $y_{s\ell}-\text{vs}-M_{A_s^0}$ plane which puts a strong upper limit on the value of $y_{s\ell} \leq 10^{-6}- 10^{-7}$. However, as we are considering the non-thermal production of dark matter, the relic density bound can still be satisfied with smaller $y_{s\ell}$ as shown in the blue line in Fig.~\ref{fig:gslMAs1}. 

\section{Collider analysis}
\label{sec:colld}
We now investigate the Collider signals as the mirror fermion interactions with the SM particles make it feasible to probe DM via the Collider sector. We would like to point out that the mirror fermion masses are assumed to be less than 1 TeV and these mass range is currently valid evading the Collider bounds under the current assumptions applicable in this framework. We focus on the existing literature (see Ref.~\cite{No:2019gvl}) outlining some possible ways to explore the DM signals. The Ref.~\cite{No:2019gvl} has primarily investigated the FIMP dark matter scenario in the context of 14 TeV LHC data with  high integrated luminosity at the MATHUSLA surface detector in the near future. In this framework, there is a possibility of charged track formations due to the mirror fermion decays into SM fermion and dark matter fields at the Colliders.
The decay width of the charged mirror fermion is given by eqn.~\ref{eq:decayMF}. The decay width ($\Gamma(f_{MF} \rightarrow f_{SM} A_s^0$) for the mirror fermions is proportional to $y_{s\ell}^2$ and the mirror fermion masses.
As discussed previously, an estimated value of the coupling $y_{s\ell} \lesssim 10^{-8}$ is needed to obtain the correct dark matter density through the Freeze-in process.
We find the decay length for these charged fermions to be $\mathcal{O}(1)$ meter for this choice of $y_{s\ell}$.
We are interested to see if we can get sufficient number of events from this charged tracks for the DM detection.
It mainly depends on the production cross-section $\sigma^{\rm LHC}_{\sqrt{s}}$ of the mother particle and luminosity $\mathcal{L}$ at the detector.
The number of events at the LHC is calculated in Ref.~\cite{No:2019gvl} and is given by
\begin{equation}
N_{events} = \sigma^{\rm LHC}_{\sqrt{s}} \, \mathcal{L}  \, \int P^{\rm MATH}_{\rm Decay},~~~~~{\rm with}~~~P^{\rm MATH}_{\rm Decay}=0.05(e^{-\frac{L_a}{\beta \, c \, \tau_{f_{MF}}}} - e^{-\frac{L_b}{\beta \, c \, \tau_{f_{MF}}}}).
\end{equation}
It has been reported in Ref.~\cite{No:2019gvl} that the number of events is $N_{events}\geq 3$ for $\sqrt{s}=13$ TeV with an integrated luminosity $\mathcal{L}=3000 ~{\rm fb^{-1}}$ using a specific model parameter space, showing that the MATHUSLA100/200 detector could detect these mother particles up to $1$ TeV mass with the dominant production of the mother particles coming from the Drell-Yan processes. Following this reference, in this framework, as the production of the mother particles (mirror fermions) are hugely suppressed due to the small mixing (see the BPs) between the mirror and standard model particles, we find the production cross-section to be less than $\mathcal{O}(10^{-10}) $ fb. Hence for this scenario, larger luminosity ($>>1~{\rm ab^{-1}}$) and energy is needed to get any significant events at the current MATHUSLA surface detector.

\section{Conclusion} 
\label{sec:summary}
In this work, we have presented a model incorporating a sub-MeV DM based on the exploration of the scalar sector of the Electroweak-scale Right-handed neutrino model. The idea of EW-$\nu_R$ model with additional GeV scale mirror fermions with large displaced vertices containing long lived particles (LLP) signatures is already highly appealing from the LHC perspective and has been extensively studied before \cite{Chakdar:2016adj,Chakdar:2015sra,Chakdar:2020igt}. The rich scalar sector of EW-$\nu_R$ includes doublets, triplets and an additional complex-singlet scalar $\Phi_s$ and the imaginary part of this complex singlet (pseudo-Nambu Goldstone (PNG) boson), $A_s^0$ is investigated to be a plausible DM candidate in the present context. The dark matter $A_s^0$ acquires a sub-MeV mass from the explicit breaking term in the scalar potential; this explicit breaking term is characterized by some mass scale assumed to be much smaller than the scale of spontaneous symmetry breaking (SSB). The various model parameters present in the scalar sector of this framework are investigated to generate possible benchmark points in the context of a sub-MeV dark matter, satisfying the current  $125 $ GeV Higgs branching ratio and signal strength constraints from the LHC. 
In this work, we have focused on the limitations of the well established Freeze out mechanism, for which the observed abundance is set almost exclusively by the annihilation cross-section and is largely insensitive to unknown details of early Universe and to the mass, producing overabundance for the sub-MeV DM particle $A_s^0$ we are interested to study. Null results at direct detection experiments have currently put tight constraints on the WIMP paradigm and alternative possibilities like ALP, axions, SIMPs, FIMPS have become relevant in this context. We have implemented the Freeze-in mechanism to obtain the correct order of relic density for the chosen dark matter masses $< 1 $ MeV. 
For such feebly interacting massive particles or FIMPs, we can invoke the non-thermal Freeze-in mechanism that necessitates feeble interactions making it one of the reasons to have such a tiny fine-tuned coupling present in this EW-$\nu_R$ model. 
We find that the non-thermally produced PSS (pseudo-scalar singlet) can serve as a 
viable $\leq 1 $ MeV dark metter depending on the parameters $\lambda_{5c}, \lambda_{4a},\lambda_{s}$, $y_{s}$, $y_{s\ell}$ and VEVs satisfying the dark matter relic density.\\
Using the Freeze in mechanism to investigate the scalar sector of the EW-$\nu_R$, we obtain a significant parameter space of $y_{s\ell}-M_{A_s^0}$ for the sub-MeV dark matter mass satisfying the correct relic density and successfully put bounds on the coupling strength $y_{s\ell}$ vs $M_{A_s^0}$ exclusion region from the stability (lifetime of DM > lifetime of the Universe) of the dark matter, the rare processes ($\mu \rightarrow e\gamma$, and $\mu - e$ conversion), several indirect detection experiments constraining this particular mass region (HEAO-1 and INTEGRAL) etc. We also found that indirect detection experiments, such as Fermi-LAT data are currently unable to successfully constrain the parameter space of $y_{s\ell}-M_{A_s^0}$ for the mass range of $< 1 $ MeV. Also, due to such feeble interactions, it is challenging to get handle on the signatures of this light dark matter from the direct-detection experiments through nucleon-dark matter scattering as well as dark matter-electron interactions via the magnon excitation.\\
Through our investigation, we found that $y_{s\ell} \sim 10^{-8}$ is needed for the correct relic abundance and have pointed out the parameter space available for sub-MeV FIMP dark matter ready to be explored by the future experiments. We have discussed in detail the possible future implications of this scenario in the Collider searches, in specific the MATHUSLA detector. From a particle physics point of view, this scenario is highly interesting as the model framework has already been successful incorporating the non sterile right handed neutrinos with Electroweak scale Majorana masses. Having a substantial parameter space available to explore after implementing relevant constraints for a natural sub-MeV FIMP Dark matter particle in the current and future experiments makes this scenario even more relevant and exciting. The current framework casts light on the feebly explored sub-MeV dark sector frontier, and offers many opportunities for exciting and profound discoveries in the future.

\section{Acknowledgements}
SC is supported by the College of Holy Cross Bachelor Ford Summer fellowship '20-'21. DKG, NK and DN would like to thank Prof. Satyanarayan Mukhopadhyay and Dr. Anirban Biswas for useful discussions. NK would like to acknowledge support from the DAE, Government of India and the Regional Centre for Accelerator-based Particle Physics (HRI-RECAPP-2023-11).

\appendix

\section{ Model}

\label{app:A}
The model framework and motivations  including the gauge structure, particle content~\cite{Hoang:2014pda} has already been introduced previously in this work.  
Here we include a summary of the details on the extended scalar sector of the Electro-weak Right handed neutrino model that is vital for studying the dark matter portion of this framework. 
The present framework includes a rich scalar sector incorporating four doublets (two for the THDM like, two for mirror sector), two triplets and a singlet given by
\allowdisplaybreaks
\begin{eqnarray}
\Phi_1 &=& \begin{pmatrix}
 \phi_{1}^{0,*} & \phi_{1}^+ \\ 
 \phi_{1}^- &  \phi_{1}^{0} 
\end{pmatrix},~
\Phi_{1M}= \begin{pmatrix}
 \phi_{1M}^{0,*} & \phi_{1M}^+ \\ 
 \phi_{1M}^- &  \phi_{1M}^{0} 
\end{pmatrix},~ 
\Phi_2 = \begin{pmatrix}
 \phi_{2}^{0,*} & \phi_{2}^+ \\ 
 \phi_{2}^- &  \phi_{2}^{0} 
\end{pmatrix},~
\Phi_{2M}= \begin{pmatrix}
 \phi_{2M}^{0,*} & \phi_{2M}^+ \\ 
 \phi_{2M}^- &  \phi_{2M}^{0} 
\end{pmatrix},\nn\\
~ \tilde{\chi} &=& \begin{pmatrix}
\chi^+/\sqrt 2 & \chi^{++}\\
\chi^0 & -\chi^+/\sqrt 2
\end{pmatrix}, ~ \xi = \begin{pmatrix}
 \xi^+, \xi^0,\xi^-
\end{pmatrix},~
{\rm Complex ~singlet ~scalar}=\Phi_s
\label{eq:fieldS}
\end{eqnarray}
The transformation of these scalar multiplet under $U(1)_{\rm SM} \times U(1)_{\rm MF}$ symmetry are as follows: 
$\Phi_{1,2}\rightarrow e^{-2 i \alpha_{\rm SM}}\,\Phi_{1,2},\Phi_{1M,2M} \rightarrow e^{2 i \alpha_{\rm MF}}\,\Phi_{1M,2M},~  \tilde{\chi} \rightarrow e^{-2 i \alpha_{\rm MF}}\, \tilde{\chi} ,\,\xi \rightarrow \xi$ and $\Phi_s \rightarrow e^{- i (\alpha_{\rm SM} + \alpha_{\rm MF})}\,\Phi_s$. Additionally,
the Higgs potential has a global $SU(2)_{\rm L} \times SU(2)_{\rm R}$ symmetry.
The triplet and doublet scalars transform as $(3,3)$ and $(2,2)$ under that 
global symmetry.
The combination of these triplet can be written as~\cite{Hoang:2014pda}
\begin{eqnarray}
 \chi = \begin{pmatrix}
\chi^0 & \xi^+ & \chi^{++}\\
\chi^{-} & \xi^0 &\chi^{+}\\
\chi^{--} & \xi^- &\chi^{0,*}
\end{pmatrix},
\label{eq:fieldS2}
\end{eqnarray}
Proper vacuum alignment gives
\begin{eqnarray}
<\Phi_1> = \begin{pmatrix}
 v_1/\sqrt 2 & 0 \\ 
 0 &  v_1/\sqrt 2
\end{pmatrix},~
<\Phi_{1M}>= \begin{pmatrix}
 v_{1M}/\sqrt 2 & 0 \\ 
 0 &  v_{1M}/\sqrt 2
\end{pmatrix},\\
<\Phi_2> = \begin{pmatrix}
 v_2/\sqrt 2 & 0 \\ 
 0 &  v_2/\sqrt 2
\end{pmatrix},~
<\Phi_{2M}>= \begin{pmatrix}
 v_{2M}/\sqrt 2 & 0 \\ 
 0 &  v_{2M}/\sqrt 2
\end{pmatrix},\\
<\chi> = \begin{pmatrix}
v_M & 0 & 0\\
0 & v_M  &0\\
0 & 0 &v_M 
\end{pmatrix},~~<\Phi_s>=v_s~~~~~~~~~~~~~~~~~~~~~~
\label{eq:fieldS3}
\end{eqnarray}
The generic scalar potential for these scalars can now be written as
\begin{eqnarray}
\label{eq:pot}
V&=& \lambda_{1a} \Big[ Tr \Phi_1^\dagger \Phi_1 -v_1^2 \Big]^2 +\lambda_{2a} \Big[ Tr \Phi_{1M}^\dagger \Phi_{1M} -v_{1M}^2 \Big]^2 + \lambda_{1b} \Big[ Tr \Phi_2^\dagger \Phi_2 \nn\\
&&-v_2^2 \Big]^2+\lambda_{2b} \Big[ Tr \Phi_{2M}^\dagger \Phi_{2M}-v_{2M}^2 \Big]^2 + \lambda_3 \Big[ Tr \chi^\dagger \chi -3 v_M^2 \Big]^2  \nn\\
&&+ \lambda_s \Big[  \Phi_s^\dagger \Phi_s - v_s^2 \Big]^2+ \lambda_4 \Big[ Tr \Phi_1^\dagger \Phi_1 -v_1^2 + Tr \Phi_{1M}^\dagger \Phi_{1M} -v_{1M}^2 \nn\\
&&+ Tr \Phi_2^\dagger \Phi_2 -v_2^2 + Tr \Phi_{2M}^\dagger \Phi_{2M} -v_{2M}^2 +  Tr \chi^\dagger \chi -3 v_M^2 \Big]^2 \nn\\
&&+ \lambda_{4a} \Big[ Tr \Phi_1^\dagger \Phi_1 -v_1^2 + Tr \Phi_{1M}^\dagger \Phi_{1M} -v_{1M}^2 + Tr \Phi_2^\dagger \Phi_2 -v_2^2 \nn\\
&&+ Tr \Phi_{2M}^\dagger \Phi_{2M} -v_{2M}^2 +  Tr \chi^\dagger \chi -3 v_M^2\Big]\Big[  \Phi_s^\dagger \Phi_s - v_s^2\Big]\nn\\
&&+ \lambda_{5a}    \Big[ (Tr \Phi_1^\dagger \Phi_1) (Tr \chi^\dagger \chi)-  2  (Tr \Phi_1^\dagger \frac{\tau^a}{2} \Phi_1 \frac{\tau^b}{2}) (Tr \chi^\dagger T^a \chi T^b) \Big] \nn\\
&& +\lambda_{6a} \Big[ (Tr \Phi_{1M}^\dagger \Phi_{1M}) (Tr \chi^\dagger \chi)-  2  (Tr \Phi_{1M}^\dagger \frac{\tau^a}{2} \Phi_{1M} \frac{\tau^b}{2}) (Tr \chi^\dagger T^a \chi T^b) \Big]\nn\\
&&+ \lambda_{5b}    \Big[ (Tr \Phi_2^\dagger \Phi_2) (Tr \chi^\dagger \chi)-  2  (Tr \Phi_2^\dagger \frac{\tau^a}{2} \Phi_2 \frac{\tau^b}{2}) (Tr \chi^\dagger T^a \chi T^b) \Big] \\
&& +\lambda_{6b} \Big[ (Tr \Phi_{2M}^\dagger \Phi_{2M}) (Tr \chi^\dagger \chi)-  2  (Tr \Phi_{2M}^\dagger \frac{\tau^a}{2} \Phi_{2M} \frac{\tau^b}{2}) (Tr \chi^\dagger T^a \chi T^b) \Big]\nn\\
&&+ \lambda_{5c} \Big[ \{ \Phi_s^\dagger \Phi_s (Tr \Phi_{1}^\dagger \Phi_{1M} + Tr \Phi_{1}^\dagger \Phi_{2M} +Tr \Phi_{2}^\dagger \Phi_{1M} +Tr \Phi_{2}^\dagger \Phi_{2M}) +h.c.\}\nn\\
&&-  2 \, \Phi_s^\dagger \Phi_s (Tr \Phi_{1}^\dagger \Phi_{1M} + Tr \Phi_{1}^\dagger \Phi_{2M} +Tr \Phi_{2}^\dagger \Phi_{1M} +Tr \Phi_{2}^\dagger \Phi_{2M}) \Big]\nn\\
&& +\lambda_{7a} \Big[(Tr \Phi_{1}^\dagger \Phi_{1}) (Tr \Phi_{1M}^\dagger \Phi_{1M})-(Tr \Phi_{1}^\dagger \Phi_{1M}) (Tr \Phi_{1M}^\dagger \Phi_{1}) \Big]\nn\\
&& +\lambda_{7b} \Big[(Tr \Phi_{2}^\dagger \Phi_{2}) (Tr \Phi_{2M}^\dagger \Phi_{2M})-(Tr \Phi_{2}^\dagger \Phi_{2M}) (Tr \Phi_{2M}^\dagger \Phi_{2}) \Big]\nn\\
&& +\lambda_{7ab} \Big[(Tr \Phi_{1}^\dagger \Phi_{1}) (Tr \Phi_{2}^\dagger \Phi_{2})-(Tr \Phi_{1}^\dagger \Phi_{2}) (Tr \Phi_{2}^\dagger \Phi_{1}) \Big]\nn\\
&& +\lambda_{7Mab} \Big[(Tr \Phi_{1M}^\dagger \Phi_{1M}) (Tr \Phi_{2M}^\dagger \Phi_{2M})-(Tr \Phi_{1M}^\dagger \Phi_{2M}) (Tr \Phi_{2M}^\dagger \Phi_{1M}) \Big]\nn\\
&& + \lambda_{7aMb} \Big[(Tr \Phi_{1}^\dagger \Phi_{1}) (Tr \Phi_{2M}^\dagger \Phi_{2M})-(Tr \Phi_{1}^\dagger \Phi_{2M}) (Tr \Phi_{2M}^\dagger \Phi_{1}) \Big]\nn\\
&& + \lambda_{7abM} \Big[(Tr \Phi_{2}^\dagger \Phi_{2}) (Tr \Phi_{1M}^\dagger \Phi_{1M})-(Tr \Phi_{2}^\dagger \Phi_{1M}) (Tr \Phi_{1M}^\dagger \Phi_{2}) \Big]\nn\\
&& + \lambda_8 \Big[ Tr \chi^\dagger \chi \chi^\dagger \chi - (Tr \chi^\dagger \chi)^2\Big]\nn,
\end{eqnarray}
where $a,b=1,2,3$ and from~\cite{Hartling:2014zca}
\begin{eqnarray}
{T}^1=\frac{1}{\sqrt{2}}\left(
\begin{array}{ccc}
 0 & 1 & 0 \\
 1 & 0 & 1 \\
 0 & 1 & 0 \\
\end{array}
\right);~~
{T}^2=\frac{1}{\sqrt{2}}\left(
\begin{array}{ccc}
 0 & -i & 0 \\
 i & 0 & -i \\
 0 & i & 0 \\
\end{array}
\right);~~
{T}^3=\left(
\begin{array}{ccc}
 1 & 0 & 0 \\
 0 & 0 & 0 \\
 0 & 0 & -1 \\
\end{array}
\right);\\
\tau^1=\left(
\begin{array}{cc}
 0 & 1 \\
 1 & 0 \\
\end{array}
\right);~~
\tau^2=\left(
\begin{array}{cc}
 0 & -i \\
 i & 0 \\
\end{array}
\right);~~
\tau^3=\left(
\begin{array}{cc}
 1 & 0 \\
 0 & -1 \\
\end{array}
\right);~~~~~~~~~~
\label{eq:Ttau}
\end{eqnarray}
Please note the transformation
$\Phi_{1,2}\rightarrow e^{-2 i \alpha_{\rm SM}}\,\Phi_{1,2},\Phi_{1M,2M} 
\rightarrow e^{2 i \alpha_{\rm MF}}\,\Phi_{1M,2M},~  \tilde{\chi} 
\rightarrow e^{-2 i \alpha_{\rm MF}}\, \tilde{\chi} ,\,\xi \rightarrow \xi$ 
and $\Phi_s \rightarrow e^{- i (\alpha_{\rm SM} + \alpha_{\rm MF})}\,\Phi_s$. 
Hence the $\lambda_{5,6}$'s terms break explicitly the $U(1)_{\rm SM} 
\times U(1)_{\rm MF}$ symmetries. The three `massless' Nambu-Goldstone bosons 
can be obtained after spontaneous breaking of $SU(2)_L \times U (1)_Y 
\rightarrow U (1)_{em}$, with the condition $\lambda_{5a}=\lambda_{5b}=
\lambda_{6a}=\lambda_{6b}=\lambda_{7a}=\lambda_{7b}=\lambda_{7ab}=
\lambda_{7Mab}=\lambda_{7aMb}=\lambda_{7abM}=\lambda_5$ imposed on the 
potential above. The first line of $\lambda_{5c}$ term is $U(1)_{\rm SM} 
\times U(1)_{\rm MF}$ conserving, and the second line explicitly violates 
these symmetries. Both of them will help us to get exact minimization of the 
scalar potential and non-zero mass for the singlet-type complex scalar field.
There are eighteen physical scalars grouped into $5 + 3 +3+3+ 3 + 1$ of the 
custodial $SU(2)_D$ with $6$ real singlets.
Here, we would like to mention that the dedicated study of
vacuum stability condition for a multi-Higgs model like ours
is extremely complicated and is beyond the scope of this paper.

To express the Nambu-Goldstone bosons and the physical scalars let us adopt the following convenient notation:
\begin{eqnarray}
{v_{\rm SM}}&=&\sqrt{v_{1}^2+v_{1M}^2+v_{2}^2+v_{2M}^2+8 v_{M}^2}\approx 246 ~{\rm GeV}\nn\\
{s_1}&=&\frac{{v_{1}}}{{v_{\rm SM}}},~{c_1}=\frac{\sqrt{v_{1M}^2+v_{2}^2+v_{2M}^2+8 v_{M}^2}}{{v_{\rm SM}}},\nn\\
{s_2}&=&\frac{{v_{2}}}{{v_{\rm SM}}},~{c_2}=\frac{\sqrt{v_{1M}^2+v_{1}^2+v_{2M}^2+8 v_{M}^2}}{{v_{\rm SM}}},\nn\\
{s_{1m}}&=&\frac{{v_{1M}}}{{v_{\rm SM}}},~{c_{1m}}=\frac{\sqrt{v_{1}^2+v_{2}^2+v_{2M}^2+8 v_{M}^2}}{{v_{\rm SM}}},\\
{s_{2m}}&=&\frac{{v_{2M}}}{{v_{\rm SM}}},~ {c_{2m}}=\frac{\sqrt{v_{1}^2+v_{1M}^2+v_{2}^2+8 v_{M}^2}}{{v_{\rm SM}}},\nn\\
{s_m}&=&\frac{2 \sqrt{2} {v_{M}}}{{v_{\rm SM}}},~{c_m}=\frac{\sqrt{v_{1}^2+v_{1M}^2+v_{2}^2+v_{2M}^2}}{{v_{\rm SM}}}.\nn
\end{eqnarray}
Thus, $s_1^2+c_1^2=s_2^2+c_2^2=s_{1m}^2+c_{1m}^2=s_{2m}^2+c_{2m}^2=s_m^2+c_m^2=1.$ We also defined 
\begin{eqnarray}
\phi_{1}^0=\frac{1}{\sqrt{2}}(\phi_{1}^{0r}+v_{1}+i \phi_{1}^{0i}),~~ && \phi_{2}^0=\frac{1}{\sqrt{2}}( \phi_{2}^{0r}+v_{2}+i \phi_{2}^{0i}), \nn\\
\phi_{1M}^0=\frac{1}{\sqrt{2}}(\phi_{1M}^{0r}+v_{1M}+i \phi_{1M}^{0i}),~~ && \phi_{2M}^0=\frac{1}{\sqrt{2}}( \phi_{2M}^{0r}+v_{2M}+i \phi_{2M}^{0i}),\nn\\
\chi^0=\frac{1}{\sqrt{2}}(\chi^{0r}+v_{M}+i \chi^{0i}),~~ && \xi^0=( \xi^{0}+v_{M}), \Phi_s=\phi_s^{0r}+v_s+i \phi_s^{0i} \nn\\
{\rm and}~\psi^\pm=\frac{1}{\sqrt{2}}(\chi^\pm+\xi^\pm),~~ && \zeta^\pm=\frac{1}{\sqrt{2}}(\chi^\pm-\xi^\pm)\nn.
\end{eqnarray}
for the complex neutral and charged fields respectively. With these fields the Nambu-Goldstone bosons are given by
\begin{eqnarray}
y_1^\pm &=& s_{1} \phi_{1}^\pm + s_{2} \phi_{2}^\pm + s_{1M} \phi_{1M}^\pm + s_{2M} \phi_{2M}^\pm + s_M \psi^\pm,\nn\\
y_1^0=&=& -i (s_{1} \phi_{1}^{0i} + s_{2} \phi_{2}^{0i} + s_{1M} \phi_{1M}^{0i} + s_{2M} \phi_{2M}^{0i}) + i s_M \chi^{0i}.
\label{eq:gold}
\end{eqnarray}

The physical scalars can be grouped, as stated in the previous section, based on their transformation properties under $SU(2)_D$ as follows:
\begin{eqnarray}
{\rm five\text{-}plet ~(quintet)} &\rightarrow & H^{\pm\pm}_5,H^{\pm}_5,~{\rm and}~H^{0}_5\nn\\
{\rm triplet} &\rightarrow & H_{3}^{\pm},~{\rm and}~H_{3}^{0}\nn\\
{\rm triplet} &\rightarrow & H_{3}^{'\pm},~{\rm and}~H_{3}^{'0}\nn\\
{\rm triplet} &\rightarrow & H_{M}^{\pm},~{\rm and}~H_{M}^{0}\\
{\rm triplet} &\rightarrow & H_{M}^{'\pm},~{\rm and}~H_{M}^{'0}\nn\\
{\rm Real~singlet} &\rightarrow & H_{1}^{0},\, H_2^0,\, H_{1M}^0, \, H_{2M}^0,H_{1}^{0'},~{\rm and}~H_{s}^{0}\nn\\
\label{eq:unpstate}
\end{eqnarray}
where,
\begin{eqnarray}
\label{eq:state}
H^{\pm\pm}_5&=&\chi^{\pm\pm},~~
H^{+}_5=\zeta^\pm,~~
H^{0}_5=\frac{1}{\sqrt{6}} (2 \zeta^0 - \sqrt{2} \chi^{0r}),\nn\\
\nn\\
H^{+}_3&=& - \frac{s_{1} s_{M}}{c_{M}} \phi_{1}^+ - \frac{s_{2} s_{M}}{c_{M}} \phi_{2}^+ - \frac{s_{1M} s_{M}}{c_{M}} \phi_{1M}^+  - \frac{s_{2M} s_{M}}{c_{M}} \phi_{2M}^+ + c_{M} \psi_M\nn\\
H^0_3&=& i \left( \frac{s_{1} s_{M}}{c_{M}} \phi_{1}^{0i} + \frac{s_{2} s_{M}}{c_{M}} \phi_{2}^{0i} + \frac{s_{1M} s_{M}}{c_{M}} \phi_{1M}^{0i} + \frac{s_{2M} s_{M}}{c_{M}} \phi_{2M}^{0i} + c_{M} \chi^{0i} \right) \nn\\
\nn\\
H'^{+}_3&=&  \frac{ s_{1}}{c_{M}} \phi_{2}^+ - \frac{s_{2}}{c_{M}} \phi_{1}^+  + \frac{ s_{1M}}{c_{M}} \phi_{2M}^+ - \frac{s_{2M}}{c_{M}} \phi_{1M}^+ \nn\\
H'^0_3&=& i \left( \frac{s_{1} }{c_{M}} \phi_{2}^{0i} - \frac{s_{2}}{c_{M}} \phi_{1}^{0i} + \frac{s_{1M} }{c_{M}} \phi_{2M}^{0i} - \frac{s_{2M}}{c_{M}} \phi_{1M}^{0i}  \right) \\
\nn\\
H^{+}_{3M}&=& \frac{ s_{1M}}{c_{M}} \phi_{2}^+ - \frac{s_{2M}}{c_{M}} \phi_{1}^+  + \frac{ s_{1}}{c_{M}} \phi_{2M}^+ - \frac{s_{2}}{c_{M}} \phi_{1M}^+  \nn\\
H^0_{3M}&=&  i \left( \frac{s_{1M} }{c_{M}} \phi_{2}^{0i} - \frac{s_{2M}}{c_{M}} \phi_{1}^{0i} + \frac{s_{1} }{c_{M}} \phi_{2}^{0i} - \frac{s_{2}}{c_{M}} \phi_{1M}^{0i}  \right) \nn\\
\nn\\
H'^{+}_{M3}&=& \frac{ s_{2M}}{c_{M}} \phi_{2}^+ - \frac{s_{1M}}{c_{M}} \phi_{1}^+  - \frac{ s_{2}}{c_{M}} \phi_{2M}^+ + \frac{s_{1}}{c_{M}} \phi_{1M}^+ \nn\\
H'^0_{M3}&=&   i \left( \frac{s_{2M} }{c_{M}} \phi_{2}^{0i} - \frac{s_{1M}}{c_{M}} \phi_{1}^{0i} - \frac{s_{2} }{c_{M}} \phi_{2}^{0i} + \frac{s_{1}}{c_{M}} \phi_{1M}^{0i}  \right) \nn\\
\nn\\
H_1^0&=&\phi_{1}^{0r},~H_2^0=\phi_{2}^{0r},~H_{1M}^0=\phi_{1M}^{0r},~H_{2M}^0=\phi_{2M}^{0r},\nn\\
H_s^0&=&\phi_s^{0r},~~{\rm and}~~H_{1}^{0'}=\sqrt{\frac{2}{3}} \, \chi^{0r}+ \sqrt{\frac{1}{3}} \, \zeta^0,\nn
\end{eqnarray}
where, $H^{--}=(H^{++})^*$, $H^{-}_{\rm All}=-(H^{+}_{\rm All})^*$, , $H^{0}_{\rm All}=-(H^{0}_{\rm All})^*$. The masses of these physical scalars can easily be obtained from  \eqref{eq:pot}. Since, the potential preserves the $SU(2)_D$ custodial symmetry, members of the physical scalar multiplets have degenerate masses. These masses are
\begin{eqnarray}
m_5^2&=&3 \left({\lambda_5} \left(v_{1}^2+v_{1M}^2+v_{2}^2+v_{2M}^2\right)+8 {\lambda_8} v_{3M}^2\right)\equiv 3 \left({\lambda_5} c_m^2+8 {\lambda_8} s_{m}^2\right) v_{\rm SM}^2,\nn\\
m_{3,H^\pm,H^0_3}^2&=&  {\lambda_5} \left(v_{1}^2+v_{1M}^2+v_{2}^2+v_{2M}^2+8 v_{3M}^2\right) \equiv  {\lambda_5}  v_{\rm SM}^2 \nn\\
m_{3,\rm All~others}^2&=& 2 {\lambda_5} \left(v_{1}^2+v_{1M}^2+v_{2}^2+v_{2M}^2+4 v_{3M}^2\right) \equiv {\lambda_5}  (1+c_m^2) v_{\rm SM}^2.
\end{eqnarray}

In general, $H_1^0$,$~H_2^0$,$~H_{1M}^0$,$~H_{2M}^0$,$~ H_s^0$ and $~H_{1}^{0'}$ can mix according to the mass-squared matrix
\begin{equation}
\mathcal{M}_{{\cal H}}^{2} =  { \fontsize{8}{10} v_{{\rm  SM}}^2  \left(
\begin{array}{cccccc}
 8 (\lambda_{1a}+\lambda_{4}) s_1^2   & 8 \lambda_{4} s_1 s_2   & 8 \lambda_{4} s_1 s_m   & 8 \lambda_{4} s_1 s_{2 m}   & 8 \lambda_{4a} \frac{ v_s  s_1 } {v_{{\rm  SM}}} & 2 \sqrt{6} \lambda_{4} s_1 s_m   \\
 8 \lambda_{4} s_1 s_2   & 8 (\lambda_{1b}+\lambda_{4}) s_2^2   & 8 \lambda_{4} s_2 s_m   & 8 \lambda_{4} s_2 s_{2 m}   & 8 \lambda_{4a} \frac{ v_s  s_2 } {v_{{\rm  SM}}} & 2 \sqrt{6} \lambda_{4} s_2 s_m   \\
 8 \lambda_{4} s_1 s_m   & 8 \lambda_{4} s_2 s_m   & 8 (\lambda_{2a}+\lambda_{4}) s_m^2   & 8 \lambda_{4} s_m s_{2 m}   & 8 \lambda_{4a} \frac{ v_s  s_{m} } {v_{{\rm  SM}}} & 2 \sqrt{6} \lambda_{4} s_m^2   \\
 8 \lambda_{4} s_1 s_{2 m}   & 8 \lambda_{4} s_2 s_{2 m}   & 8 \lambda_{4} s_m s_{2 m}   & 8 (\lambda_{2b}+\lambda_{4}) s_{2 m}^2   & 8 \lambda_{4a} \frac{ v_s  s_{2m} } {v_{{\rm  SM}}} & 2 \sqrt{6} \lambda_{4} s_m s_{2 m}   \\
 8 \lambda_{4a} \frac{ v_s  s_1 } {v_{{\rm  SM}}} & 8 \lambda_{4a} \frac{ v_s  s_2 } {v_{{\rm  SM}}} & 8 \lambda_{4a} \frac{ v_s  s_{m} } {v_{{\rm  SM}}} & 8 \lambda_{4a} \frac{ v_s  s_{2m} } {v_{{\rm  SM}}} & 8 (\lambda_{4a}+\lambda_{s}) \frac{ v_s^2 }{v_{{\rm  SM}}^2} & 2 \sqrt{6} \lambda_{4a} \frac{ v_s  s_{m} } {v_{{\rm  SM}}} \\
 2 \sqrt{6} \lambda_{4} s_1 s_m   & 2 \sqrt{6} \lambda_{4} s_2 s_m   & 2 \sqrt{6} \lambda_{4} s_m^2   & 2 \sqrt{6} \lambda_{4} s_m s_{2 m}   & 2 \sqrt{6} \lambda_{4a} \frac{ v_s  s_{m} } {v_{{\rm  SM}}} & 3 (\lambda_{3}+\lambda_{4}) s_m^2   \\
\end{array}
\right)}
\label{eq:MhhMx}
\end{equation}

We denote the mass eigenstates by $\widetilde{H}^{''''},\, \widetilde{H}^{'''},\, \widetilde{H}^{''},\, \widetilde{H}^{'},\, \widetilde{H}, \, \widetilde{H}_s$. We adopt a convention of denoting the lightest of the six by $\widetilde{H}_s$. The next heavier one by $\widetilde{H}$, the next one is $\widetilde{H}^{'}$ and so on. The heaviest state is $\widetilde{H}^{''''}$. The descending order of mass of the physical eigenstates is 
$(M_{\widetilde {H}^{''''}} > M_{\widetilde {H}^{'''}} >
M_{\widetilde {H}^{''}} > M_{\widetilde {H}^{'}} > M_{\widetilde {H}} > 
M_{\widetilde {H}_S} )$. The diagonalizing $6\times6$ orthogonal 
matrix element is denoted by ${\Fontskrivan O}_H^{ij} \, (i,j=1...6)$. The 125-GeV Higgs-like scalar component can be written as
\begin{equation}
\widetilde{H} = {\Fontskrivan O}_H^{51} \, H_1^0 + {\Fontskrivan O}_H^{52} \, H_2^0 + {\Fontskrivan O}_H^{53} \, H_{1M}^0 + {\Fontskrivan O}_H^{54}\, H_{2M}^0 + {\Fontskrivan O}_H^{55} \, H_s^0 + {\Fontskrivan O}_H^{56}\,  H_{1}^{0'} 
\end{equation}
In this analysis, multiple scalar fields alongside the Standard Model (SM) Higgs boson raise concerns regarding the experimentally measured SM Higgs boson mass under radiative corrections. The scalar sector of our model is also very similar to various versions of 2HDMs and Triplet scalar models, and we expect that the lightness of the SM Higgs boson mass will be ensured here, too. 
The SM fermion sector in this EW-$\nu_R$ model are given by~\cite{Hung:2006ap,Chakdar:2016adj,Hung:2017exy,Hung:2017voe}
\begin{eqnarray}
{\psi}_L&=&\begin{pmatrix}
 \nu_\ell \\ 
 \ell
\end{pmatrix}_L, ~~~{\rm and}~~\ell_R,~
{Q}_L=\begin{pmatrix}
 u \\ 
 d 
\end{pmatrix}_L, ~u_R~~{\rm and}~~d_R,
\label{eq:smferm}
\end{eqnarray}
where $\ell=e,\mu,\tau$ and $u$ stands for the $up-type$ quarks ($u,c,t$) and $d$ denotes the $down-type$ quarks ($d,s,b$). $L$ indicates the left-chirality and $R$ is right-chirality. In the EW-$\nu_R$  model, right-handed neutrinos are parts of $SU(2)$ doublets along with their charged partners (the mirror charged leptons). Anomaly freedom dictates the existence of doublets of right-handed and singlets left-handed mirror quarks~\cite{Hung:2006ap,Chakdar:2016adj,Hung:2017exy,Hung:2017voe}
\begin{eqnarray}
{\psi}_R^M&=&\begin{pmatrix}
 \nu_\ell^M \\ 
 \ell^M 
\end{pmatrix}_R, ~~~{\rm and}~~\ell_L^M,~~~
{Q}_R^M=\begin{pmatrix}
 u^M \\ 
 d^M 
\end{pmatrix}_R, ~u_L^M~~{\rm and}~~d_L^M
\label{eq:mirrorferm}
\end{eqnarray}

It is noted that the left-handed SM fermions and righ-handed Mirror fermions are doublet under $SU(2)_L$ whereas right-handed SM fermions and left-handed Mirror fermions are singlet under $SU(2)_L$ transformations. 
Again $SU (2)$-singlet right-handed SM
fermions, left-handed mirror fermions are taken to be singlet under $U(1)_{\rm SM} \times U(1)_{\rm MF}$ transformation in Ref.~\cite{Hung:2006ap,Chakdar:2016adj}. To solve the strong CP problem, we follow different transformation as in Ref.~\cite{Hung:2017exy}. The SM $SU(2)$ singlet fermions transform as $(\ell_R,u_R, d_R) \rightarrow e^{i \alpha_{\rm SM}}\, (\ell_R,u_R, d_R)$ whereas $SU(2)$ singlet fermions go as $(\ell_R^M,u_R^M, d_R^M) \rightarrow e^{i \alpha_{\rm MF}}\, (\ell_R^M,u_R^M, d_R^M)$.

The Higgs doublet $\Phi_2$ only couples to SM $up-quarks$ while another doublet $\Phi_1$ couples to $down-quarks$ and $leptons$. It behaves like type-II two Higgs doublet model. Similar interactions are also there in the mirror sector with $\Phi_{1M}$ and $\Phi_{2M}$ scalar doublets.
The $\tilde{\chi}$ in the Higgs triplet fields with hypercharge $Y=2$ and $\chi$ is a real Higgs triplet with $Y=0$. $\Phi_s$ is a complex singlet scalar which helps to generate the neutrino observables and strong CP problems.
The total Yukawa part of the Lagrangian is  given  by~\cite{Hung:2006ap,Chakdar:2016adj,Hung:2017exy,Hung:2017voe}
\begin{eqnarray}
\mathcal{L}_y &=& y_\ell  \, \overline{\psi}_L  \, \Phi_1 \, \ell_R  + y_\ell^M  \, \overline{\psi}_R^M  \, \Phi_{1M}  \, \ell_R^M + y_{s\ell}  \, \overline{\psi}_L  \, \psi_R^M  \,  \Phi_s + y_{M} \, \psi_R^{M,T}  \, i C \sigma_2  \, \tilde{\chi}  \, \psi_R^M \nn\\
&& + y_d  \, \overline{Q}_L  \, \Phi_1 \, d_R  + y_d^M  \, \overline{Q}_R^M  \, \Phi_{1M}  \, d_R^M - y_u  \, \overline{Q}_L  \, i \sigma_2 \, \Phi_2 \, u_R  - y_u^M  \, \overline{Q}_R^M  \, i \sigma_2 \, \Phi_{2M}  \, u_R^M  \nn\\
&& + y_{sd}  \,  \overline{d}_R d_L^M \,  \Phi_s + y_{sq}  \, \overline{Q}_L Q_R^M \,  \Phi_s + y_{su}  \,  \overline{u}_R u_R^M  \,  \Phi_s + ~{\rm h.c} 
\label{eq:LYukawa}
\end{eqnarray}
where $C$ is the charge conjugation operator, $\sigma_2$ being the second Pauli's spin matrix.

Now we will calculate the various mass-mixing matrix after electroweak symmetry breaking, physical mass eigenstates of the fermions in this model. 
The charged-lepton mixing matrix can be found as~\cite{Hung:2006ap}
\begin{eqnarray}
\mathcal{M}_l = \begin{pmatrix}
 m_l & m_\nu^D \\ 
 m_\nu^D &  m_l^M
\end{pmatrix},
\label{eq:chargel}
\end{eqnarray}
where, $m_\ell= y_\ell v_1/\sqrt{2}$, $m_\ell^M= y_\ell^M v_{1M} / \sqrt{2}$ and $m_\nu^D=y_{s\ell} v_s$. The $l$ and $\ell^M$ stand for flavor eigenstates whereas $\tilde{\ell}$ and $\tilde{\ell}^M$ stand for the mass eigenstates. The mixing angle between $\ell$ and $\ell^M$ is $\theta_\ell$, hence $\tan 2 \theta_\ell=\frac{2 m_\nu^D }{m_\ell^M-m_\ell}$. The mixing matrix is $R_\ell=\{\{ \cos \theta_\ell , \,\sin \theta_\ell \},\{-\sin \theta_\ell,\, \cos \theta_\ell\}\}$. For $m_\ell^M \gg m_\ell, m_\nu^D$, one can write $\tan \theta_\ell \approx \sin \theta_\ell \approx \theta_\ell \approx \frac{ m_\nu^D }{m_\ell^M}= \frac{\sqrt{2} y_{s\ell} v_s}{ y_\ell^M v_{1M}}$.  The mass eigenstates can be written as 
\bea
\tilde{\ell} = \ell \cos \theta_\ell   + \ell^M \sin \theta_\ell\nn\\
\tilde{\ell}^M= -\ell  \sin \theta_\ell  + \ell^M \cos \theta_\ell
\eea 
As there is no singlet right-handed neutrino in this model, there is no such mixing in the neutrino sector and hence the pseudoscalar $A_s^0$ could not decay into two light neutrinos. 
The up and down sector mixing matrix are given by~\cite{Hung:2017exy,Hung:2017pss}
\begin{eqnarray}
\mathcal{M}_u = \begin{pmatrix}
 m_u & m_{sq} \\ 
 m_{su} &  m_u^M
\end{pmatrix},~~~~~{\rm and}~~~~~~
\mathcal{M}_d = \begin{pmatrix}
 m_d & m_{sq} \\ 
 m_{sd} &  m_d^M
\end{pmatrix},
\label{eq:chargeud}
\end{eqnarray}
where, $m_d= y_d v_1/\sqrt{2}$, $m_d^M= y_d^M v_{1M} / \sqrt{2}$, $m_u= y_d v_2/\sqrt{2}$, $m_u^M= y_u^M v_{2M} / \sqrt{2}$, $m_{sq} \approx y_{sq} v_s $, $m_{su} \approx y_{su} v_s $ and $m_{su} \approx y_{sd} v_s $.
The mixing matrix are $R_{u,d}=\{\{ \cos \theta_{u,d} , \,\sin \theta_{u,d} \},\{-\sin \theta_{u,d},\, \cos \theta_{u,d}\}\}$, where
\beq
\sin \theta_{u}=\sqrt{\frac{\left(m_u^M-m_u\right) \left(m_u^M-m_u -\sqrt{\left(m_u^M -m_u\right)^2+4 m_{{sq}} m_{{su}}}\right)+2 m_{{sq}} m_{{su}}+2 m_{{su}}^2}{2 \left(\left(m_u^M-m_u\right)^2+2 m_{{sq}} m_{{su}}+m_{{sq}}^2+m_{{su}}^2\right)}}
\label{eq:thetau}
\eeq
We can also get the similar analytical form of $\sin \theta_d$ by replacing $u$ to $d$ in eqn.~\ref{eq:thetau}. Let $u,d$ and $u^M,d^M$ stand for flavor eigenstates and $\tilde{u},\tilde{d}$ and $\tilde{u}^M, \tilde{d}^M$ stand for the mass eigenstates. Thus one can write
\bea
\tilde{u} = u \cos \theta_u   + u^M \sin \theta_u\nn\\
\tilde{u}^M= -u  \sin \theta_u  + u^M \cos \theta_u\\
\tilde{d} = d \cos \theta_d   + d^M \sin \theta_d\nn\\
\tilde{d}^M= -d  \sin \theta_d  + d^M \cos \theta_d
\eea 
The terms $ (y_{s\ell}  \, \overline{\psi}_L  \, \psi_R^M  + y_{sd}  \,  \overline{d}_R d_L^M  + y_{sq}  \, \overline{Q}_L Q_R^M  + y_{su}  \,  \overline{u}_R u_R^M ) \, \Phi_s$ in eqn~\ref{eq:LYukawa} can help us to get the $A_s^0 \, \bar{\tilde{f}}_i \tilde{f}_i$ coupling strengths. $A_s^0$ to two leptons (both the SM and MF charged and neutral leptons) can be written as  
\bea
y_f^\ell=y_{s\ell} \sin \theta_\ell \cos \theta_\ell  \gamma_5 \approx y_{s\ell} \sin \theta_\ell  \gamma_5 \approx \sqrt{2}  \frac{ y_{s\ell}^2 v_s}{ y_\ell^M v_{1M}} \gamma_5.
\label{eq:Asll}
\eea
For $y_{sq} \sim y_{su} \sim y_{sd} = y_s $, $v_{1M}=v_{2M}$ and $y_d^M=y_u^M$, we can write the mixing angle $\theta_u=\theta_d\approx \frac{\sqrt{2} y_{s} v_s}{ y_u^M v_{1M}}$. Similarly $A_s^0$ to two quarks (both the SM and MF  up and down quarks) coupling strengths are given by
\beq
y_f^q=y_{s} \sin \theta_u \cos \theta_u \gamma_5 \approx y_{s} \sin \theta_u \gamma_5 \approx \sqrt{2}  \frac{ y_{s}^2 v_s}{ y_u^M v_{1M}} \gamma_5.
\label{eq:Asqq}
\eeq
%
\providecommand{\href}[2]{#2}\begingroup\raggedright\endgroup

\begin{thebibliography}{10}

\bibitem{Zwicky:1933gu}
F.~Zwicky, ``{Die Rotverschiebung von extragalaktischen Nebeln},''
  \href{http://dx.doi.org/10.1007/s10714-008-0707-4}{{\em Helv. Phys. Acta}
  {\bfseries 6} (1933) 110--127}.

\bibitem{Rubin:1970zza}
V.~C. Rubin and W.~K. Ford, Jr., ``{Rotation of the Andromeda Nebula from a
  Spectroscopic Survey of Emission Regions},''
  \href{http://dx.doi.org/10.1086/150317}{{\em Astrophys. J.} {\bfseries 159}
  (1970) 379--403}.

\bibitem{Clowe:2006eq}
D.~Clowe, M.~Bradac, A.~H. Gonzalez, M.~Markevitch, S.~W. Randall, C.~Jones,
  and D.~Zaritsky, ``{A direct empirical proof of the existence of dark
  matter},'' \href{http://dx.doi.org/10.1086/508162}{{\em Astrophys. J. Lett.}
  {\bfseries 648} (2006) L109--L113},
  \href{http://arxiv.org/abs/astro-ph/0608407}{{\ttfamily
  arXiv:astro-ph/0608407}}.

\bibitem{Massey:2010hh}
R.~Massey, T.~Kitching, and J.~Richard, ``{The dark matter of gravitational
  lensing},'' \href{http://dx.doi.org/10.1088/0034-4885/73/8/086901}{{\em Rept.
  Prog. Phys.} {\bfseries 73} (2010) 086901},
  \href{http://arxiv.org/abs/1001.1739}{{\ttfamily arXiv:1001.1739
  [astro-ph.CO]}}.

\bibitem{Aghanim:2018eyx}
{\bfseries Planck} Collaboration, N.~Aghanim {\em et~al.}, ``{Planck 2018
  results. VI. Cosmological parameters},''
  \href{http://dx.doi.org/10.1051/0004-6361/201833910}{{\em Astron. Astrophys.}
  {\bfseries 641} (2020) A6}, \href{http://arxiv.org/abs/1807.06209}{{\ttfamily
  arXiv:1807.06209 [astro-ph.CO]}}.

\bibitem{Giagu:2019fmp}
S.~Giagu, ``{WIMP Dark Matter Searches With the ATLAS Detector at the LHC},''
  \href{http://dx.doi.org/10.3389/fphy.2019.00075}{{\em Front. in Phys.}
  {\bfseries 7} (2019) 75}.

\bibitem{Aaboud:2018xdl}
{\bfseries ATLAS} Collaboration, M.~Aaboud {\em et~al.}, ``{Search for dark
  matter in events with a hadronically decaying vector boson and missing
  transverse momentum in $pp$ collisions at $\sqrt{s} = 13$ TeV with the ATLAS
  detector},'' \href{http://dx.doi.org/10.1007/JHEP10(2018)180}{{\em JHEP}
  {\bfseries 10} (2018) 180}, \href{http://arxiv.org/abs/1807.11471}{{\ttfamily
  arXiv:1807.11471 [hep-ex]}}.

\bibitem{Aaboud:2017phn}
{\bfseries ATLAS} Collaboration, M.~Aaboud {\em et~al.}, ``{Search for dark
  matter and other new phenomena in events with an energetic jet and large
  missing transverse momentum using the ATLAS detector},''
  \href{http://dx.doi.org/10.1007/JHEP01(2018)126}{{\em JHEP} {\bfseries 01}
  (2018) 126}, \href{http://arxiv.org/abs/1711.03301}{{\ttfamily
  arXiv:1711.03301 [hep-ex]}}.

\bibitem{Aaboud:2017dor}
{\bfseries ATLAS} Collaboration, M.~Aaboud {\em et~al.}, ``{Search for dark
  matter at $\sqrt{s}=13$ TeV in final states containing an energetic photon
  and large missing transverse momentum with the ATLAS detector},''
  \href{http://dx.doi.org/10.1140/epjc/s10052-017-4965-8}{{\em Eur. Phys. J. C}
  {\bfseries 77} no.~6, (2017) 393},
  \href{http://arxiv.org/abs/1704.03848}{{\ttfamily arXiv:1704.03848
  [hep-ex]}}.

\bibitem{Aaboud:2017bja}
{\bfseries ATLAS} Collaboration, M.~Aaboud {\em et~al.}, ``{Search for an
  invisibly decaying Higgs boson or dark matter candidates produced in
  association with a $Z$ boson in $pp$ collisions at $\sqrt{s} =$ 13 TeV with
  the ATLAS detector},''
  \href{http://dx.doi.org/10.1016/j.physletb.2017.11.049}{{\em Phys. Lett. B}
  {\bfseries 776} (2018) 318--337},
  \href{http://arxiv.org/abs/1708.09624}{{\ttfamily arXiv:1708.09624
  [hep-ex]}}.

\bibitem{Akerib:2016vxi}
{\bfseries LUX} Collaboration, D.~S. Akerib {\em et~al.}, ``{Results from a
  search for dark matter in the complete LUX exposure},''
  \href{http://dx.doi.org/10.1103/PhysRevLett.118.021303}{{\em Phys. Rev.
  Lett.} {\bfseries 118} no.~2, (2017) 021303},
  \href{http://arxiv.org/abs/1608.07648}{{\ttfamily arXiv:1608.07648
  [astro-ph.CO]}}.

\bibitem{Cui:2017nnn}
{\bfseries PandaX-II} Collaboration, X.~Cui {\em et~al.}, ``{Dark Matter
  Results From 54-Ton-Day Exposure of PandaX-II Experiment},''
  \href{http://dx.doi.org/10.1103/PhysRevLett.119.181302}{{\em Phys. Rev.
  Lett.} {\bfseries 119} no.~18, (2017) 181302},
  \href{http://arxiv.org/abs/1708.06917}{{\ttfamily arXiv:1708.06917
  [astro-ph.CO]}}.

\bibitem{Aprile:2018dbl}
{\bfseries XENON} Collaboration, E.~Aprile {\em et~al.}, ``{Dark Matter Search
  Results from a One Ton-Year Exposure of XENON1T},''
  \href{http://dx.doi.org/10.1103/PhysRevLett.121.111302}{{\em Phys. Rev.
  Lett.} {\bfseries 121} no.~11, (2018) 111302},
\href{http://arxiv.org/abs/1805.12562}{{\ttfamily arXiv:1805.12562
  [astro-ph.CO]}}.

\bibitem{Amole:2019fdf}
{\bfseries PICO} Collaboration, C.~Amole {\em et~al.}, ``{Dark Matter Search
  Results from the Complete Exposure of the PICO-60 C$_3$F$_8$ Bubble
  Chamber},'' \href{http://dx.doi.org/10.1103/PhysRevD.100.022001}{{\em Phys.
  Rev. D} {\bfseries 100} no.~2, (2019) 022001},
  \href{http://arxiv.org/abs/1902.04031}{{\ttfamily arXiv:1902.04031
  [astro-ph.CO]}}.

\bibitem{Daylan:2014rsa}
T.~Daylan, D.~P. Finkbeiner, D.~Hooper, T.~Linden, S.~K.~N. Portillo, N.~L.
  Rodd, and T.~R. Slatyer, ``{The characterization of the gamma-ray signal from
  the central Milky Way: A case for annihilating dark matter},''
  \href{http://dx.doi.org/10.1016/j.dark.2015.12.005}{{\em Phys. Dark Univ.}
  {\bfseries 12} (2016) 1--23},
  \href{http://arxiv.org/abs/1402.6703}{{\ttfamily arXiv:1402.6703
  [astro-ph.HE]}}.

\bibitem{Ahnen:2016qkx}
{\bfseries MAGIC, Fermi-LAT} Collaboration, M.~L. Ahnen {\em et~al.}, ``{Limits
  to Dark Matter Annihilation Cross-Section from a Combined Analysis of MAGIC
  and Fermi-LAT Observations of Dwarf Satellite Galaxies},''
  \href{http://dx.doi.org/10.1088/1475-7516/2016/02/039}{{\em JCAP} {\bfseries
  02} (2016) 039}, \href{http://arxiv.org/abs/1601.06590}{{\ttfamily
  arXiv:1601.06590 [astro-ph.HE]}}.

\bibitem{Lin:2019uvt}
T.~Lin, ``{Dark matter models and direct detection},''
  \href{http://dx.doi.org/10.22323/1.333.0009}{{\em PoS} {\bfseries 333} (2019)
  009}, \href{http://arxiv.org/abs/1904.07915}{{\ttfamily arXiv:1904.07915
  [hep-ph]}}.

\bibitem{Hall:2009bx}
L.~J. Hall, K.~Jedamzik, J.~March-Russell, and S.~M. West, ``{Freeze-In
  Production of FIMP Dark Matter},''
  \href{http://dx.doi.org/10.1007/JHEP03(2010)080}{{\em JHEP} {\bfseries 03}
  (2010) 080}, \href{http://arxiv.org/abs/0911.1120}{{\ttfamily arXiv:0911.1120
  [hep-ph]}}.

\bibitem{Bernal:2017kxu}
N.~Bernal, M.~Heikinheimo, T.~Tenkanen, K.~Tuominen, and V.~Vaskonen, ``{The
  Dawn of FIMP Dark Matter: A Review of Models and Constraints},''
  \href{http://dx.doi.org/10.1142/S0217751X1730023X}{{\em Int. J. Mod. Phys. A}
  {\bfseries 32} no.~27, (2017) 1730023},
  \href{http://arxiv.org/abs/1706.07442}{{\ttfamily arXiv:1706.07442
  [hep-ph]}}.

\bibitem{Essig:2011nj}
R.~Essig, J.~Mardon, and T.~Volansky, ``{Direct Detection of Sub-GeV Dark
  Matter},'' \href{http://dx.doi.org/10.1103/PhysRevD.85.076007}{{\em Phys.
  Rev. D} {\bfseries 85} (2012) 076007},
  \href{http://arxiv.org/abs/1108.5383}{{\ttfamily arXiv:1108.5383 [hep-ph]}}.

\bibitem{Essig:2017kqs}
R.~Essig, T.~Volansky, and T.-T. Yu, ``{New Constraints and Prospects for
  sub-GeV Dark Matter Scattering off Electrons in Xenon},''
  \href{http://dx.doi.org/10.1103/PhysRevD.96.043017}{{\em Phys. Rev. D}
  {\bfseries 96} no.~4, (2017) 043017},
  \href{http://arxiv.org/abs/1703.00910}{{\ttfamily arXiv:1703.00910
  [hep-ph]}}.

\bibitem{Emken:2017erx}
T.~Emken, C.~Kouvaris, and I.~M. Shoemaker, ``{Terrestrial Effects on Dark
  Matter-Electron Scattering Experiments},''
  \href{http://dx.doi.org/10.1103/PhysRevD.96.015018}{{\em Phys. Rev. D}
  {\bfseries 96} no.~1, (2017) 015018},
  \href{http://arxiv.org/abs/1702.07750}{{\ttfamily arXiv:1702.07750
  [hep-ph]}}.

\bibitem{Green:2017ybv}
D.~Green and S.~Rajendran, ``{The Cosmology of Sub-MeV Dark Matter},''
  \href{http://dx.doi.org/10.1007/JHEP10(2017)013}{{\em JHEP} {\bfseries 10}
  (2017) 013}, \href{http://arxiv.org/abs/1701.08750}{{\ttfamily
  arXiv:1701.08750 [hep-ph]}}.

\bibitem{Essig:2015cda}
R.~Essig, M.~Fernandez-Serra, J.~Mardon, A.~Soto, T.~Volansky, and T.-T. Yu,
  ``{Direct Detection of sub-GeV Dark Matter with Semiconductor Targets},''
  \href{http://dx.doi.org/10.1007/JHEP05(2016)046}{{\em JHEP} {\bfseries 05}
  (2016) 046}, \href{http://arxiv.org/abs/1509.01598}{{\ttfamily
  arXiv:1509.01598 [hep-ph]}}.

\bibitem{Essig:2012yx}
R.~Essig, A.~Manalaysay, J.~Mardon, P.~Sorensen, and T.~Volansky, ``{First
  Direct Detection Limits on sub-GeV Dark Matter from XENON10},''
  \href{http://dx.doi.org/10.1103/PhysRevLett.109.021301}{{\em Phys. Rev.
  Lett.} {\bfseries 109} (2012) 021301},
  \href{http://arxiv.org/abs/1206.2644}{{\ttfamily arXiv:1206.2644
  [astro-ph.CO]}}.

\bibitem{Bernal:2017mqb}
N.~Bernal, X.~Chu, and J.~Pradler, ``{Simply split strongly interacting massive
  particles},'' \href{http://dx.doi.org/10.1103/PhysRevD.95.115023}{{\em Phys.
  Rev. D} {\bfseries 95} no.~11, (2017) 115023},
  \href{http://arxiv.org/abs/1702.04906}{{\ttfamily arXiv:1702.04906
  [hep-ph]}}.

\bibitem{Hung:2006ap}
P.~Q. Hung, ``{A Model of electroweak-scale right-handed neutrino mass},''
  \href{http://dx.doi.org/10.1016/j.physletb.2007.03.067}{{\em Phys. Lett. B}
  {\bfseries 649} (2007) 275--279},
  \href{http://arxiv.org/abs/hep-ph/0612004}{{\ttfamily arXiv:hep-ph/0612004}}.

\bibitem{Hung:2020vuo}
P.~Q. Hung, ``{Topologically stable, finite-energy electroweak-scale
  monopoles},'' \href{http://dx.doi.org/10.1016/j.nuclphysb.2020.115278}{{\em
  Nucl. Phys. B} {\bfseries 962} (2021) 115278},
  \href{http://arxiv.org/abs/2003.02794}{{\ttfamily arXiv:2003.02794
  [hep-ph]}}.

\bibitem{Ellis:2020bpy}
J.~Ellis, P.~Q. Hung, and N.~E. Mavromatos, ``{An electroweak monopole, Dirac
  quantization and the weak mixing angle},''
  \href{http://dx.doi.org/10.1016/j.nuclphysb.2021.115468}{{\em Nucl. Phys. B}
  {\bfseries 969} (2021) 115468},
  \href{http://arxiv.org/abs/2008.00464}{{\ttfamily arXiv:2008.00464
  [hep-ph]}}.

\bibitem{Hoang:2014pda}
V.~Hoang, P.~Q. Hung, and A.~S. Kamat, ``{Non-sterile electroweak-scale
  right-handed neutrinos and the dual nature of the 125-GeV scalar},''
  \href{http://dx.doi.org/10.1016/j.nuclphysb.2015.05.007}{{\em Nucl. Phys. B}
  {\bfseries 896} (2015) 611--656},
  \href{http://arxiv.org/abs/1412.0343}{{\ttfamily arXiv:1412.0343 [hep-ph]}}.

\bibitem{Hartling:2014zca}
K.~Hartling, K.~Kumar, and H.~E. Logan, ``{The decoupling limit in the
  Georgi-Machacek model},''
  \href{http://dx.doi.org/10.1103/PhysRevD.90.015007}{{\em Phys. Rev. D}
  {\bfseries 90} no.~1, (2014) 015007},
  \href{http://arxiv.org/abs/1404.2640}{{\ttfamily arXiv:1404.2640 [hep-ph]}}.

\bibitem{Chakdar:2016adj}
S.~Chakdar, K.~Ghosh, V.~Hoang, P.~Q. Hung, and S.~Nandi, ``{The search for
  electroweak-scale right-handed neutrinos and mirror charged leptons through
  like-sign dilepton signals},''
  \href{http://dx.doi.org/10.1103/PhysRevD.95.015014}{{\em Phys. Rev.}
  {\bfseries D95} no.~1, (2017) 015014},
\href{http://arxiv.org/abs/1606.08502}{{\ttfamily arXiv:1606.08502 [hep-ph]}}.

\bibitem{Hung:2017exy}
P.~Q. Hung, ``{Mirror fermions and the strong CP problem: A new axionless
  solution and experimental implications},''
\href{http://arxiv.org/abs/1712.09701}{{\ttfamily arXiv:1712.09701 [hep-ph]}}.

\bibitem{Hung:2017voe}
P.~Hung, T.~Le, V.~Q. Tran, and T.-C. Yuan, ``{Muon-to-Electron Conversion in
  Mirror Fermion Model with Electroweak Scale Non-Sterile Right-handed
  Neutrinos},'' \href{http://dx.doi.org/10.1016/j.nuclphysb.2018.05.020}{{\em
  Nucl. Phys. B} {\bfseries 932} (2018) 471--504},
  \href{http://arxiv.org/abs/1701.01761}{{\ttfamily arXiv:1701.01761
  [hep-ph]}}.

\bibitem{Hung:2017pss}
P.~Hung, ``{A non-vanishing neutrino mass and the strong CP problem: A new
  solution from the perspective of the EW-$\nu_R$ model},'' in {\em {Meeting of
  the APS Division of Particles and Fields}}.
\newblock 2017.
\newblock \href{http://arxiv.org/abs/1710.00498}{{\ttfamily arXiv:1710.00498
  [hep-ph]}}.

\bibitem{Peskin:1991sw}
M.~E. Peskin and T.~Takeuchi, ``{Estimation of oblique electroweak
  corrections},'' \href{http://dx.doi.org/10.1103/PhysRevD.46.381}{{\em Phys.
  Rev. D} {\bfseries 46} (1992) 381--409}.

\bibitem{Hoang:2013jfa}
V.~Hoang, P.~Q. Hung, and A.~S. Kamat, ``{Electroweak precision constraints on
  the electroweak-scale right-handed neutrino model},''
  \href{http://dx.doi.org/10.1016/j.nuclphysb.2013.10.002}{{\em Nucl. Phys. B}
  {\bfseries 877} (2013) 190--232},
  \href{http://arxiv.org/abs/1303.0428}{{\ttfamily arXiv:1303.0428 [hep-ph]}}.

\bibitem{Miyazaki:2007zw}
{\bfseries Belle} Collaboration, Y.~Miyazaki {\em et~al.}, ``{Search for Lepton
  Flavor Violating tau Decays into Three Leptons},''
  \href{http://dx.doi.org/10.1016/j.physletb.2007.12.046}{{\em Phys. Lett. B}
  {\bfseries 660} (2008) 154--160},
  \href{http://arxiv.org/abs/0711.2189}{{\ttfamily arXiv:0711.2189 [hep-ex]}}.

\bibitem{Baldini:2018nnn}
{\bfseries MEG II} Collaboration, A.~M. Baldini {\em et~al.}, ``{The design of
  the MEG II experiment},''
  \href{http://dx.doi.org/10.1140/epjc/s10052-018-5845-6}{{\em Eur. Phys. J. C}
  {\bfseries 78} no.~5, (2018) 380},
  \href{http://arxiv.org/abs/1801.04688}{{\ttfamily arXiv:1801.04688
  [physics.ins-det]}}.

\bibitem{Dohmen:1993mp}
{\bfseries SINDRUM II} Collaboration, C.~Dohmen {\em et~al.}, ``{Test of lepton
  flavor conservation in mu ---\ensuremath{>} e conversion on titanium},''
  \href{http://dx.doi.org/10.1016/0370-2693(93)91383-X}{{\em Phys. Lett. B}
  {\bfseries 317} (1993) 631--636}.

\bibitem{TheMEG:2016wtm}
{\bfseries MEG} Collaboration, A.~M. Baldini {\em et~al.}, ``{Search for the
  lepton flavour violating decay $\mu ^+ \rightarrow \mathrm {e}^+ \gamma $
  with the full dataset of the MEG experiment},''
  \href{http://dx.doi.org/10.1140/epjc/s10052-016-4271-x}{{\em Eur. Phys. J. C}
  {\bfseries 76} no.~8, (2016) 434},
  \href{http://arxiv.org/abs/1605.05081}{{\ttfamily arXiv:1605.05081
  [hep-ex]}}.

\bibitem{Hung:2015hra}
P.~Q. Hung, T.~Le, V.~Q. Tran, and T.-C. Yuan, ``{Lepton Flavor Violating
  Radiative Decays in EW-Scale $\nu_R$ Model: An Update},''
  \href{http://dx.doi.org/10.1007/JHEP12(2015)169}{{\em JHEP} {\bfseries 12}
  (2015) 169}, \href{http://arxiv.org/abs/1508.07016}{{\ttfamily
  arXiv:1508.07016 [hep-ph]}}.

\bibitem{Hung:2007ez}
P.~Q. Hung, ``{Electroweak-scale mirror fermions, mu ---\ensuremath{>} e gamma
  and tau ---\ensuremath{>} mu gamma},''
  \href{http://dx.doi.org/10.1016/j.physletb.2007.12.005}{{\em Phys. Lett. B}
  {\bfseries 659} (2008) 585--592},
  \href{http://arxiv.org/abs/0711.0733}{{\ttfamily arXiv:0711.0733 [hep-ph]}}.

\bibitem{Lee:1973iz}
T.~D. Lee, ``{A Theory of Spontaneous T Violation},''
  \href{http://dx.doi.org/10.1103/PhysRevD.8.1226}{{\em Phys. Rev. D}
  {\bfseries 8} (1973) 1226--1239}.

\bibitem{WahabElKaffas:2007xd}
A.~Wahab El~Kaffas, P.~Osland, and O.~M. Ogreid, ``{Constraining the
  Two-Higgs-Doublet-Model parameter space},''
  \href{http://dx.doi.org/10.1103/PhysRevD.76.095001}{{\em Phys. Rev. D}
  {\bfseries 76} (2007) 095001},
  \href{http://arxiv.org/abs/0706.2997}{{\ttfamily arXiv:0706.2997 [hep-ph]}}.

\bibitem{Branco:2011iw}
G.~C. Branco, P.~M. Ferreira, L.~Lavoura, M.~N. Rebelo, M.~Sher, and J.~P.
  Silva, ``{Theory and phenomenology of two-Higgs-doublet models},''
  \href{http://dx.doi.org/10.1016/j.physrep.2012.02.002}{{\em Phys. Rept.}
  {\bfseries 516} (2012) 1--102},
  \href{http://arxiv.org/abs/1106.0034}{{\ttfamily arXiv:1106.0034 [hep-ph]}}.

\bibitem{Sirunyan:2018koj}
{\bfseries CMS} Collaboration, A.~M. Sirunyan {\em et~al.}, ``{Combined
  measurements of Higgs boson couplings in proton\textendash{}proton collisions
  at $\sqrt{s}=13\,\text {Te}\text {V} $},''
  \href{http://dx.doi.org/10.1140/epjc/s10052-019-6909-y}{{\em Eur. Phys. J. C}
  {\bfseries 79} no.~5, (2019) 421},
  \href{http://arxiv.org/abs/1809.10733}{{\ttfamily arXiv:1809.10733
  [hep-ex]}}.

\bibitem{Rott:2014kfa}
C.~Rott, K.~Kohri, and S.~C. Park, ``{Superheavy dark matter and IceCube
  neutrino signals: Bounds on decaying dark matter},''
  \href{http://dx.doi.org/10.1103/PhysRevD.92.023529}{{\em Phys. Rev. D}
  {\bfseries 92} no.~2, (2015) 023529},
  \href{http://arxiv.org/abs/1408.4575}{{\ttfamily arXiv:1408.4575 [hep-ph]}}.

\bibitem{Fiorentin:2016avj}
M.~Re~Fiorentin, V.~Niro, and N.~Fornengo, ``{A consistent model for
  leptogenesis, dark matter and the IceCube signal},''
  \href{http://dx.doi.org/10.1007/JHEP11(2016)022}{{\em JHEP} {\bfseries 11}
  (2016) 022}, \href{http://arxiv.org/abs/1606.04445}{{\ttfamily
  arXiv:1606.04445 [hep-ph]}}.

\bibitem{Audren:2014bca}
B.~Audren, J.~Lesgourgues, G.~Mangano, P.~D. Serpico, and T.~Tram, ``{Strongest
  model-independent bound on the lifetime of Dark Matter},''
  \href{http://dx.doi.org/10.1088/1475-7516/2014/12/028}{{\em JCAP} {\bfseries
  12} (2014) 028}, \href{http://arxiv.org/abs/1407.2418}{{\ttfamily
  arXiv:1407.2418 [astro-ph.CO]}}.

\bibitem{Aartsen:2014gkd}
{\bfseries IceCube} Collaboration, M.~Aartsen {\em et~al.}, ``{Observation of
  High-Energy Astrophysical Neutrinos in Three Years of IceCube Data},''
  \href{http://dx.doi.org/10.1103/PhysRevLett.113.101101}{{\em Phys. Rev.
  Lett.} {\bfseries 113} (2014) 101101},
  \href{http://arxiv.org/abs/1405.5303}{{\ttfamily arXiv:1405.5303
  [astro-ph.HE]}}.

\bibitem{Plehn:2017fdg}
M.~Bauer and T.~Plehn, \href{http://dx.doi.org/10.1007/978-3-030-16234-4}{{\em
  {Yet Another Introduction to Dark Matter}: {The Particle Physics Approach}}},
  vol.~959 of {\em Lecture Notes in Physics}.
\newblock Springer, 2019.
\newblock \href{http://arxiv.org/abs/1705.01987}{{\ttfamily arXiv:1705.01987
  [hep-ph]}}.

\bibitem{Biswas:2016bfo}
A.~Biswas and A.~Gupta, ``{Freeze-in Production of Sterile Neutrino Dark Matter
  in U(1)$_{\rm B-L}$ Model},''
  \href{http://dx.doi.org/10.1088/1475-7516/2016/09/044}{{\em JCAP} {\bfseries
  09} (2016) 044}, \href{http://arxiv.org/abs/1607.01469}{{\ttfamily
  arXiv:1607.01469 [hep-ph]}}. [Addendum: JCAP 05, A01 (2017)].

\bibitem{Gondolo:1990dk}
P.~Gondolo and G.~Gelmini, ``{Cosmic abundances of stable particles: Improved
  analysis},'' \href{http://dx.doi.org/10.1016/0550-3213(91)90438-4}{{\em Nucl.
  Phys. B} {\bfseries 360} (1991) 145--179}.

\bibitem{Zyla:2020zbs}
{\bfseries Particle Data Group} Collaboration, P.~Zyla {\em et~al.}, ``{Review
  of Particle Physics},'' \href{http://dx.doi.org/10.1093/ptep/ptaa104}{{\em
  PTEP} {\bfseries 2020} no.~8, (2020) 083C01}.

\bibitem{Djouadi:1995gv}
A.~Djouadi, J.~Kalinowski, and P.~Zerwas, ``{Two and three-body decay modes of
  SUSY Higgs particles},'' \href{http://dx.doi.org/10.1007/s002880050121}{{\em
  Z. Phys. C} {\bfseries 70} (1996) 435--448},
  \href{http://arxiv.org/abs/hep-ph/9511342}{{\ttfamily arXiv:hep-ph/9511342}}.

\bibitem{Borah:2018gjk}
D.~Borah, B.~Karmakar, and D.~Nanda, ``{Common Origin of Dirac Neutrino Mass
  and Freeze-in Massive Particle Dark Matter},''
  \href{http://dx.doi.org/10.1088/1475-7516/2018/07/039}{{\em JCAP} {\bfseries
  07} (2018) 039}, \href{http://arxiv.org/abs/1805.11115}{{\ttfamily
  arXiv:1805.11115 [hep-ph]}}.

\bibitem{PeymanZakeri:2018zaa}
S.~Peyman~Zakeri, S.~Mohammad Moosavi~Nejad, M.~Zakeri, and S.~Yaser~Ayazi,
  ``{A Minimal Model For Two-Component FIMP Dark Matter: A Basic Search},''
  \href{http://dx.doi.org/10.1088/1674-1137/42/7/073101}{{\em Chin. Phys. C}
  {\bfseries 42} no.~7, (2018) 073101},
  \href{http://arxiv.org/abs/1801.09115}{{\ttfamily arXiv:1801.09115
  [hep-ph]}}.

\bibitem{Herms:2019mnu}
J.~Herms and A.~Ibarra, ``{Probing multicomponent FIMP scenarios with gamma-ray
  telescopes},'' \href{http://dx.doi.org/10.1088/1475-7516/2020/03/026}{{\em
  JCAP} {\bfseries 03} (2020) 026},
  \href{http://arxiv.org/abs/1912.09458}{{\ttfamily arXiv:1912.09458
  [hep-ph]}}.

\bibitem{DEramo:2020gpr}
F.~D'Eramo and A.~Lenoci, ``{Lower Mass Bounds on FIMPs},''
  \href{http://arxiv.org/abs/2012.01446}{{\ttfamily arXiv:2012.01446
  [hep-ph]}}.

\bibitem{Das:2021zea}
P.~Das, M.~K. Das, and N.~Khan, ``{The FIMP-WIMP dark matter and Muon g-2 in
  the extended singlet scalar model},''
  \href{http://arxiv.org/abs/2104.03271}{{\ttfamily arXiv:2104.03271
  [hep-ph]}}.

\bibitem{Das:2021qqr}
P.~Das, M.~K. Das, and N.~Khan, ``{Extension of Hyperchargeless Higgs Triplet
  Model},'' \href{http://arxiv.org/abs/2107.01578}{{\ttfamily arXiv:2107.01578
  [hep-ph]}}.

\bibitem{Pandey:2017quk}
M.~Pandey, D.~Majumdar, and K.~P. Modak, ``{Two Component Feebly Interacting
  Massive Particle (FIMP) Dark Matter},''
  \href{http://dx.doi.org/10.1088/1475-7516/2018/06/023}{{\em JCAP} {\bfseries
  06} (2018) 023}, \href{http://arxiv.org/abs/1709.05955}{{\ttfamily
  arXiv:1709.05955 [hep-ph]}}.

\bibitem{Biswas:2017tce}
A.~Biswas, S.~Choubey, and S.~Khan, ``{Neutrino mass, leptogenesis and FIMP
  dark matter in a $\mathrm{U}(1)_{B-L}$ model},''
  \href{http://dx.doi.org/10.1140/epjc/s10052-017-5436-y}{{\em Eur. Phys. J. C}
  {\bfseries 77} no.~12, (2017) 875},
  \href{http://arxiv.org/abs/1704.00819}{{\ttfamily arXiv:1704.00819
  [hep-ph]}}.

\bibitem{Yaguna:2011qn}
C.~E. Yaguna, ``{The Singlet Scalar as FIMP Dark Matter},''
  \href{http://dx.doi.org/10.1007/JHEP08(2011)060}{{\em JHEP} {\bfseries 08}
  (2011) 060}, \href{http://arxiv.org/abs/1105.1654}{{\ttfamily arXiv:1105.1654
  [hep-ph]}}.

\bibitem{Borah:2019bdi}
D.~Borah, D.~Nanda, and A.~K. Saha, ``{Common origin of modified chaotic
  inflation, nonthermal dark matter, and Dirac neutrino mass},''
  \href{http://dx.doi.org/10.1103/PhysRevD.101.075006}{{\em Phys. Rev. D}
  {\bfseries 101} no.~7, (2020) 075006},
  \href{http://arxiv.org/abs/1904.04840}{{\ttfamily arXiv:1904.04840
  [hep-ph]}}.

\bibitem{Gruber:1999yr}
D.~Gruber, J.~Matteson, L.~Peterson, and G.~Jung, ``{The spectrum of diffuse
  cosmic hard x-rays measured with heao-1},''
  \href{http://dx.doi.org/10.1086/307450}{{\em Astrophys. J.} {\bfseries 520}
  (1999) 124}, \href{http://arxiv.org/abs/astro-ph/9903492}{{\ttfamily
  arXiv:astro-ph/9903492}}.

\bibitem{Bouchet:2008rp}
L.~Bouchet, E.~Jourdain, J.~Roques, A.~Strong, R.~Diehl, F.~Lebrun, and
  R.~Terrier, ``{INTEGRAL SPI All-Sky View in Soft Gamma Rays: Study of Point
  Source and Galactic Diffuse Emissions},''
  \href{http://dx.doi.org/10.1086/529489}{{\em Astrophys. J.} {\bfseries 679}
  (2008) 1315}, \href{http://arxiv.org/abs/0801.2086}{{\ttfamily
  arXiv:0801.2086 [astro-ph]}}.

\bibitem{Dev:2013yza}
P.~S. Bhupal~Dev, A.~Mazumdar, and S.~Qutub, ``{Constraining Non-thermal and
  Thermal properties of Dark Matter},''
  \href{http://dx.doi.org/10.3389/fphy.2014.00026}{{\em Front. in Phys.}
  {\bfseries 2} (2014) 26}, \href{http://arxiv.org/abs/1311.5297}{{\ttfamily
  arXiv:1311.5297 [hep-ph]}}.

\bibitem{Choi:2020kch}
G.~Choi, T.~T. Yanagida, and N.~Yokozaki, ``{Feebly interacting $U(1)_{\rm
  {B-L}} $ gauge boson warm dark matter and XENON1T anomaly},''
  \href{http://dx.doi.org/10.1016/j.physletb.2020.135836}{{\em Phys. Lett. B}
  {\bfseries 810} (2020) 135836},
  \href{http://arxiv.org/abs/2007.04278}{{\ttfamily arXiv:2007.04278
  [hep-ph]}}.

\bibitem{Essig:2013goa}
R.~Essig, E.~Kuflik, S.~D. McDermott, T.~Volansky, and K.~M. Zurek,
  ``{Constraining Light Dark Matter with Diffuse X-Ray and Gamma-Ray
  Observations},'' \href{http://dx.doi.org/10.1007/JHEP11(2013)193}{{\em JHEP}
  {\bfseries 11} (2013) 193}, \href{http://arxiv.org/abs/1309.4091}{{\ttfamily
  arXiv:1309.4091 [hep-ph]}}.

\bibitem{Ackermann:2012qk}
{\bfseries Fermi-LAT} Collaboration, M.~Ackermann {\em et~al.}, ``{Fermi LAT
  Search for Dark Matter in Gamma-ray Lines and the Inclusive Photon
  Spectrum},'' \href{http://dx.doi.org/10.1103/PhysRevD.86.022002}{{\em Phys.
  Rev. D} {\bfseries 86} (2012) 022002},
  \href{http://arxiv.org/abs/1205.2739}{{\ttfamily arXiv:1205.2739
  [astro-ph.HE]}}.

\bibitem{Cirelli:2009dv}
M.~Cirelli, P.~Panci, and P.~D. Serpico, ``{Diffuse gamma ray constraints on
  annihilating or decaying Dark Matter after Fermi},''
  \href{http://dx.doi.org/10.1016/j.nuclphysb.2010.07.010}{{\em Nucl. Phys. B}
  {\bfseries 840} (2010) 284--303},
  \href{http://arxiv.org/abs/0912.0663}{{\ttfamily arXiv:0912.0663
  [astro-ph.CO]}}.

\bibitem{kappadath}
K.~S. C., {\em {Ph. D. Thesis}}.
\newblock PhD thesis, University of New Hampshire, USA, 1998.

\bibitem{Strong:2004de}
A.~W. Strong, I.~V. Moskalenko, and O.~Reimer, ``{Diffuse galactic continuum
  gamma rays. A Model compatible with EGRET data and cosmic-ray
  measurements},'' \href{http://dx.doi.org/10.1086/423193}{{\em Astrophys. J.}
  {\bfseries 613} (2004) 962--976},
  \href{http://arxiv.org/abs/astro-ph/0406254}{{\ttfamily
  arXiv:astro-ph/0406254}}.

\bibitem{No:2019gvl}
J.~M. No, P.~Tunney, and B.~Zaldivar, ``{Probing Dark Matter freeze-in with
  long-lived particle signatures: MATHUSLA, HL-LHC and FCC-hh},''
  \href{http://dx.doi.org/10.1007/JHEP03(2020)022}{{\em JHEP} {\bfseries 03}
  (2020) 022}, \href{http://arxiv.org/abs/1908.11387}{{\ttfamily
  arXiv:1908.11387 [hep-ph]}}.

\bibitem{Chakdar:2015sra}
S.~Chakdar, K.~Ghosh, V.~Hoang, P.~Q. Hung, and S.~Nandi, ``{Search for mirror
  quarks at the LHC},''
  \href{http://dx.doi.org/10.1103/PhysRevD.93.035007}{{\em Phys. Rev. D}
  {\bfseries 93} no.~3, (2016) 035007},
  \href{http://arxiv.org/abs/1508.07318}{{\ttfamily arXiv:1508.07318
  [hep-ph]}}.

\bibitem{Chakdar:2020igt}
S.~Chakdar and P.~Q. Hung, ``{Prospect of the Electroweak Scale Right-handed
  neutrino model in the Lifetime Frontier},'' in {\em {International Conference
  on Neutrinos and Dark Matter}}.
\newblock 5, 2020.
\newblock \href{http://arxiv.org/abs/2006.00381}{{\ttfamily arXiv:2006.00381
  [hep-ph]}}.

\end{thebibliography}

\end{document}